# Toward nonthermal control of excited quantum materials: framework and investigations by ultrafast electron scattering and imaging

Xiaoyi Sun, Shuaishuai Sun, and Chong-Yu Ruan*

Department of Physics and Astronomy, Michigan State University, East Lansing, MI 48825, USA

*Corresponding author: ruanc@msu.edu

## Abstract

Quantum material systems upon applying ultrashort laser pulses provide a rich platform to access excited material phases and their transformations that are not entirely like their equilibrium counterparts. The addressability and potential controls of metastable or long-trapped out-of-equilibrium phases have motivated interests both for the purposes of understanding the nonequilibrium physics and advancing the quantum technologies. Thus far, the dynamical spectroscopic probes eminently focus on microscopic electronic and phonon responses. For characterizing the long-range dynamics, such as order parameter fields and fluctuation effects, the ultrafast scattering probes offer direct sensitivity. Bridging the connections between the microscopic dynamics and macroscopic responses is central toward establishing the nonequilibrium physics behind the light-induced phases. Here, we present a path toward such understanding by cross-examining the structure factors associated with different dynamical states obtained from ultrafast electrons scattering, imaging, and modeling. We give the basic theoretical framework on describing the non-equilibrium scattering problems and briefly describe how such framework relates to the out-of-equilibrium phenomena. We give effective models outlining the emergences of nonthermal critical points, hidden phases, and non-equilibrium relaxational responses from vacuum-suspended rare-earth tritellurides, tantalum disulfides thin films, and vanadium dioxide nanocrystalline materials upon light excitations.

## 1. Introduction

The interests for exploring light-induced new functional phases or properties of matters are motivated by practical endeavors, dubbed as materials on demand[1, 2], as a new research direction of quantum materials. The desire to harness the functionalities beyond the conventional metals or semiconductors has drawn significant attentions for exploring quantum materials[3]. The quantum materials, also referred to as strongly correlated electron materials, are featured by their complex phase diagrams where multiple electronic phases often occur adjacently. While such complexity is rooted in the active interactions between multiple microscopic degrees of freedom — lattice, charge, spin, and orbital — competing ground states are of macroscopic nature[2, 4] and their evolutions over the external control parameters can be discussed much without the detailed knowledge at the microscopic scales[5]. The physics of phase transitions can often be encapsulated in generalized order parameters that either explicitly breaks or implicitly connects to the symmetry-breaking processes. Much of the success in developing predicative models owes to phenomenological models based on the order parameter concept[6-8], which has been demonstrated in contexts[9] ranging from condensed matter physics to cosmology[10].

Much anticipated is the successful models for nonequilibrium phase transitions in the quantum materials. A number of important developments have emerged in recent years concerning this topic. One is the growing capabilities of the ultrafast techniques that gave increasingly more details of the nonequilibrium transformations between quantum phases; see for example the recent reviews[11-14]. Indeed, many recent ultrafast pump-probe studies of quantum matters also led to surprising results that cannot be identified from the equilibrium states[15-23], often referred to as the hidden state problem[24, 25]. Second, studying the nonequilibrium collective state evolution in quantum materials involves fundamental concept of nonequilibrium many-body physics[26-30]. Especially, understanding how a nonequilibrium system self-organizes into a broken-symmetry phase is a problem of broad interests from condensed matter[31-33] to high-energy physics[34-36]. It is widely believed that the generic responses of an isolated many-body system upon quench is to evolve towards the equilibrium state. However, before the system could fully equilibrate, novel behaviors may occur. The confluence of new experiments and theoretical frameworks has prompt synergistic developments. For example, the nonequilibrium phase transitions have also been intensively researched under controlled settings using the trapped cold atoms as the quantum material simulators[37-42].

In this article, we will attempt to establish a unified framework to treat ultrafast scattering from the nonequilibrium states of quantum materials. In this case, the system we refer to is the broken-symmetry order expressed in the lattice field with distinct order parameter that can be measured by the scattering approach. The pump-probe platform offers new opportunities for studying nonequilibrium physics. In the nonequilibrium physics context, the ultrafast light excitation couples to the system through a perturbation that changes the system parameters. More specifically, we ask how a quantum material containing long-range broken-symmetry states may effectively switch under an ultrafast "quench"[13, 43] enforced by laser pulse in routes distinctively different from a thermal state[44, 45]. In this central aspect, excited quantum material transformation is akin to the femtochemistry problem[46] where the ultrafast electronic excitation sets the new bonding landscape before the heavier molecular nuclear dynamics can follow. Given the separation of the timescale, the impulsive unveiling of the new potential energy landscape sets forth the ensuing molecule conformational dynamics where the dynamics of electrons follow those of the nuclei adiabatically. This scenario is referred to as the impulse-adiabatic approximation.

Scattering from a nonequilibrium system contains information on both the microscopic dynamics and macroscopic state evolution, and properly extracting such information in a controlled nonequilibrium quantum material will offer valuable insight on the symmetry-breaking properties and the nonequilibrium effects. Three prototypical quantum material systems: rare-earth tritellurides, tantalum disulfides, and vanadium dioxide will be discussed under this framework. We also examine the pump and material settings required for the controlled experiments. Facing the challenging issues with the multiscale dynamics, the technological aspects of the multi-messenger approaches based on a unified framework of ultrafast electron microscopy system will also be discussed. Our goals here are twofold. One is to understand the still mysterious hidden phase phenomena in nonequilibrium quantum materials. The second is to explore the ideas of using the light-excited quantum material as a platform to study the nonequilibrium physics.

## 2. Description of non-equilibrium phase transition

Quantum phases are macroscopic states that exist at a finite temperature with quantum mechanism in origin but often behave semi-classically[2, 4]. They often emerge by breaking the existing symmetry of the underpinning Hamilton defining the microscopic states[8]. In doing so, they distinguish themselves from the microstate evolution, and the collective state properties typically are described by a very small number of long-scale order parameters, in which the microscopic dynamics are coarse-grained[9]. Ultrafast pump-probe approaches utilize the temporal resolution and pump control to explore hidden or transient metastable phases as a means to unveil the underpinning complex landscape of macroscopic quantum phases and to study the nonequilibrium physics.

Based on defining a local order parameter, the Landau-Ginzburg mean-field theory accurately captures a system undergoing a phase transition in which some symmetry is broken[6, 7]. The scattering techniques, which can provide direct evidence of the symmetry change and the properties of the order parameters, have been instrumental for the success of developing Landau-Ginzburg theory for quantum materials[6, 47-49]. The main focus of this article is to examine how nonequilibrium quantum materials involving multiple broken-symmetry ground states evolve upon applying laser quench. The process typically involves nonequilibrium state of the order parameters. Ultrafast electron scattering is used as a sensitive probe to characterize the dynamical order parameter fields.

For the experimental examples discussed in this paper, the macroscopic systems we refer to are the broken-symmetry orders expressed in the lattice field with distinct order parameter. In these cases, the phase evolution is characterized by the long-wave responses on much greater length scale than the periodicity of the mean atomic positions. The scale difference allows different physical principles governing the order parameters and the microscopic dynamics to be separated. For example, macrostate evolution is decoupled dynamically from the local vibrational excitations around the mean positions. Therefore, it is natural for the impulse-adiabatic description of the femtochemistry problems to apply to the nonequilibrium phase transition of quantum materials under the ultrafast quench – a paradigmatic scenario has been given for the photoinduced phase transition (PIPT) problems concerning the strongly correlated molecular solids; see the review[25]. Here, the Landau-Ginzburg free-energy equation serves to describe the long-scale property, i.e., a general functional of the coarse-grained order parameter, which is registered in coherent scattering structural factor; whereas the microscopic processes are encoded in the diffuse and inelastic scattering.

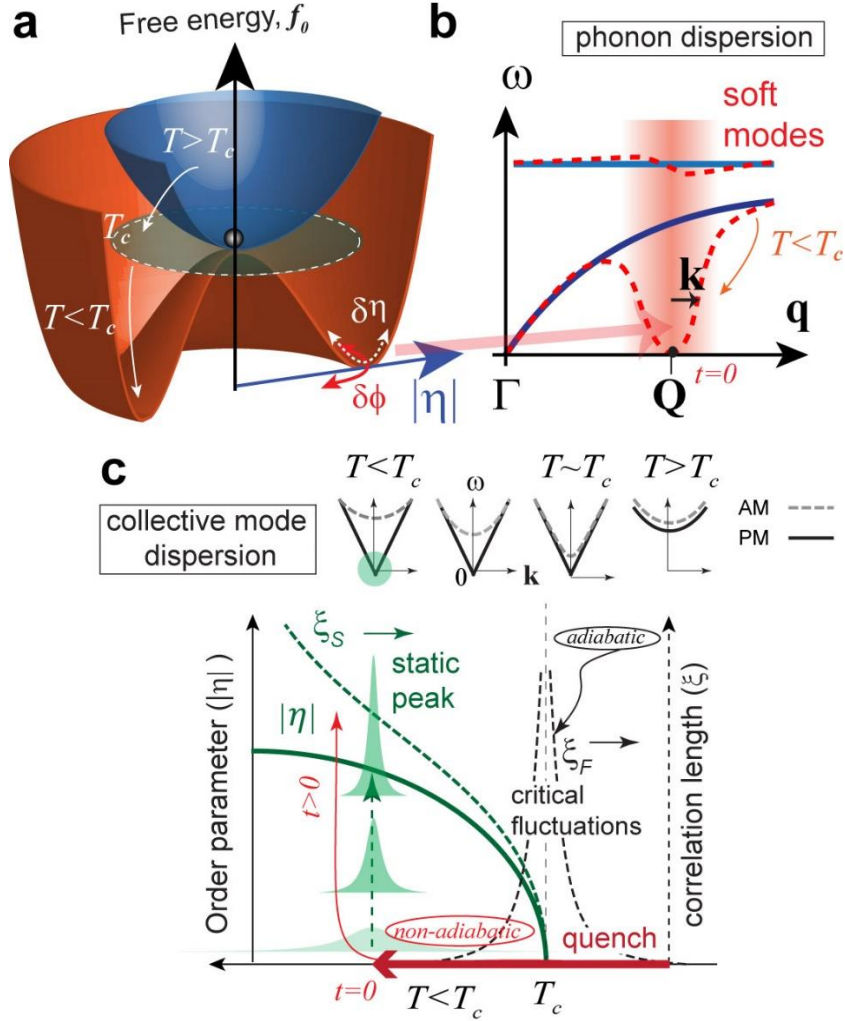

**Figure 1. Landau-Ginzburg free-energy surface for spontaneous symmetry breaking and the order parameter dynamics. a**. The 2D free-energy landscape that defines the order parameter field at different temperature. The arrow directions show temperature quench. The field fluctuations are depicted in the change of the amplitude ($\delta\eta$) and the phase ($\delta\phi$). **b**. The corresponding lattice phonon dispersion curves that couple to the landscape modification. The lattice softening, driven by the temperature quench, occurs at momentum wavevector **Q** of the long-range state. **c**. The phase ordering kinetics orchestrated by the fluctuation fields. The near-equilibrium scenario is depicted in black curves. The nonequilibrium one, depicted in green, is driven by a deep quench, where $T \ll T_c$. The dispersion curves for the amplitude (AM) and phase mode (PM) of the CDW state are depicted at the top. Adapted with permission from ref. [23].

For illustrating the correspondence between the order-parameter field and the scattering functions, we start with the case of a singular order parameter $\eta = |\eta|e^{i\theta}$ associated with a continuous phase transition. Typically, only the amplitude and its gradient will change the free-energy density, written as:

$$f = \frac{1}{2}a(T - T_c)|\eta|^2 + \frac{1}{4}A_4|\eta|^4 + \frac{1}{2}\alpha_{ij}\frac{\partial \eta}{\partial x_i}\frac{\partial \eta}{\partial x_j}. \quad (1)$$

For a uniform order parameter, the potential surface described by the first two terms on the righthand side (RHS) has the stationary points that give the coordinate for the broken-symmetry state ($|\eta| > 0$). For the broken-symmetry state typically residing at low temperature, $a$ and $A_4$ are positive. With the phase rigidity tensor $\alpha_{ij} > 0$, the third term raises the free energy for a system being inhomogeneous. This simple mean-field depiction captures the universal laws governing the system when it approaches the critical point, with the temperature difference $|T_c - T|$ as the control parameter. For $T < T_c$, $f_{\eta\eta} = 2a(T - T_c)$, i.e., approaching the broken-symmetry state from above the free energy changes from a parabolic uphill to a double well potential. The nonanalyticity at $T_c$ and the ensuing BCS-type onset of the order parameter are given by following the stationary point as a function of temperature:

$$|\eta(T)| = \sqrt{\frac{a}{A_4}}(T_c - T)^{1/2}. \quad (2)$$

The Landau-Ginzburg equation also gives the right description for the pre-transitional instabilities probed by the diffuse scattering, which provides additional information about the phase rigidity. To see this, we express the long-range parameter variations with the fluctuation waves components:

$$\delta\eta(r) = \sum_{\mathbf{k}} \eta_{\mathbf{k}} e^{i\mathbf{k}\cdot\mathbf{r}}, \quad (3)$$

where $\eta_{\mathbf{k}}$ and $\mathbf{k}$ represent the amplitude and the momentum wavevector. The least work required to produce the variation $\delta\eta$ according to Eqn. (1) is

$$\delta\eta(\mathbf{r}) = \frac{1}{2}f_{\eta\eta}(\delta\eta)^2 + \frac{1}{2}\alpha_{ij}\frac{\partial f}{\partial x_i}\frac{\partial f}{\partial x_j}. \quad (4)$$

We can calculate the mean-square Fourier component of the fluctuation wave at momentum $\mathbf{k}$:

$$\langle \eta_{\mathbf{k}} \eta_{\mathbf{k}}^* \rangle = \frac{k_B T}{V}\left(\frac{1}{f_{\eta\eta} + \alpha_{ij}k^2}\right). \quad (5)$$

Eqn. (5) gives the Lorentzian line shape of the anomalous spectrum of diffuse scattering near $T_c$. From Eqn. (1) and (2), we establish for $T > T_c$, $f_{\eta\eta} = a(T - T_c)$. Rearranging Eqn. (5), we obtain $\langle \eta_{\mathbf{k}} \eta_{\mathbf{k}}^* \rangle = \frac{k_B T}{V\alpha_{ij}}\left(\frac{1}{(\Delta k_L)^2 + k^2}\right)$ for $T > T_c$, with the Lorentz line width $\Delta k_L = \left(\frac{a(T - T_c)}{\alpha_{ij}}\right)^{1/2}$.

Thus far, only the statics of the Landau-Ginzburg equation is discussed. For considering the nonequilibrium phase transition, connections to the microscopic physics must be made. To give a microscopic picture, we consider a single-wavevector charge-density wave (CDW) system[50]. Here, the specific ordering formation is by breaking the translational symmetry over a specific wave-vector $\mathbf{Q}$, typically driven by the electronic instabilities at the relevant length scale ($2\pi/\mathbf{Q}$) coupled to the lattice field[51]. In this case, the order parameter has the amplitude and phase components and can be written as $\eta = |\eta|e^{i(\mathbf{Q}\cdot\mathbf{r} + \phi)}$ with $|\eta|$ and $\phi$ representing the amplitude and phase fields. The order parameter fields can be probed via their connection to the distorted lattice, or lattice periodic distortion wave (LDW). At each lattice site $\mathbf{L}$, the distorted amplitude is described by

$$\mathbf{u}_{\mathbf{L}}(r) = u_0 \hat{\mathbf{e}} \sin(\mathbf{Q}\cdot\mathbf{L} + \phi), \quad (6)$$

where $\hat{\mathbf{e}}$ and $u_0$ represent the polarization vector and amplitude. The order parameter and LDW are related by $u_0 = |\eta| A_{\eta u}$ with $A_{\eta u}$ a constant that can be probed under the steady state prior to applying quench, where the order parameter $|\eta|$ is typically set to be 1 by convenience.

We first give a hypothetical scenario of a swift system-wide temperature quench to below $T_c$ to drive spontaneous phase ordering. For a complex order parameter, this process is now described on a two-dimensional (2D) landscape of the order parameter fields (amplitude and phase); see Fig. 1a. Driving the order parameter evolution is the free-energy potential which unfolds from the uphill into the Mexican-hat shape of the broken-symmetry phase. After choosing a phase, the amplitude ($|\eta|$) dynamics is deterministic as a ball falling from the top of the hill to a location in the trough, i.e. a spontaneous symmetry breaking (SSB). However, nonequilibrium scenarios appear with the underlying separation of scales in the physical CDW system. This can be understood by mapping the unfolding energy landscape involving the long-wave state evolution to the corresponding changes in the lattice field. The relevant lattice dynamics are governed by the momentum-dependent lattice potential expressed in the phonon dispersion curves (Fig. 1b), which shift from those of the normal ($T > T_c$; solid lines) state to the broken-symmetry phase ($T < T_c$; dashed lines) with a mode softening at phonon momentum wavevector $\mathbf{q} \sim \mathbf{Q}$ which may be probed via the inelastic and diffuse scattering spectrum[52-59]. As the lattice potential changes leveled at the electronic scale is assumed to be nearly instantaneous, the soft modes in the critical regime (colored in red where the dispersion curves drop in frequency, ω) cannot respond adiabatically. This inherent nonadiabicity between the potential energy shift and the long-wave soft collective mode response is prominent for any photoinduced phase transition driven by a quench[23].

Two types of collective modes are involved in the phase change dynamics[50, 60]. One is the amplitude modes (AM), represented by the amplitude fluctuation $\delta|\eta|$ as depicted in Fig. 1a. The AM would be generally gapped within the broken-symmetry ground state as it costs energy to increase or decrease the amplitude $|\eta|$. Meanwhile, the incommensurate CDW state also hosts the ungapped phase mode (PM)[50], as also depicted in Fig. 1a where the level of the free energy surface does not change over phase variation, $\delta\phi$; in this case PM is referred to as the Goldstone mode[50, 61-63]. In Fig. 1c, we depict the dispersion curves for the AM and PM modes as a function of the temperature. It is easy to see that the AM would be always gapped with the exception at the critical point, whereas the PM becomes ungapped after establishing the new broken symmetry phase. These soft modes are dynamically connected with the field instabilities only at the bottom of the free energy. The asymptotic slope of the dispersion curve determines the 'sound speed' of the fluctuation wave.

In the ultrafast scattering experiments, the order parameter is mapped into the LDW amplitude and probed via the time-dependent two-point equal-time correlation function:

$$S_\eta(\mathbf{r} - \mathbf{r}'; t) \sim \langle u(\mathbf{r}; t) u(r'; t) \rangle, \tag{7}$$

where the bracket denotes the spatial and ensemble averaging over the probed volume and the acquisition time window. The Fourier transformation of the Eqn. (7) gives the structure factor $S_\eta(\mathbf{q}; t)$ which results in satellites for the ordered states in the reciprocal space. As will be discussed in Sec. 4, Eqn. (5) gives the diffuse scattering component of $S_\eta(\mathbf{q}; t)$. Hence, we can extract the $\xi_F$, the correlation length for the critical fluctuations, from the line shape of the diffuse scattering, i.e. $\xi_F = \frac{1}{\Delta k_L}$, and deduce the rigidity based on the $T$-dependence:

$$\frac{\alpha_{ij}}{a} = (T - T_c)\xi_F^2 \tag{8}$$

It is instructive to point out that the bandwidth of the phonon softening probed by inelastic scattering is typically much larger than the anomalous linewidth of the diffuse scattering. This is because only those stochastic soft-phonon modes near the CDW **Q** vector will eventually condense into the static CDW order. The order parameter field fluctuations may be described over the Mexican hat energy surface.

These long-wave and low-energy excitations are considered as the hydrodynamic modes[61] or fluctuation waves[49], with wavelength $\lambda = 2\pi/|\mathbf{k}|$, where typically the momentum wavevector $|\mathbf{k}| \ll |\mathbf{Q}|$, the wave vector of the CDW/LDW. Such field fluctuations must be created by joining two soft phonon modes from the lattice field. It has been shown by Overhauser[50, 62] that to form a collective mode with momentum **k**, either AM or PM, two soft phonons with momentum $\mathbf{q} = \mathbf{k} \pm \mathbf{Q}$, must be coherently jointed[50]; conversely, to quench the fluctuating order parameters, the excited collective mode decay by dissociating into a pair of phonons. The dissociation process would be most relevant in discussing the overdamped dynamics following the impulsive suppression of a pre-existing CDW order.

Now we concern how the order parameter of the nonequilibrium state will manifest physically. We expect in potential-driven dynamics the system will be attracted to a minimum energy basin. The simplest nonequilibrium dynamical model is relaxational for a non-conserved order parameter $\eta(r; t)$, such as the CDW system and with a stochastic contribution, often referred to as time-dependent Ginzburg-Landau equation [9, 27, 64]:

$$\frac{\partial \eta}{\partial t} = -\Gamma \frac{\delta f}{\delta \eta} + \varsigma = -\Gamma \left[ \frac{\delta f_0}{\delta \eta} - \alpha \nabla^2 \eta \right] + \varsigma(\mathbf{r}, t). \tag{9}$$

Here $f_0$ is the effective potential energy (excluding the gradient term), $\Gamma$ is a relaxation constant and $\varsigma(\mathbf{r}, t)$ is the noise source. One can often take the noise as random and for a Gaussian white noise $\langle \varsigma(\mathbf{r}, t) \rangle = 0$ and the noise correlator $\langle \varsigma(\mathbf{r}, t) \varsigma(\mathbf{r}', t') \rangle = 2T\Gamma \delta(\mathbf{r} - \mathbf{r}') \delta(t - t')$ [9, 27].

We can now connect this expression to the phenomenological description of SSB. Under a deep quench with the eventual $T \ll T_c$, a large shift of the local curvature of the potential is induced as depicted in Fig. 1a. In the initial amplitude dynamics, the first term on the RHS is most important. However, the phase space of the order parameter is broadened by its coupling to the stochastic background (the 3rd term), but only in the soft phonon mode regions due to the scale matching. The nonlinear coupling to phonons leads to a rectification allowing the amplitude mode to build up as the system move from the hill to the low-energy basin. When reaching the bottom of the basin where $\frac{\delta f_0}{\delta \eta} \sim 0$, the second term will now become more important. This means the nonequilibrium system will undergo pattern formation, driven by the positive rigidity favoring long-range ordering. The Eqn. (9) can be simplified to[65]

$$\frac{\partial \eta}{\partial t} = D \nabla^2 \eta + \varsigma, \tag{10}$$

where $D = \Gamma \alpha$ is the diffusion constant[27] and defines a characteristic timescale $t_D \sim \frac{\xi_s^2}{D}$ for the coarsening where $\xi_s$ is the size of the coherent domain created. It is easy to see that in this case the relaxational dynamics for the coarsening follows a universal scaling law[66, 67]. The size of the domain increases at velocity $\sim \frac{\partial \xi_s(t)}{\partial t} = \frac{2D}{\xi_s(t)}$. This implies that the characteristic length scale $\xi_s(t) = \sqrt{2Dt}$ grows as $t^{1/2}$[66].

In the following, we will go beyond the hypothetic scenario of temperature quench. The temperature quench is inherently impractical to implement as such a process is physically slow and is prone to generate inhomogeneous phase ordering due to the presence of interfaces for cooling the system from outside. The more effective approach to implement the physical quench is instead via driving the system interaction parameters rather than changing the temperature of the system[43]. This will require additional order parameter(s) to couple with the present one – a scenario of cooperativity or competitions described in a multi-parameter free-energy equation, which is in fact a characteristic feature of quantum material phase transitions[31, 32, 68, 69]. As will be discussed in the experimental studies, rich scenarios as those involve competitive broken-symmetry orders, vestigial orders, and the intertwined ground states could be identified, leading to intriguing nonequilibrium states and hidden phases; see Sec. 6.

## *3. Light-induced hidden phases through competitions*

While one expects that there should always be a corresponding critical threshold directly linked to a thermal phase transition according to Eqn. (2), there have been many reports that the experimentally identified thresholds are significantly less than the thermodynamic requirements. In many cases, an additional low energy threshold is linked to the appearance of a light-induced hidden state without entirely melting the pre-existing ordered phase. Indeed, the threshold behavior with a sub-thermal activation energy density has been the hallmarks of PIPT phenomena[25].

We now consider a multi-component free-energy surface in which the order parameters are coupled to account for phase transitions where different broken-symmetry orders coexist or compete in quantum materials. In particular, in strongly correlated electron systems, the competing orbital, spin, and lattice interactions yield a multiplicity of nearly degenerate broken-symmetry phases and complex phase diagrams[2, 11, 70, 71]. For example, within the high-critical temperature superconductors, such as iron-based and cuprates, superconductivity is often found to compete with a density-wave order. A prototypical case concerning the competitive SSB within the density-wave systems is recently discussed and observed in the rare-earth telluride compounds[22, 23].

A generic Landau-Ginzburg expression that applies to the broad range of physical phenomena concerning competitions is [31, 72, 73]

$$f_0(\eta_1, \eta_2,...) = \frac{1}{2}\sum_i a_i(T^{(i)} - T_{c,i})|\eta_i|^2 + \frac{1}{4}\sum_i A_{4,i}|\eta_i|^4 + \frac{1}{2}\sum_{ij}\tilde{A}_{ij}|\eta_i|^2|\eta_j|^2, \quad (11)$$

where the first two terms represent the free energy from the individual phase that will undergo SSB at the respective critical temperature $T_{c,i}$. The third term gives the coupling energy where a competitive scenario has $\tilde{A}_{ij} > 0$. One can apply this multi-component potential energy surface to Eqn. (9) to determine the order parameter dynamics. Of concern here is the nonequilibrium phase competitions driven predominantly by the laser interaction quench. Generally, the stochastic effects propagating to the long-range degree of freedom are slow to manifest. The microscopic physics depends on 'local' momentum-dependent coupling between the excited hot carriers and the lattice modes, to establish the initial bath that is expected to be inhomogeneous in effective temperatures. Nonetheless, we expect the post-quench immediate response to be potential-driven. We first discuss here the truly competitive SSB scenario, applying to the rare-earth tritelluride system[23, 73], in which $T_c$ is shared among the two order parameters $\eta_1$ and $\eta_2$.

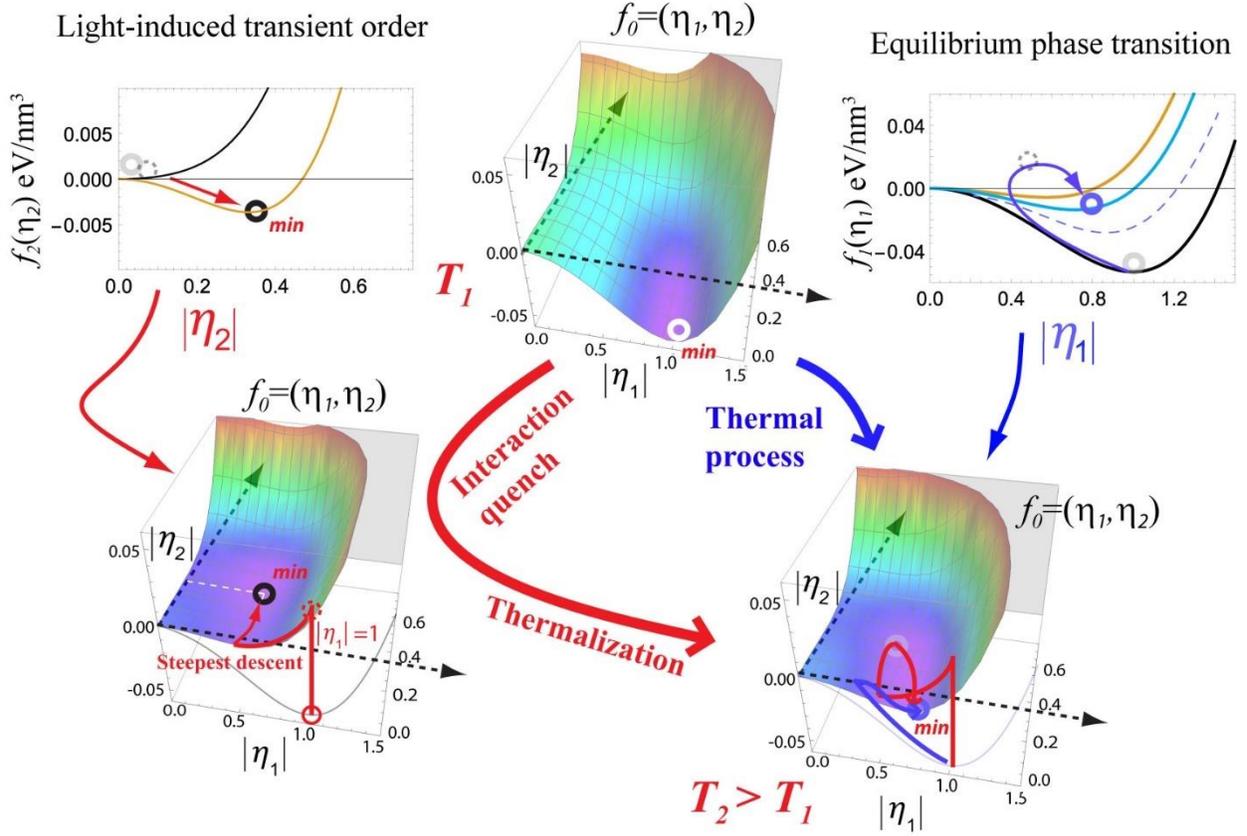

**Figure 2. Light-induced changes of the Landau-Ginzburg free-energy landscape involving two order parameters $\eta_1$ and $\eta_2$.** The non-equilibrium pathway (depicted in the left route) involves the formation of a bi-directional hidden state with order parameter $\eta_2$ when the repulsive coupling potential is suppressed by a swift reduction of the preeminent order parameter $\eta_1$. In contrast, in the thermal pathway (right) the global free energy minimum remains in the condition where $|\eta_2| = 0$, namely the CDW is uniaxial. Adapted with permission from ref. [23].

While much of the ordering dynamics involving complex order parameters will depend on the microscopic details of the experimental settings, here we focus on the phenomenology of amplitude dynamics, i.e. $|\eta_1|$ and $|\eta_2|$. The competition here precludes the thermal phase of $|\eta_2|$ from appearing below $T_c$ if $|\eta_1|$ is selected by SSB. This is described in Fig. 2, where $f_0$ is expressed in two dimensions along $|\eta_1|$ and $|\eta_2|$. For a model simulation the parameters here are chosen for $|\eta_1| = 1$ to be the sole minimum at an initial temperature $T_1$ (black curve), before applying laser quench. Within the reversible thermodynamic pathway, i.e. via adiabatically tuning the temperature that is homogeneous system-wide, the $|\eta_1|$ remains the dominate phase as represented by the sole minimum basin in the global energy surface. This can be seen by taking the derivative over $|\eta_1|$, which we assume to be the thermodynamically preferred state at the low temperature. We then obtain $|\eta_1| = \sqrt{(a_1(T_c - T) - \tilde{A}|\eta_2|^2)/A_{4,1}}$, i.e. $|\eta_1|$ naturally takes a non-zero value when $|\eta_2| = 0$ at $T < T_c$. This thermal route is depicted in blue on the right side of Fig. 2. In the model calculation, $|\eta_1|$ reaches a new minimum at 0.6 (brown curve) due to heating. Conversely, a stable new static broken-symmetry order (i.e., $\langle \eta_2 \rangle > 0$) could appear at $T < T_c$ if $a_2(T - T_c) + \tilde{A}|\eta_1|^2$ becomes negative. To achieve this naturally, one needs to supply an amplitude quench of the initial order to a level beyond:

$$|\eta_{th}| = \sqrt{\frac{a_2(T_c - T)}{\tilde{A}}}, \tag{12}$$

namely a threshold behavior emerges. However, by requiring $|\eta_1|$ to be dominant, i.e. $\frac{a}{A_4} > \frac{a'}{\tilde{A}}$, one can see this is prohibited if the quenched system were to maintain the thermal equilibrium condition. Therefore, the light-induced hidden state scenario shall occur in a nonequilibrium process in which one can suppress $|\eta_1|$ transiently through an interaction quench that initially couples only to a subset of lattice modes.

This scenario is described on the left side of Fig. 2. A nonthermal quench of $|\eta_1|$ to ~ 0.5 leads to a minimum in the subspace of the free energy for $|\eta_2|$ at ~0.36 (see the black circle). Here, we assume the field of the new order parameter, being decoupled from the high-temperature bath created by quench, has a temperature of the ambient. Meanwhile, the order parameter field of the present order, being driven directly by the quench, has a higher local temperature, i.e., $T^{(2)} \ll T^{(1)}$ in the initial period. Following this, it is possible for the new order parameter to undergo a threshold onset at a critical laser fluence $F_c$, which is much smaller than what is required to completely melt the existing order [19, 23]. However, if the system were to regain thermal equilibrium, i.e. $T^{(2)} \sim T^{(1)}$, then one can show the free energy again has the global minimum over a non-zero $|\eta_1|$, i.e. the hidden state is removed from the order parameter field when the system thermalizes.

## 4. Mapping the dynamical order parameter with ultrafast scattering

Optical excitations can result in a departure from the thermal responses making the modeling of the scattering factor, Debye-Waller effects, and the diffuse signatures quite different from the conventional approaches for the equilibrium states. In this section, we focus on understanding the different lattice responses to the nonequilibrium excitation that go into the structure factor function. The analyses here are based on momentum-resolved structure factor $S(\mathbf{q}, t)$ or their momentum integration $m(\mathbf{Q}, t)$, which can be directly compared with the ultrafast electron scattering experiments. The goals are to derive directly from the experimental measurements the nonequilibrium systems, including the evolution of the static order parameters and the associated order-parameter field fluctuations of the quantum phases. However, in a nonequilibrium phase transition, such as SSB driven by a quench, these long-range parameters are rooted in the microscopic excitations. While there has been extended literature dedicated to the scattering by the phonon processes[74, 75] and by the long-ranged broken-symmetry state with periodic lattice distortions[50, 52, 76], the discussions were often partial to one aspect only and aimed at the near-equilibrium processes. The goal here is to lay out the scattering formalism to treat the two as well as the quasi-static order parameter evolution in a self-contained framework.

The nonequilibrium phases created by the impulsive excitation involve two types of lattice dynamical effects: the incoherent microscopic responses of the lattice, i.e. the phonons, and the long-range parameter fluctuation waves of the cooperative states. To be more specific, we refer to these two types of excited lattice wave manifolds as $\{\mathbf{q}\}$ and $\{\mathbf{k}\}$ based on their respective momentum wavevectors. The signatures from the soft modes that belong in $\{\mathbf{q}\}$ and the those of the collective modes that belongs in $\{\mathbf{k}\}$ can be differentiated experimentally. In the spectroscopy experiments, because of the momentum conservation, the probe photon does not directly couple to the soft mode at a finite momentum $\mathbf{q}$. However, it can couple to the collective modes at the long-wavelength limit: $\mathbf{k} = \mathbf{q} - \mathbf{Q} \to 0$. In contrast, in the scattering experiments, one detects both effects at different parts of the momentum space. We shall attempt to capture these two types of lattice excitations in the scattering formalism. In ultrafast scattering, one probes the time-dependent structural factor around the reciprocal lattice $\mathbf{G}_{hkl}$ at the momentum transfer wavevector, $\mathbf{s} = \mathbf{q} + \mathbf{G}_{hkl}$:

$$S(\mathbf{s}, t) = \int e^{-i\mathbf{s}\cdot(\mathbf{r}-\mathbf{r}')} \langle \rho(\mathbf{r},t)\rho(\mathbf{r}',t) \rangle \, d\mathbf{r}d\mathbf{r}', \qquad (13)$$

which is the Fourier counterpart of the correlation function $\langle \rho(\mathbf{r},t)\rho(\mathbf{r}',t) \rangle$ $\langle ... \rangle$ denoting spatial and ensemble-averaging of the probed volume set by the experimental conditions. One can show for a sufficiently long-ranged order parameter on a periodic lattice $\mathbf{L} = n_1\mathbf{a}_1 + n_2\mathbf{a}_2 + n_3\mathbf{a}_3$ with $n_1, n_2, n_3$ being integers $(0, \pm 1, \pm 2 \ldots)$, the integral is simplified into

$$S(\mathbf{s}, t) = \langle F(\mathbf{s}, t) F^*(\mathbf{s}, t) \rangle,$$

where $F(\mathbf{s}, t) = \int \sum_\mathbf{L} f_\mathbf{L} \delta(\mathbf{r} - \mathbf{L} - \mathbf{u}_L(t)) e^{-i\mathbf{s}\cdot\mathbf{r}} d\mathbf{r} = \sum_\mathbf{L} f_\mathbf{L} e^{-i\mathbf{s}\cdot(\mathbf{L} + \mathbf{u}_L(t))}$ is the Fourier spectrum of the lattice. The term $f_\mathbf{L}$ equal to $\sum_i \rho_i e^{-i\mathbf{s}\cdot\boldsymbol{\varrho}_i}$ represents the unit cell scattering form factor where $\boldsymbol{\varrho}_i$ is the mean unmodified relative position of the atom in the unit cell. Here, $\mathbf{u}_L(t)$ is the displacement from the mean position at each lattice site, which may originate either from the long-range parameter or fluctuations due to phonons or collective modes.

More specifically here, we consider a single long-range parameter $\eta$ present in the system, e.g. introduced by a charge-density or orbital order (Eqn. (6)). We write down the general form of $\mathbf{u}_\mathbf{L}$ considering phonons $\mathbf{u}_q(\mathbf{r},t) = \sum_\mathbf{q} u_{\mathbf{q},0} \hat{\mathbf{e}}_\mathbf{q} sin(\mathbf{q}\cdot\mathbf{r} - \omega_\mathbf{q} t + \phi_\mathbf{q})$ and the contributions from the LDW $\mathbf{u}_\eta(\mathbf{r}) = u_{0,\eta} \hat{\mathbf{e}}_\eta sin[\mathbf{Q}\cdot\mathbf{r} + \phi_\eta]$, with $u_{i,0}$, $\hat{\mathbf{e}}_i$ and $\phi_i$ denoting the respective amplitude, polarization directional vector and phase. For describing the fluctuations of the order parameter, we include in $\mathbf{u}_\eta$ the associated collective excitations in terms of the phase and amplitude fluctuations in $\delta\phi(r,t) = \sum_k \phi_{0,k} sin(\mathbf{k}\cdot\mathbf{r} - \omega_\mathbf{k} t)$ and $\delta\hat{A}(\mathbf{r},t) = \sum_{\mathbf{k}'} \hat{A}_{0,\mathbf{k}'} sin(\mathbf{k}'\cdot\mathbf{r} - \omega_{\mathbf{k}'} t)$. We then have $\mathbf{u}_q(\mathbf{r},t) = \sum_\mathbf{q} u_{\mathbf{q},0} \hat{\mathbf{e}}_\mathbf{q} sin(\mathbf{q}\cdot\mathbf{r} - \omega_\mathbf{q} t + \phi_\mathbf{q})$ for the phonons, and $\mathbf{u}_\eta(r,t) = u_{0,\eta} \hat{\mathbf{e}}_\eta \left(1 + \delta\hat{A}(\mathbf{r},t)\right) sin[\mathbf{Q}\cdot\mathbf{r} + \delta\phi(\mathbf{r},t)]$ for the displacement associated with long-range parameter. Following this, the overall displacement at each lattice site is

$$\mathbf{u}_\mathbf{L}(t) = \sum_\mathbf{q} \mathbf{u}_\mathbf{q}(\mathbf{L},t) + u_{0,\eta}\hat{\mathbf{e}}_\eta \left(1 + \delta\hat{A}(\mathbf{L},t)\right) sin[\mathbf{Q}\cdot\mathbf{L} + \delta\phi(\mathbf{L},t)] \qquad (14)$$

Eqn. (14) has all key ingredients to describe the dynamics of the excited states with soft modes and collective modes in a broken-symmetry state. To simplify the derivation without losing generality, we first drop the amplitude fluctuations $\delta\hat{A}$ and expand the equation based on the momentum-dependent displacement $\mathbf{u} = \mathbf{u}_\mathbf{q} + \mathbf{u}_\mathbf{k}$, with the order-parameter fluctuations dominated by the phase modes[50]. In this case, the distribution function is written as:

$$\rho(\mathbf{r},t) = \sum_\mathbf{L} \delta\{\mathbf{r} - \mathbf{L} - \sum_\mathbf{q} u_\mathbf{q} \hat{\mathbf{e}}_\mathbf{q}(\mathbf{L},t) - u_\eta \hat{\mathbf{e}}_\eta sin(\mathbf{Q}\cdot\mathbf{L} + \sum_\mathbf{k} \phi_\mathbf{k} sin(\mathbf{k}\cdot\mathbf{L} - \omega_\mathbf{k} t))\} \quad (15)$$

The system form factor is given:

$$F(\mathbf{s},t) = \int \rho(\mathbf{r},t) e^{-i\mathbf{s}\cdot\mathbf{r}} d\mathbf{r}$$

$$= \sum_{\mathbf{L}} f_{\mathbf{L}} e^{-i\mathbf{s}\cdot\left(\mathbf{L}+\sum_{\mathbf{q}} u_q \hat{\mathbf{e}}_{\mathbf{q}}(\mathbf{L})+u_\eta \hat{\mathbf{e}}_\eta sin(\mathbf{Q}\cdot\mathbf{L}+\sum_{\mathbf{k}} \phi_{\mathbf{k}} sin(\mathbf{k}\cdot\mathbf{L}-\omega_{\mathbf{k}}t))\right)}$$

$$= \sum_{\mathbf{L}} f_{\mathbf{L}} e^{-i\mathbf{s}\cdot\mathbf{L}} \left\{ e^{-i\mathbf{s}\cdot\left(\sum_{\mathbf{q}} u_q \hat{\mathbf{e}}_{\mathbf{q}}(\mathbf{L})\right)} \right\} \left\{ e^{-i\mathbf{s}\cdot u_\eta \hat{\mathbf{e}}_\eta sin(\mathbf{Q}\cdot\mathbf{L}+\sum_{\mathbf{k}} \phi_{\mathbf{k}} sin(\mathbf{k}\cdot\mathbf{L}-\omega_{\mathbf{k}}t))} \right\}. \quad (16)$$

The first bracket on RHS is simply the scattering by lattice phonons. We have

$$e^{-i\mathbf{s}\cdot\left(\sum_{\mathbf{q}} u_q \hat{\mathbf{e}}_{\mathbf{q}}(L)\right)} = \prod_{\mathbf{q}} e^{-i\mathbf{s}\cdot\mathbf{u}_{\mathbf{q},0} sin(\mathbf{q}\cdot\mathbf{L}-\omega_q t)}$$

Using the Jacobi-Anger generating function:

$$e^{-izsin\phi} = \sum_{n=0}^{\infty} e^{-in\phi} J_n(z), \quad (17)$$

we have

$$e^{-i\mathbf{s}\cdot\left(\sum_{\mathbf{q}} \mathbf{u}_{\mathbf{q}} \hat{\mathbf{e}}_{\mathbf{q}}(\mathbf{L})\right)} = \prod_{\mathbf{q}} \sum_{l} e^{-il(\mathbf{q}\cdot\mathbf{L}-\omega_q t)} J_l(\mathbf{s}\cdot\mathbf{u}_{\mathbf{q},0})$$

The second bracket on RHS gives the scattering from the order parameter static wave with phase fluctuations. By twice applying the Jacobi-Anger generating function, we can derive

$$e^{-i\mathbf{s}\cdot u_{\eta,0}\hat{\mathbf{e}}_\eta sin(\mathbf{Q}\cdot\mathbf{L}+\sum_{\mathbf{k}}\phi_{\mathbf{k}} sin(\mathbf{k}\cdot\mathbf{L}-\omega_{\mathbf{k}}t))} = \left\{\sum_m e^{-im\mathbf{Q}\cdot\mathbf{L}} J_m(\mathbf{s}\cdot\mathbf{u}_{\eta,0})\right\} e^{-i\sum_{\mathbf{k}} m\phi_{\mathbf{k}} sin(\mathbf{k}\cdot\mathbf{L}-\omega_{\mathbf{k}}t)}$$

$$= \left\{\sum_m e^{-im\mathbf{Q}\cdot\mathbf{L}} J_m(\mathbf{s}\cdot\mathbf{u}_{\eta,0})\right\} \left\{\prod_{\mathbf{k}} \sum_n e^{-in(\mathbf{k}\cdot\mathbf{L}-\omega_{\mathbf{k}}t)} J_n(m\phi_{\mathbf{k}})\right\}$$

$$= \prod_{\mathbf{k}} \sum_{m,n} e^{-i(m\mathbf{Q}+n\mathbf{k})\cdot\mathbf{L}} e^{in\omega_{\mathbf{k}}t} J_m(\mathbf{s}\cdot\mathbf{u}_{\eta,0}) J_n(m\phi_{\mathbf{k}})$$

Putting together, we obtain

$$F(\mathbf{s},t) = \sum_{\mathbf{L}} f_{\mathbf{L}} \left\{ \prod_{\mathbf{q}} \sum_l e^{-i(\mathbf{s}+l\mathbf{q})\cdot\mathbf{L}} e^{il\omega_q t} J_l(\mathbf{s}\cdot\mathbf{u}_{\mathbf{q},0}) \prod_{\mathbf{k}} \sum_{m,n} e^{-i(m\mathbf{Q}+n\mathbf{k})\cdot\mathbf{L}} e^{in\omega_{\mathbf{k}}t} J_m(\mathbf{s}\cdot\mathbf{u}_{\eta,0}) J_n(m\phi_{\mathbf{k}}) \right\}$$

With the understanding that each $\mathbf{q}$ and $\mathbf{k}$ component is distinct and incoherent with respect to each other, we can describe them separately in the form factor:

$$F(\mathbf{s},t;\mathbf{q},\mathbf{k}) = \sum_{\mathbf{L},l,m,n} f_{\mathbf{L}}\{e^{-i(\mathbf{s}+l\mathbf{q}+m\mathbf{Q}+n\mathbf{k})\cdot\mathbf{L}} e^{il\omega_q t} e^{im\omega_{\mathbf{k}} t} J_l(\mathbf{s}\cdot\mathbf{u}_{\mathbf{q},0}) J_m(\mathbf{s}\cdot\mathbf{u}_{\eta,0}) J_n(m\phi_{\mathbf{k}})\}$$

$$= \sum_{l,m,n} f_{\mathbf{L}} \delta(\mathbf{s}-\mathbf{G}_{hkl}-l\mathbf{q}-m\mathbf{Q}-n\mathbf{k}) e^{il\omega_q t} e^{im\omega_{\mathbf{k}} t} J_l(\mathbf{s}\cdot\mathbf{u}_{\mathbf{q},0}) J_m(\mathbf{s}\cdot\mathbf{u}_{\eta,0}) J_n(m\phi_{\mathbf{k}}) \quad (18)$$

Now, we are in a position to consider scattering weight transfer under different excitation scenarios.

*Case A: Debye-Waller effect and diffusive scattering from lattice phonons.*

Here, we consider the effective 'heating' of the lattice, resulting in the Debye-Waller effect and the diffuse scattering around the main lattice Bragg peak at $\mathbf{s} = \mathbf{G}_{hkl}$ and simplify Eqn. (7) by assuming there is no static distortion wave present i.e. $l=m=n=0$, however, the effect of phonons is included. Then the structure factor for the main lattice Bragg peak

$$S_0^{(0)}(\mathbf{G}_{hkl}) = FF^* = \delta(\mathbf{s} - \mathbf{G}_{hkl})|f_L J_0(\mathbf{s} \cdot \mathbf{u}_{\mathbf{q},0})|^2. \tag{19}$$

One can simplify the equation with $J_0(\mathbf{s} \cdot \mathbf{u}_{\mathbf{q},0})$ well approximated by $1 - \frac{(\mathbf{s} \cdot \mathbf{u}_{\mathbf{q},0})^2}{4}$, justified since $\mathbf{s} \cdot \mathbf{u}_{\mathbf{q},0} \ll 1$ with $\mathbf{u}_{\mathbf{q},0} \sim 0.01$ Å. Then by taking logarithm on both sides, one can derive

$$e^{-i\mathbf{s} \cdot (\sum_{\mathbf{q}} u_{\mathbf{q}} \hat{\mathbf{e}}_{\mathbf{q}}(\mathbf{L}))} = e^{-\sum_{\mathbf{q}} \frac{1}{4}(\mathbf{s} \cdot \mathbf{u}_{\mathbf{q},0})^2} = e^{-M_q} \;,$$

where $M_q = \sum_{\mathbf{q}} \frac{1}{4}(\mathbf{s} \cdot \mathbf{u}_{\mathbf{q},0})^2$. This gives $S_0^{(0)} = \delta(\mathbf{s} - \mathbf{G}_{hkl})|f_L|^2 e^{-2M_q}$, with $e^{-2M_q}$ simply the conventional phonon Debye-Waller factor (DWF), although in this form one does not require the system to be in thermal equilibrium.

Experimentally, the DWF is deduced for individual Bragg peak at $\mathbf{G}_{hkl}$. Eqn. (19) gives the projected mean-square (ms) value of lattice vibration from all independent vibrational modes

$$u_{hkl}^2 = 2M_q/|\mathbf{G}_{hkl}|^2, \tag{20}$$

according to the projection of the respective polarization vector $\hat{\mathbf{e}}_{\mathbf{q}}$ onto $\mathbf{G}_{hkl}$. The excitation of phonons that leads to the suppression of the main Bragg peak intensity as described in Eqn. (18) also gives rise to diffuse scattering around the main Bragg peak at $\mathbf{s} = \mathbf{G}_{hkl} + \mathbf{q}$: Here, one may assume that the scattering by phonon at different $\mathbf{q}$ is incoherent, hence allowing the cross terms to be dropped. Considering the 1st order diffuse scattering, the structure factor becomes

$$S_1^{(0)}(\mathbf{q}) = FF^* = |f_L|^2 e^{-2M_q} \sum_{\mathbf{q}} \delta(\mathbf{s} - \mathbf{G}_{hkl} - \mathbf{q})|J_1(\mathbf{s} \cdot \mathbf{u}_{\mathbf{q},0})|^2. \tag{21}$$

$|J_1(\mathbf{s} \cdot \mathbf{u}_{\mathbf{q},0})|^2$ can be similarly well approximated by $(\mathbf{s} \cdot \mathbf{u}_{\mathbf{q},0}/2)^2 = 1/2 \, G_{\mathbf{q}}$, where $G_{\mathbf{q}} = 1/2(\mathbf{s} \cdot \mathbf{u}_{\mathbf{q},0})^2$. Hence, one obtains

$$S_1^{(0)}(\mathbf{q}) \cong |f_L|^2 e^{-2M_q} \sum_{\mathbf{q}} G_{\mathbf{q}} \delta(\mathbf{s} - \mathbf{G}_{hkl} - \mathbf{q}). \tag{22}$$

Similarly, from the 2nd-order phonon diffuse scattering ($l=2$)

$$S_2^{(0)}(\mathbf{q}) = |f_L|^2 e^{-2M_q} \sum_{\mathbf{q}} \delta(\mathbf{s} - \mathbf{G}_{hkl} - 2\mathbf{q})|J_2(\mathbf{s} \cdot \mathbf{u}_{\mathbf{q},0})|^2$$

$$\cong |f_L|^2 e^{-2M_q} \sum_{\mathbf{q}} 1/2(G_{\mathbf{q}})^2 \delta(\mathbf{s} - \mathbf{G}_{hkl} - 2\mathbf{q}). \tag{23}$$

By substituting $\sum_{\mathbf{q}} G_{\mathbf{q}} = 2M_q$ and including contribution from $S_0^{(0)}$, a conservation law is obtained for the integrated intensity near $\mathbf{G}_{hkl}$: $m_{\mathbf{G}_{hkl}} = \sum_l S_l^{(0)} = |f_L|^2 e^{-2M_q}\left(1 + 2M_q + \frac{1}{2}(2M_q)^2 + \cdots\right) = |f_L|^2$. Hence, the effect from scattering by phonons can be regarded as transferring the scattering weight from $\mathbf{G}_{hkl}$ to $\mathbf{G}_{hkl} + l\mathbf{q}$, resulting in the creation of the diffuse background. On one hand, this allows one to sum up the effects from all vibrational modes into the lattice DWF. On the other hand, the diffuse scattering techniques measure the momentum distribution of the phonon structure factors[52, 75] and in a wide angle setting such measurements can be conducted along with coherent scattering experiments to unpack the dynamics of phonons in the ultrafast X-ray and electron diffraction; for example, see Refs.[57-59]. We point out the scattering formalism discussed here applies to both thermal and nonthermal scenarios.

Taking only the dominant one-phonon contribution, the diffuse scattering $S_{diff}^{(0)}(\mathbf{q})$ is given by summing the phonon structure factor from different vibrational state occupancy $n(\mathbf{q})$:

$$S_{diff}^{(0)}(\mathbf{q}) = \sum_\mathbf{q} \frac{n(\mathbf{q})+1}{\omega(\mathbf{q})} \left| S_1^{(0)}(\mathbf{q}, \hat{\mathbf{e}}_\mathbf{q}) \right| \quad (24)$$

Normally, in the equilibrium experiment, $n(\mathbf{q}) = coth(\hbar\omega_\mathbf{q}/2k_B T)$ and a Gaussian vibrational state from the equilibrium Boltzmann statistics[75] and hence allowing one to map DWF into temperature assuming the deposited energy is equipartitioned among all active modes[52, 74]. Whereas this assumption is no longer guaranteed in the experimental conditions at far-from-equilibrium regime[77], neither Eqn. (22) nor Eqn. (23) is restricted to the Boltzmann statistics, hence they will apply to the nonequilibrium experiments as a way to track the kinetics of vibrational energy flow from the initially strongly coupled modes into the rest to establish the lattice phonon baths.

*Case B: Evaluating order parameter evolution and fluctuation effects by phase modes*

Now, we look at the scattering by the superlattice, namely the LDW, at $\mathbf{s} = \mathbf{G}_{hkl} + \mathbf{Q}$, but also recognizing the existence of phonons in the system. Here, the scenario has $m=1$, $n=0$, and Eqn. (18) gives the structure factor for the 1st-order satellite associated with the static wave:

$$S_0^{(1)}(\mathbf{k}) = |f_\mathbf{L}|^2 e^{-2M_q} \delta(\mathbf{s} - \mathbf{G}_{hkl} - \mathbf{Q} - \mathbf{k}) |J_1(\mathbf{s} \cdot \mathbf{u}_{\eta,0})|^2 \prod_\mathbf{k} |J_0(\phi_\mathbf{k})|^2. \quad (25)$$

Applying the same argument in *case A*, one can write down the equivalence of the DWF for the scattering at $\mathbf{Q}$ as $e^{-2M_\phi} = \prod_\mathbf{k} |J_0(\phi_\mathbf{k})|^2$, where $2M_\phi = \sum_\mathbf{k} \frac{1}{2}\phi_\mathbf{k}^2$. This gives the scattering by the superlattice

$$S_0^{(1)}(\mathbf{Q}) = |f_\mathbf{L}|^2 e^{-2M_q} e^{-2M_\phi} \delta(\mathbf{s} - \mathbf{G}_{hkl} - \mathbf{Q}) |J_1(\mathbf{s} \cdot \mathbf{u}_{\eta,0})|^2. \quad (26)$$

Again we consider each phase fluctuation component as independent and write the structure factor for the 1st-order fluctuation wave: $S_1^{(1)} = |f_\mathbf{L}|^2 \sum_\mathbf{k} e^{-2M_q} \delta(\mathbf{s} - \mathbf{G}_{hkl} - \mathbf{Q} - \mathbf{k}) |J_1(\mathbf{s} \cdot \mathbf{u}_{\mathbf{q},0})|^2 |J_1(\phi_\mathbf{k})|^2$. The scenario is very similar to the main lattice diffuse scattering case, and by considering also higher-order fluctuation waves, the total satellite scattering at $\mathbf{Q}$: $\sum_m S_m^{(1)} = |f_\mathbf{L}|^2 e^{-2M_q} |J_1(\mathbf{s} \cdot \mathbf{u}_{\mathbf{q},0})|^2 e^{-2M_\phi}(1 + 2M_\phi + \frac{1}{2}(2M_\phi)^2 + \cdots) = |f_\mathbf{L}|^2 e^{-2M_q} |J_1(\mathbf{s} \cdot \mathbf{u}_{\mathbf{q},0})|^2$, which is conserved; the effect of phase fluctuations can be considered as transferring the scattering weight from $\mathbf{Q}$ to $\mathbf{Q} + l\mathbf{k}$. But unlike in the case of main lattice DWF[52, 74], the phase fluctuations play a much more significant role here given the phase modes are ungapped (or low-energy modes), especially when considering an incommensurate wave state[50, 62]. This, from Eqn. (18), leads to peak broadening in the structure factor $S_\mathbf{Q}(\mathbf{k})$. Nonetheless, from the conservation law, fully integrating contributions from both the static and the fluctuational components, $m_\mathbf{Q} = \int S_\mathbf{Q}(\mathbf{k})d\mathbf{k}$, allows one to still retrieve the order parameter amplitude $u_{\eta,0}$ from an evolving CDW structure factor:

$$m_\mathbf{Q} = |f_\mathbf{L}|^2 e^{-2M_q} |J_1(\mathbf{s} \cdot \mathbf{u}_{\eta,0})|^2. \quad (27)$$

The situation is more complex for the main lattice Bragg peak if there are more than one CDW states present in the system (denoted by *l*). The momentum integration of $S_\mathbf{G}$ gives:

$$m_{\mathbf{G}_{hkl}} = |f_\mathbf{L}|^2 e^{-2M_q} \prod_l \left| J_0\left(\mathbf{G}_{hkl} \cdot \hat{\mathbf{e}}_{\eta_l} u_{0,\eta_l}(t)\right) \right|^2. \quad (28)$$

*Case C: Fluctuation effects by amplitude modes*

Here we consider diffuse scattering by amplitude fluctuations of the distortion wave at $\mathbf{s} = \mathbf{G}_{hkl} + n\mathbf{Q} \pm \mathbf{k}$. Given that the AM is in general gapped (except at the critical point of a continuous phase transition), it has a much less impact when compared to the PM. To see this, we again assume that the AM and the PM are independent. The effect from amplitude fluctuations is formulated in the density modulations

$$\rho(\mathbf{r},t) = \sum_{\mathbf{L}} \delta\left\{\mathbf{r} - \mathbf{L} - \sum_{\mathbf{q}} u_{\mathbf{q}} \hat{\mathbf{e}}_{\mathbf{q}}(\mathbf{L}) - u_{\eta}(1+\delta\eta)\hat{\mathbf{e}}_{\eta} sin(\mathbf{Q}\cdot\mathbf{L}+\phi)\right\},$$

where $\delta\eta$ is expanded with the Fourier components $\delta\eta = \sum_{\mathbf{k}} \eta_{\mathbf{k}} sin(\mathbf{k}\cdot\mathbf{L} - \omega_{\mathbf{k}}t)$. Let us look at the contribution from only one "$\mathbf{k}$" component: $J_n[\mathbf{s}\cdot\mathbf{u}_{\eta}(1+\eta_{\mathbf{k}} sin(\mathbf{k}\cdot\mathbf{L}-\omega_{\mathbf{k}}t))]$. In the limit of small distortion, i.e. $\mathbf{s}\cdot\mathbf{u}_{\eta} \ll 1$, the expression is simplified using the recursion relation:

$$J_n[\mathbf{s}\cdot\mathbf{u}_{\eta}(1+\eta_{\mathbf{k}} sin(\mathbf{k}\cdot\mathbf{L}-\omega_{\mathbf{k}}t))]$$
$$= J_n(\mathbf{s}\cdot\mathbf{u}_{\eta})\left\{1 + |n|\eta_{\mathbf{k}} sin(\mathbf{k}\cdot\mathbf{L}-\omega_{\mathbf{k}}t) + \frac{|n|(|n|-1)}{2}\eta_{\mathbf{k}}^2 sin^2(\mathbf{k}\cdot\mathbf{L}-\omega_{\mathbf{k}}t)\right\}$$

We expand $sin(\mathbf{k}\cdot\mathbf{L}-\omega_{\mathbf{k}}t) = \frac{1}{2i}\{e^{i(\mathbf{k}\cdot\mathbf{L}-\omega_{\mathbf{k}}t)} + e^{-i(\mathbf{k}\cdot\mathbf{L}-\omega_{\mathbf{k}}t)}\}$, and similarly for $sin^2(\mathbf{k}\cdot\mathbf{L}-\omega_{\mathbf{k}}t)$ and obtain the form factor for the fluctuation wave with momentum wavevector $\mathbf{k}$ to the second order:

$$F(\mathbf{s};\mathbf{k}) = e^{-M_q} \sum_{\mathbf{L},n} e^{-i(\mathbf{s}+n\mathbf{Q})\cdot\mathbf{L}} J_n(\mathbf{s}\cdot\mathbf{u}_{\eta})\left\{1 + \frac{|n|}{2i}\eta_{\mathbf{k}}\left[e^{i(\mathbf{k}\cdot\mathbf{L}-\omega_{\mathbf{k}}t)} + e^{-i(\mathbf{k}\cdot\mathbf{L}-\omega_{\mathbf{k}}t)}\right]\right.$$
$$\left. - \frac{|n|(|n|-1)}{8}\eta_{\mathbf{k}}^2[2 + e^{i(2\mathbf{k}\cdot\mathbf{L}-2\omega_{\mathbf{k}}t)} + e^{-i(2\mathbf{k}\cdot\mathbf{L}-2\omega_{\mathbf{k}}t)}]\right\}$$

$$= e^{-M_q}\delta(\mathbf{s}-\mathbf{G}_{hkl}-n\mathbf{Q})J_n(\mathbf{s}\cdot\mathbf{u}_{\eta})\left(1 - \frac{|n|(|n|-1)}{4}\eta_{\mathbf{k}}^2\right)$$
$$+ e^{-M_q}\delta(\mathbf{s}-\mathbf{G}_{hkl}-n\mathbf{Q}-\mathbf{k})e^{-i\omega_{\mathbf{k}}t}J_n(\mathbf{s}\cdot\mathbf{u}_{\eta})\left(1 - \frac{|n|}{2i}\eta_{\mathbf{k}}\right)$$
$$+ e^{-M_q}\delta(\mathbf{s}-\mathbf{G}_{hkl}-n\mathbf{Q}+\mathbf{k})e^{+i\omega_{\mathbf{k}}t}J_n(\mathbf{s}\cdot\mathbf{u}_{\eta})\left(1 - \frac{|n|}{2i}\eta_{\mathbf{k}}\right)$$
$$- e^{-M_q}\delta(\mathbf{s}-\mathbf{G}_{hkl}-n\mathbf{Q}-2\mathbf{k})e^{-i2\omega_{\mathbf{k}}t}J_n(\mathbf{s}\cdot\mathbf{u}_{\eta})\frac{|n|(|n|-1)}{8}\eta_{\mathbf{k}}^2$$
$$- e^{-M_q}\delta(\mathbf{s}-\mathbf{G}_{hkl}-n\mathbf{Q}+2\mathbf{k})e^{+i2\omega_{\mathbf{k}}t}J_n(\mathbf{s}\cdot\mathbf{u}_{\eta})\frac{|n|(|n|-1)}{8}\eta_{\mathbf{k}}^2.$$

When $n=0$, the first term on RHS: $S_0^{(0)} = |f_{\mathbf{L}}|^2 e^{-2M_q}|J_0(\mathbf{s}\cdot\mathbf{u}_{\eta})|^2$ gives the intensity at the lattice Bragg peak, which is the same as before the collective excitations; both PM and AM do not alter the main lattice peak structure factor. Note, when $n=1$, the excitations of AM do not change the 1st-order satellite intensity at $\mathbf{s} = \mathbf{G}_{hkl} + \mathbf{Q}$, , which is unlike the case for PM. The amplitude fluctuations do contribute to the diffuse scattering around the satellite intensity at $\mathbf{G}_{hkl} + n\mathbf{Q}$, for $n \geq 1$; see the remaining terms on the RHS. While the AM excitation adds to the diffuse background, its gapped nature makes such contribution much smaller than the effect from PM, away from the critical point.

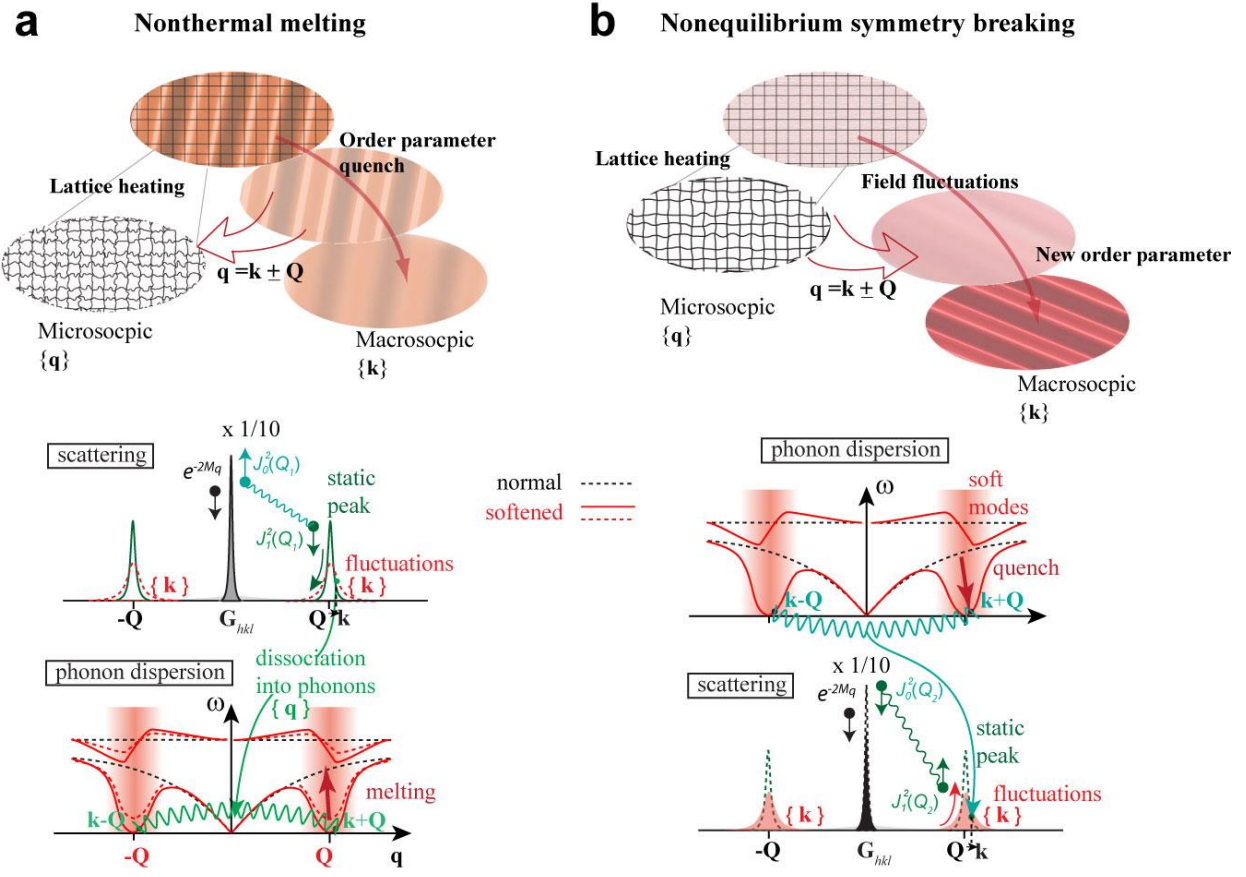

**Figure 3. Nonequilibrium phonon dynamics and fluctuation waves in the CDW system during light-induced melting and order formation.** The connection between the two dynamics upon the laser interaction quench is viewed through the changes in the phonon dispersion curves, which are coupled to the change of the order-parameter free energy landscape driving the fluctuation waves of the system. The two signatures are manifested in the scattering structure factors of the main lattice ($S_G$) and the CDW superstructures ($S_Q$) at $G_{hkl}$ and $\mathbf{Q}$ respectively. Here, *h*, *k* and *l* represent the Miller indices. The transfer of the scattering weight between $S_G$ and $S_Q$ occurs as a CDW order is created or destroyed. The excitations of the lattice modes contribute to the dynamical DWF, expressed in $e^{-2M_q}$. On the other hand, the symmetry breaking or recovery (i.e. the melting of CDW) modifies $S_G$ and $S_Q$ respectively with the Bessel functions: $J_1$ and $J_2$, which are anti-correlated with each other. The manifolds of fluctuation waves and phonons involved in the phase transitions are denoted by their momentum wavevector in {**k**} and {**q**}.

*Case D: Scattering weight transfer between static wave fluctuations and lattice phonons during phase transitions*

Tracking the dynamical transfer of scattering weight between the microscopic and macroscopic systems driven by an ultrafast quench is at the core of discussing the nonequilibrium physics of phase transitions[26, 35, 67, 78-80]. For the prototypical case of SSB upon quench, the free energy landscape with a nonzero $\langle \eta \rangle$ would involve phase and amplitude modes; see Fig. 1c. Here, in a physical text we consider the Peierls-type SSB to form an incommensurate CDW, in which the unfolding of the order parameter landscape is preceded by the lattice softening in the normal state. The CDW wavevector is determined by a maximum of the full electronic susceptibility $\chi$ enhanced by the Fermi surface nesting and the associated electron-lattice interactions at $\mathbf{Q}_\chi$. The same conditions also lead to the softening in the phonon dispersion curve[81], with the phonon frequency $\omega_\mathbf{q} \to 0$ as $\mathbf{q} \to \mathbf{Q}_\chi$. The fluctuation waves and soft phonons are intricately connected by $\mathbf{q} = \mathbf{k} \pm \mathbf{Q}$. One can see this by considering a phase fluctuational wave (phase mode) at $\mathbf{k}$. It manifests in the lattice vibration $\mathbf{u}_{\phi,k}(\mathbf{L}, t) = \mathbf{u}_{Q,0} sin(\mathbf{Q} \cdot \mathbf{L} + \delta\phi_k(t))$, with $\delta\phi_k(t) = \phi_{k,0} sin(\mathbf{k} \cdot \mathbf{L} - \omega t)$ . For a small amplitude, $\phi_{k,0} \ll 1$ , $sin(\mathbf{Q} \cdot \mathbf{L} + \delta\phi_k(t)) = sin(\mathbf{Q} \cdot \mathbf{L})cos(\delta\phi_k(t)) + sin(\delta\phi_k(t))cos(\mathbf{Q} \cdot \mathbf{L}) \sim sin(\mathbf{Q} \cdot \mathbf{L}) + \delta\phi_k(t)cos(\mathbf{Q} \cdot \mathbf{L})$, and one arrives at $\mathbf{u}_{\phi,k}(\mathbf{L}, t) = \mathbf{u}_{Q,0} sin(\mathbf{Q} \cdot \mathbf{L}) + \boldsymbol{\phi}_{k,0} sin(\mathbf{k} \cdot \mathbf{L} - \omega t)cos(\mathbf{Q} \cdot \mathbf{L})$. The first term is simply the unperturbed static distortion wave. The second term can be rewritten as $\delta\boldsymbol{\phi}_{\eta,k} = \frac{1}{2}\boldsymbol{\phi}_{k,0}\{sin[(\mathbf{k}+\mathbf{Q})\cdot\mathbf{L}-\omega t]+sin[(\mathbf{k}-\mathbf{Q})\cdot\mathbf{L}-\omega t]\}$, which constitutes a 'coherent superposition' of two phonon modes having wavevector $\mathbf{k} + \mathbf{Q}$ and $\mathbf{k} - \mathbf{Q}$. Similarly, one can look at the amplitude mode at $\mathbf{k}$ and writes $\delta\mathbf{u}_{\eta,k} = \mathbf{u}_{k,0} sin(\mathbf{k} \cdot \mathbf{L} - \omega t) sin(\mathbf{Q} \cdot \mathbf{L}) = \frac{1}{2}\mathbf{u}_{k,0}\{cos[(\mathbf{k} + \mathbf{Q}) \cdot \mathbf{L} - \omega t] + cos[(\mathbf{k} - \mathbf{Q}) \cdot \mathbf{L} - \omega t]\}$, thus comes to the same conclusion[50].

The derivation here illustrates the direct connection between the soft phonon modes and the CDW collective modes, which would necessitate an interplay between the structure factors of the CDW and the main lattice peaks. The pairing and unpairing dynamics attributed to the scattering weight transfer are embedded in the nonequilibrium dynamics of phase transitions. Figure 3 discusses the two scenarios encountered experimentally where we look at the interconversion between two types of the lattice dynamics through the changes in the dispersion curves that are coupled to the free-energy landscape changes. Of concern is how one can decouple the DWF from the symmetry-associated contribution pertaining to the phase transition. To this end, with proper consideration of multi-$\mathbf{Q}$ contributions one can independently obtain the respective order parameter dynamics via evaluating $h(t) = \frac{m_{Q_l}(t)}{m_G(t)}$ and $g(t) = \frac{m_{Q_a}(t)}{m_{Q_c}(t)}$, where the contribution from DWF is eliminated. Specifically,

$$h(t) = \frac{\left|J_1\left(\mathbf{G}_{hkl} \cdot \hat{\mathbf{e}}_{\eta_l} u_{0,\eta_l}(t)\right)\right|^2}{\prod_l \left|J_0\left(\mathbf{G}_{hkl} \cdot \hat{\mathbf{e}}_{\eta_l} u_{0,\eta_l}(t)\right)\right|^2} \tag{29}$$

and

$$g(t) = \left|\frac{J_1(\mathbf{G}_{hkl} \cdot \hat{\mathbf{e}}_{\eta_a} u_{0,\eta_a})}{J_1(\mathbf{G}_{hkl} \cdot \hat{\mathbf{e}}_{\eta_c} u_{0,\eta_c})}\right|^2 . \tag{30}$$

Given the polarization of the CDW state $\hat{\mathbf{e}}_{\eta_l}$, the order parameter $u_{0,\eta_l}(t)$ can be retrieved and used to deduce the DWF at $\mathbf{G}_{hkl}$.

## *5. Multi-messenger ultrafast electron scattering and imaging experiments*

We now consider the practical aspects of implementing these measurements through the ultrafast scattering and imaging techniques. First, we discuss the ultrafast electron diffraction (UED) approach. A central thesis for the success of using the scattering-detected order parameter dynamics to reconstruct the free-energy landscape is the separation of scales as discussed in Sec. 2. While this approach reduces the complex nonequilibrium phase transitions to problems just involving few macroscopic degrees of freedom (order parameters), the validity of the impulse-adiabatic approximation behind this approach needs to be examined in the experiments.

The event sequences from the microscopic excitations to the macroscopic transitions, as highlighted in Fig. 3, are intrinsically multi-stepped and multi-perspective, but can be efficiently probed with recent significant advances of the fs spectroscopy and X-ray scattering techniques; for recent reviews, see [11-14]. Similarly, the development of the electron-based ultrafast electron scattering[82-93] and microscopy[94-109] techniques is also in full swing in recent years. Upon applying the laser pulses, the excitation energy is initially stored in the photo-excited hot carriers, setting off the nonequilibrium microscopic dynamics through couplings to the lattice modes. Ultrafast spectroscopy techniques have investigated these initial relaxations and found clear signatures of more than one decay channel [13]. Hot carriers decay nearly instantaneously through internal relaxations establishing an effective electron temperature, $T_e$. But the electronic energy relaxation into the lattice can only efficiently occur within a small part of phonon branches, referred to as the strongly coupled phonons (SCP), often within the higher energy optical branches most connected with the electronic excitations. Then the energy is spread to other modes loosely defined as the weakly coupled phonons (WCP). The exchange of kinetic energies between the three sub-systems is typically described by a three-temperature model (3TM) [13, 110-113].

Meanwhile, the ordering over the long-range scale does not directly couple to the microscopic processes due to a large mismatch in the momentum and energy states active at the shortest time. The dynamics of the order parameter are driven by the shift of the energy landscape established by the momentum-dependent electron-phonon coupling (EPC) matrix. The EPC is shaped by electronic instabilities at the Fermi surface (FS) and electron correlation effects, which can be altered significantly upon applying optical excitations. The shifting of the energy landscape thus can occur on a very short timescale, which, from the perspective of the long-range order parameter, represents a nonthermal interaction quench.

One of the main goals of UED is to capture the dynamics of the order parameter following the quench. Only the average structure can be obtained by the intensity of the Bragg peaks whereas the structural fluctuations resulted from the symmetry-breaking and recovery processes (see Fig. 3) shall be retrieved as well from features beyond the central coherent peak. Of concern is also the soft modes and the pre-transitional phonon dynamics[114-117], properties of the lattice elastic energy landscape supporting the symmetry breaking[9, 118, 119]. These events partially overlap in time with the microscopic processes probed by the ultrafast spectroscopy techniques. The signatures of such are obtained from the **q**-resolved fine structure of the coherent structure factors as well as the diffuse scattering surrounding the Bragg peaks. Only through the combination of the scattering signals (coherent and incoherent) gathered from the different reciprocal subspaces of the Brillouin zone a deeper understanding of the nonequilibrium phenomenon of phase transition can be gained. To this end, a strong advantage of UED lies in its large Ewald sphere, making the retrieval of the different **q**-dependent features at once possible from a large momentum-scale diffraction pattern – typically as many as 10-100 Bragg peaks can be observed simultaneously in a single diffraction pattern, in contrast to the X-ray diffraction approach.

In principle, the simple geometry of the conventional UED approaches makes it well suited to compare with results obtained by ultrafast spectroscopy[11-14]. An important problem concerns the comparison between the volume excited and the parts probed by different techniques[13, 25]. When applying an optical or near-infrared laser pulse, the excitation transforms the materials on a finite absorption penetration depth, typically within 100 nm. The very large cross-section of the electron scattering means that small volume samples, such as thin films or nanocrystals in the range of 10s nm scale (depending on the beam energy) can be used. This feature partially relieves the concerns of sampling inhomogeneously excited regions and unpacking information – an issue to be addressed in Sec. 6.3.

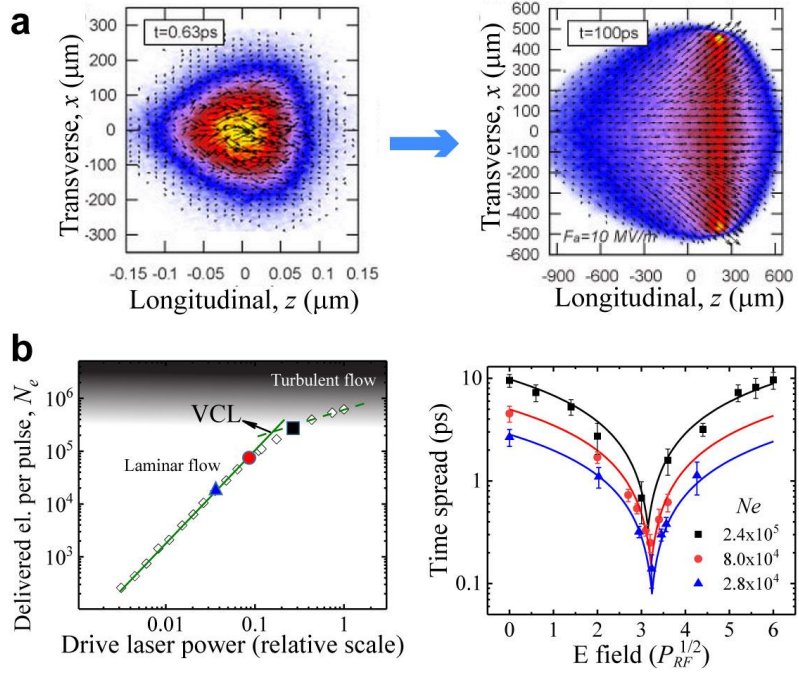

**Figure 4. The brightness-limited performance in the electron bunch compression**. **a**. The multi-level fast-multiple model (ML-FMM) simulation of the structures of the electron pulse extracted under a field $F_a$=10 MV/m from a photocathode at two different times. **b**. The number of particles in the electron bunches, $N_e$, obtained at the specimen of the UEM system as a function of the ultrafast ultraviolet drive laser power. This number is typically a small fraction (~ 5-10%) of the particle number $N_{e,0}$ generated at the cathode due to the slicing by the alignment apertures in the beamline. One obtains the virtual cathode limit (VCL) from the slope change. The right panel shows the pulse duration characteristics under the tuning of the longitudinal RF lens at different $N_e$. Panel a is adapted with permission from ref. [120].

      Meanwhile, the advantage of UED might turn into a disadvantage as the scattered signal can be dispersed into a large momentum space with a relatively poor **q** resolution. This calls for an increase of the beam flux and a decrease of the sample lateral size to improve the **q** resolution, however at the expense of signal strength. A challenge of UED as well as the ultrafast electron microscopy (UEM) approach has been to balance the requirement of the resolution versus the dose afforded under a repetition rate set by the recovery time of the excited quantum material systems, typically in the kHz and sub-kHz ranges [13, 121, 122]. There is an apparent limitation, setting the resolution affordable under a specific dose, by the space-charge-led pulse lengthening, referred to as the space-charge effect (SCE) [123-130]. Fortunately, looking deeper into this SCE issue, one can find a way to overcome this SCE even in a significant way, by properly manipulating the beams in its 6-dimensional collective phase space through electron-optical means. Key knowledge about this came from studying the dynamical phase space structures of the pulses as a function of the particle number, $N_{e,0}$, controlled by the extraction field $F_a$, both experimentally and theoretically.

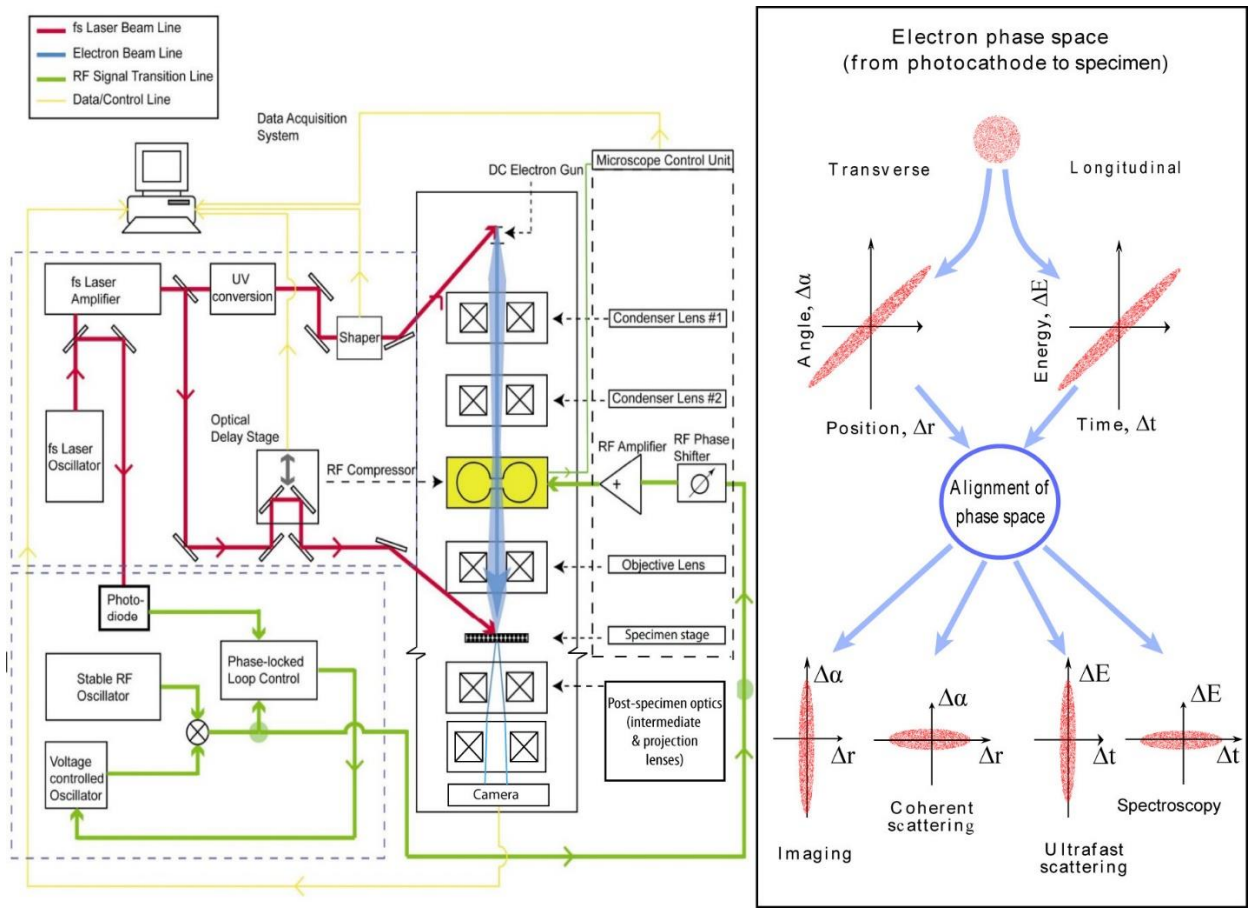

**Figure 5. Configuration of an ultrafast electron microscope system.** The main concept aims to incorporate the high temporal resolution into an existing transmission electron microscope with the rearrangement of the electron optical system to accommodate a high-density photo-electron source, which is driven by ultraviolet laser pulses. The electron pulse coming off from the cathode rapidly develops into a chirped pulse, influenced by the intrinsic strong Coulombic space-charge forces at the low energy stage. This collective space-charge effect manifests not in a blowout of the phase space, but instead in a highly correlated momentum-position phase space structures in both the longitudinal and transverse directions; see right panel where the two phase spaces are depicted in ($\Delta t$, $\Delta E$) and ($\Delta r$, $\Delta \alpha$) respectively. To remediate the resulted pulse broadening, one or more RF cavities act as the longitudinal lenses in the beamline to recompress the projected phase space along $\Delta t$ or $\Delta E$, while a similar strategy along the transverse directions is handled by the existing magnetic lenses. These two combined lead to focusing both in the longitudinal (energy and time) and transverse (crossover and coherence) dimensions. In general, different optical adjustments will allow the phase space of the incidence pulses to be realigned for the best performance of diffraction, imaging, and spectroscopy without sacrificing the electron beam dose. The successful operation of the new RF lenses relies on synchronization between the laser pulse and the cavity field, which is controlled by the phase-locked loop electronics with feedback control to counteract phase jittering within the RF cavity field for focusing; see the left panel. The physical limit of the performance is the phase space density, or brightness, of the pulse that can be delivered to the specimen. The resolution is defined in the projected sub-phase space targeted by different modalities and the relevant information encoded in the scattering process is deconvoluted by the post-specimen optics and projected onto the detector; see discussions in refs.[93, 94, 131].

The electron pulses used in the UED systems are typically created by applying the fs ultraviolet laser pulses on a cathode through the photoelectric effect[123]. From the multi-level fast multiple method (ML-FMM) calculations designed to preserve the stochastic scattering effect in the beam dynamics simulations, one shows that the collective phase space volume is conserved once the pulse is fully extracted (Liouville's theorem) from the cathode[132]. Hence, the perceived SCE associated with the pulse lengthening caused by the internal Coulombic forces can be overcome via dynamically reshaping the phase space structure of the pulse[122, 133, 134], without leading to degradation of the throughput.

This leaves the pulse brightness, defined as (particle number)/(phase space volume), as a main figure of merit in designing the photo-emission electron sources and plays a key role for improving the UEM/UED performance. In particular, a central effort has been to avoid the uncontrolled growth of phase space due to the stochastic effect[125] that leads to the degradation of the brightness, but not necessarily the (collective) SCE. Such an issue is addressed in ML-FMM calculation by studying the brightness figures and phase space structures as a function of $F_a$ [120]. Prototypical SPE-led pulse evolutions under a nominal $F_a = 10$ MV/m are shown in Fig. 4a, where the phase space structures at two stages (630 fs and 100 ps) along the transverse ($x$) and the longitudinal ($z$) directions are compared. The particle momenta ($p_x$ and $p_z$) depicted by the arrows give a certain spreading but largely are correlated with the position ($x$ and $z$) led by SCE. Accordingly, the brightness is tracked as a function of particle number $N_{e,0}$ with the extraction field ($F_a$) as the control parameter.

A main conclusion from ML-FMM simulations is that the transverse ($x - y$ plane) phase space grows sub-linearly with respective to $N_{e,0}$ until the virtual cathode limit (VCL) – one when the space-charge forces associated with the positive counter ions at the emitting surface become strong enough to reduce the efficiency of the photoemission[120, 125, 135, 136]. This is seen in the left panel of Fig. 4b where the yield over the drive laser power becomes sub-linear and the charge particle flows switch from laminar to turbulent. In the specific UEM arrangement, the observation is made at the detector where the peripheral hot electrons around the electron pulse have been sliced off with an aperture in the beamline such that the delivered particle number, $N_e$, is typical 5-10% of $N_{e,0}$ at the cathode[128]. Nonetheless, the presence of VCL is represented by an inflection point (Fig. 4b). Characterization of VCL is important as only when driven above VCL, the stochastic phase space size of the pulse will only significantly increase from the onset of turbulence within the charge particle flow; the particle flow is otherwise laminar in the regime below VCL[120]. This means that one can significantly gain transverse brightness by increasing $N_{e,0}$ up to the brink of VCL. The effect translates to improving the performance related to the transverse phase space, such as the spatial resolution of UEM or the **q**-resolution of UED. Meanwhile, one finds the phase space along the $z$ direction increases nearly linearly with respect to $N_{e,0}$, and the pulse-width ($\Delta z$) grows as $N_{e,0}^{1/2}$, also confirmed by the experiment[120]. From the right panel of Fig. 4b where the longitudinal phase space size is translated into the compressibility in time, one can conclude that even at VCL under just a fair $F_a \sim 2$ MV/m in a DC gun arrangement, a sub-ps resolution can easily be achieved from tuning the RF field to realign the phase space.

The left panel of Fig. 5 shows the setup of the electron-optical system for conducting the UED and UEM experiments. The two approaches share a similar electron-optical system before the specimen. A feature here is the incorporation of the RF cavity system, before and after the specimen for realigning the longitudinal phase space structure. Effectively here, the RF cavity acts as the longitudinal lens in a very similar role as the magnetic lens for controlling the phase space structure in the transverse direction – a feature that is fully implemented in the conventional TEM. The two lens systems combined allow the UEM/UED apparatus to achieve an optimal pulse shape targeted by the different modalities [94, 122]. The additional optical system in UEM consists of intermediate and projection lenses and the spectrometer, intended to decode the nonequilibrium physics encoded in the phase space of the scattered particles in the modality of imaging or spectroscopy[108, 109, 122, 137-144]; whereas without such optics (or set at the diffraction mode in a UEM), the scattered electrons are focused onto the screen to form the diffraction image directly.

The right panel of Fig. 5 shows schematically how the operating parameters are set based on the modality's feature-of-merit (FOM), defined by the projected phase space. For example, for coherent scattering, a key is to maximize the incident particle density projected along the transverse momentum space. This can be achieved via the condenser lens adjusted to minimize the tilt of the transverse phase space ($p_x$ vs $x$ in lab frame or $\Delta\alpha$ vs $\Delta r$ in pulse frame) as the pulse arrives at the sample. This can be considered as a pulse compression along $p_x$ (or $\Delta\alpha$). For ultrafast imaging where the resolution is typically dose-limited, the optics are frequently optimized to produce a better focusing along $x$ at the expense of beam coherence. Meanwhile, for achieving ultrashort time resolution, the adjustment is made via a compression along $z$ ($\Delta t$), whereas for spectroscopy a compression along $p_z$ ($\Delta E$) is needed to monochromatize the pulse. The post-specimen optics are typically tuned to project the proper phase space structure of scattered pulse onto the screen; see e.g. refs. [94, 122] for optimization and mode matching strategies. Their discussions are beyond the scope of this paper. As a result, the new approach can sidestep the conventional collective SPE from a high-density beam and leads to the possibility of multi-modal scattering and imaging experiments within a single platform[96]. This adaptive optics strategy has been recently employed in the UED [23] and prototype UEM[109, 137] experiments; some of which will be discussed in Sec. 6.

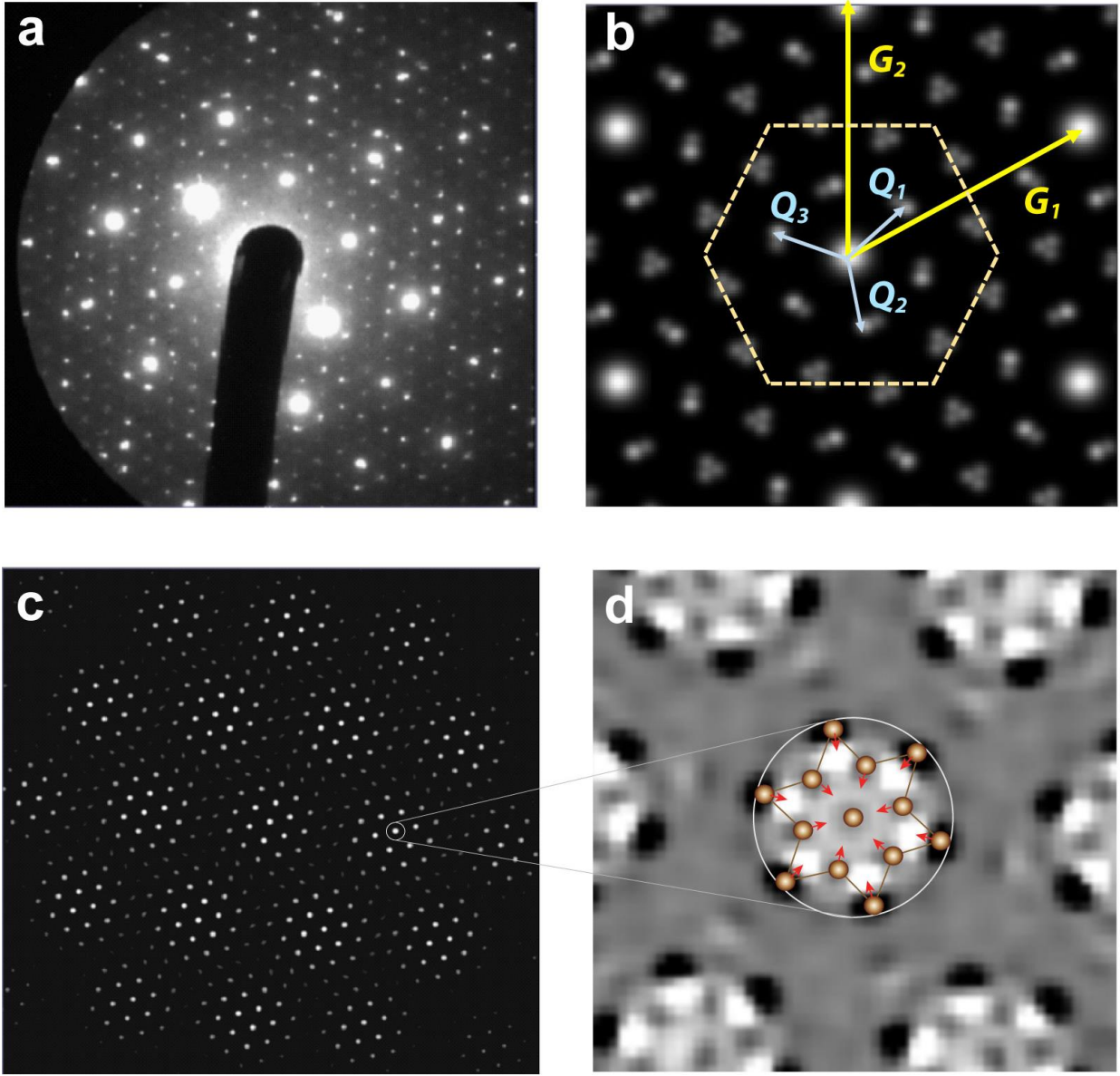

**Figure 6. Diffraction of 1T-TaS$_2$ thin film measured with an ultrafast electron diffraction setup.** **a**. The diffraction pattern obtained at the beam energy of 100 keV. **b**. A local diffraction pattern constructed from refining all relevant Bragg peaks (CDW and lattice) gives the distribution of satellites associated with the triply degenerate CDW branches and the associated high-order harmonics of the CDW from inter-CDW coupling; these features appear around each Bragg diffraction peak from the main lattice. The main CDW satellites are marked by the respective momentum wavevector $\mathbf{Q_i}$, whereas the main lattice peaks are marked with the reciprocal unit cell wavevectors $\mathbf{G_1}$ and $\mathbf{G_2}$. **c**. The real-space representation deduced from the CDW patterns showing the long-range hexagonal domain structures. **d**. The 13-atom supercell of the density wave in David-Star shape, presented in the lattice distortion map.

In Fig. 6a, we give the scattering patterns from the ultrafast coherent electron pulses delivered via the pulse compression schemes. The pulse width here is ~ 100 fs (Fig. 4b), whereas the transverse momentum compression by the condenser and objective lenses leads to a high beam coherence length (~ 40 nm; Ref. [23]). The combination of the two provides the resolutions to probe long-range cooperativity over an ultrafast time window for the nonequilibrium CDW phase transition. The order parameters of the CDW states are encoded in the satellite peaks as shown in Fig. 6b, including the higher-order satellites present due to the domain structures of the CDW[145-149]. The satellite network from three degenerate CDW branches with momentum wavevector **Q**$_i$ and their higher-order multi-**Q** components represents the Fourier spectra of the real-space hexagonal domain state[150, 151] – so-called near-commensurate CDW (NC-CDW)[145-149, 152], which can be converted from the scattering pattern; see Fig. 6c. In addition, the scattering from fluctuation waves tied to the electronic instabilities is often more spread out in the momentum space and forms the diffuse scattering background[53, 56]. The diffuse scattering gives central information on the pre-transitional phenomena dominated by the preformed short-range orders or soft phonon modes. Given the scattering intensity from these features are significantly smaller than the intensities of the Bragg peak (**G**) from the average lattice unit cells, the useable dose on the sample is a key factor for successfully probing the quantum material phase transitions[145, 149, 152]. Furthermore, ensuring no residual effect after the pumping of thermally isolated thin specimens requires a low repetition rate (≤ 1 kHz). For these reasons, the adaptive optics approach is the method of choice as it does not rely on the aperture to improve the resolution.

## *6. Example systems*

The possibility to access long-lived states with desired unconventional properties has motivated an increasing number of ultrafast experiments exploring correlated quantum materials[1, 2, 4, 11, 12, 153]. In this section, we illustrate how the measurements of the nonequilibrium order parameters by the ultrafast scattering, when connected with the Landau-Ginzburg formalism, can be used to explore the global energy landscape of the nonequilibrium quantum materials and study the nonequilibrium physics therein. We will focus on three prototypical systems, i.e., $CeTe_3$, $1T\text{-}TaS_2$, and $VO_2$, all of which exhibit long-range states ordered on the lattice with distinctive light-tunable and structurally coupled electronic phase transitions. The three systems have varying strengths of electron correlation as well as different natures of electron-lattice coupling, as depicted in Fig. 7. The charge orderings range from purely incommensurate, to near commensurate, to commensurate or even bond-ordered density waves, making them representative systems for the comparisons. The nonequilibrium platform here focuses on the ultrafast optical pump–scattering probe settings using typically isolated sub-50 nm scale thin films or nanocrystals. The entire samples are covered within pump laser and probe electron beams. The settings facilitate a condition where the pump laser with a similar penetration depth as the film thickness introduces interference effect within the film and hence may establish a uniformly excited material system in which the nonequilibrium state evolves initially from a homogeneous quench. The sample is typically suspended freely over a fine grid held under the vacuum environment. Therefore, one may safely assume that absorbed energy in the pumped system is preserved to the entire probe window (over 1-2 ns). The pump-probe repetition rate is set at 0.1 to 1 kHz, adjusted to ensure that the pumped system fully relaxes on the much longer time scales. This platform allows us to discuss nonequilibrium dynamics as an internal relaxation process. We will discuss implications to the interpretation of the experimental results when the conditions of pump-homogeneity are not met.

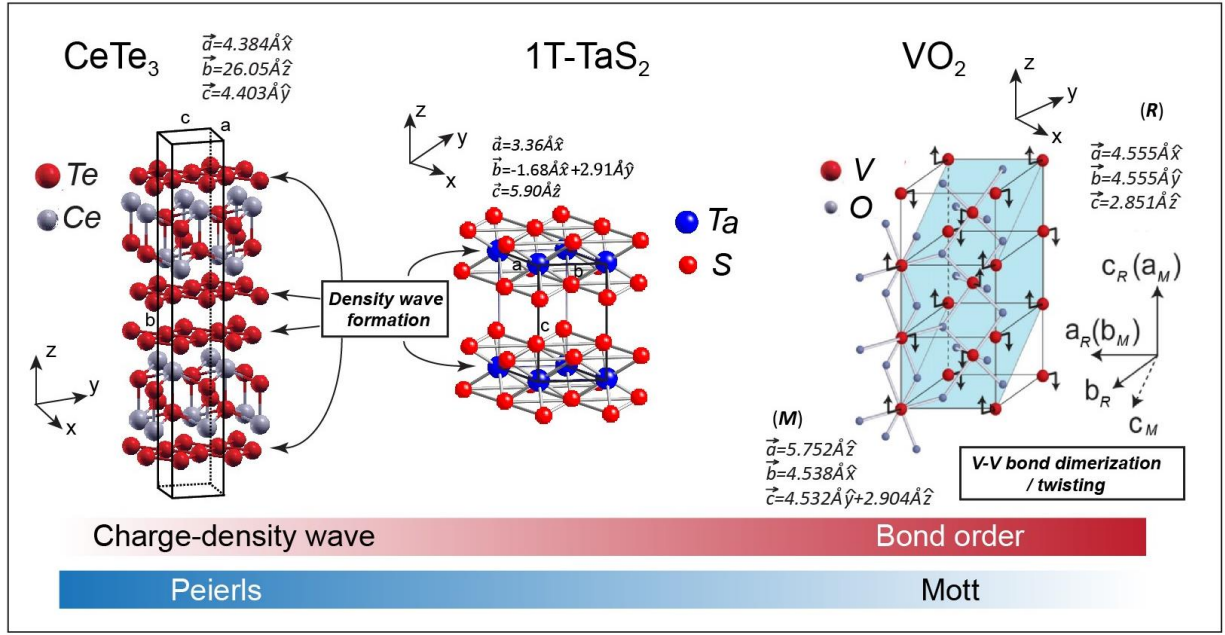

**Figure 7. The crystal structures of CeTe$_3$, 1T-TaS$_2$, and VO$_2$ investigated by the ultrafast scattering techniques**.

## 6.1. Competing degenerate broken-symmetry orders: Rare-earth tritellurides.

The rare-earth tritelluride (RTe$_3$) compound is one of the most systematically studied CDW systems to undergo continuous phase transition[154-156]. This system consists of square Te planes, alternating with weakly coupled RTe slabs (see Fig. 7). Despite the fundamental C4 symmetry in the 2D Te sheet to host CDW formation, the stripe phases are the predominant CDW orders due to spontaneous symmetry breaking. We note, bearing on a small asymmetry owing to the coupling between the two nonequivalent square Te nets[73], the dominant ground state is the stripe phase along the **c**-axis. However, the recent inelastic X-ray scattering revealed pre-transitional critical fluctuations in nearly equal strengths along both **c** and **a**-axes[157, 158] – a signature of the two order parameter fields vying for the spectroscopic weight to become a static order. Nonetheless, the SSB dictates that, upon ordering along the **c**-axis which removes a significant amount of the potential *a*-CDW spectral weight[156, 159], subsequent formation of *a*-CDW will be excluded [160].

The surprising light-induced formation of a new broken-symmetry order in the direction orthogonal to the pre-existing state (*a*-CDW) was unveiled recently in the two light rare-earth RTe$_3$ family members of LaTe$_3$[22] and CeTe$_3$[23]. Given that the *a*-CDW does not exist in the system prior to applying a fs near-infrared pulse, this is a rare scenario where the suppressed field of a new broken-symmetry phase can be created from scratch over a relatively short timescale. Therefore, from studying the real-time ultrafast dynamics of the CDW system in RTe$_3$, it is possible for one to gather crucial insight into how an SSB phase transition emerges out of equilibrium beyond the mean-field description in a condensed matter system.

In the equilibrium state phase transition, there has been already a high degree of control evidenced in the significant shift of the critical temperature $T_{c,1}$ by changing the rare-earth element and applying pressure[154-156]. The effects, such as downsizing the gap size and the CDW amplitude due to Lanthanide shrinking, are results of weakening the inter-orbital coupling strengths that shape the FS[73]. Given the two CDW order parameters are already strongly competing in the equilibrium state, it is possible to tip the balance of the competitions thus transforming the free-energy surface by shifting the orbital states occupancies [161]. The surprising introduction of *a*-CDW, even in a system deep inside the pre-existing order of *c*-CDW by applying short near-infrared pulses (here $T_b$=300 K $\ll$ $T_c$=540 K), indeed reflects this.

The possibility of optically manipulating the free energy surface allows the RTe$_3$ to be the prototype system for transient control of nonequilibrium phases. The basic physics of competitively driven transformation of the free-energy landscape and, as a result, the introduction of the hidden *a*-CDW has been given on a phenomenological ground; see Sec. 3. Here, we will validate the theoretical hypotheses and focus on elaborate the more intricate microscopic dynamics and the nonequilibrium processes enabling such phenomena. At the center stage of the discussion is the ultrafast scattering-based approaches serving to capture the nonequilibrium order parameters and the associated field fluctuations. We will show how one uses such information to reconstruct a competitive global energy landscape poised to the different orderings upon quench and the nonequilibrium physics it entails.

Transient dynamics in a system with two competing orders $\eta_c$ and $\eta_a$, in this case, are recorded using an RF-optics-augmented UED system with the transverse lenses tuned to optimize the **q**-resolution [23, 93]. In particular, the amplitude dynamics as described in Eqn. (9) are manifested in the integrated intensity of the structure factor, $m_l(\mathbf{Q}; t)$, that one can retrieve from the dynamical diffraction patterns [Eqn. (27)]. The spatially nonuniform order parameter evolution is also detected by following the correlation lengths ($\xi$) of the system encoded in the width of the structure factor [Eqn. (25)]. Taken from ref. [23], Fig. 8a shows the raw diffraction pattern where the main signature of the broken-symmetry order at $t = -1$ps is the *c*-CDW superlattice satellites at $\mathbf{G}_{hkl} \pm \mathbf{Q}_c$ with $\mathbf{Q}_c = 0.28$ $\mathbf{c}^*$ around the Bragg peaks of the square lattice at $\mathbf{G}_{hkl}$. The inset shows the patterns from the $\mathbf{G}_{401}$ region before (-1 ps) and after (+1 ps) the laser excitation. Clearly, by +1 ps the system establishes a new pair of satellites at $\pm 0.30$ $\mathbf{a}^*$ ($\mathbf{Q}_a$). The results inform the occurrence of a new broken-symmetry phase on $\approx 1$ ps timescale. The respective order parameter dynamics are plotted in Fig. 8b.

*Method to retrieve global free-energy surface through ultrafast scattering-detected order parameters.* To begin with, we give a pedagogical description on how the experimental protocol helps retrieve the Landau parameters based on controlled studies of transient metastable phases. The basic assumption is that the quenched free-energy landscape will decide the coordinates of the metastable states as its stationary points. Hence, by following the relative changes of the two competing order parameters at the metastable stages as the controlling laser fluence (*F*) is tuned the bi-dimensional free-energy surface shaped by the competitions between the two sub-systems can be evaluated. The coordinates of free energy minimum taken from the stationary points of Eqn. (11) are

$$\begin{cases} |\eta_c|^2 = \frac{-(a'(T-T_c)+\tilde{A}|\eta_a|^2)}{A'_4} \\ |\eta_a|^2 = \frac{-(a(T-T_c)+\tilde{A}|\eta_c|^2)}{A_4} \end{cases}. \quad (31)$$

These coordinates, equivalent to $\hat{m}_{Q_l}(t) \equiv \frac{m_{Q_l}(t)}{m_{Q_c}(t<0)}$ ($l = a$ or $c$), are reported as a function of *F*; see inset of Fig. 8b. We use the two critical fluences identified in Fig. 8b set the scale of the Landau parameters. First, from the established critical energy density $E_{c,c} \approx 0.64$ eV/nm$^3$ (converted from the applied fluence $F_{c,c} \approx 2.0$ mJ/cm$^2$ needed to suppress $|\eta_c|$ to 0), one derives the *c*-CDW-associated parameters: $a = \frac{4|E_{c,c}|}{T_c-T_b} = 1.05 \times 10^{-2}$ eV·nm$^{-3}$·K$^{-1}$, and $A_4 = a(T_c - T_b) = 2.54$ eV·nm$^{-3}$. In addition, the successfully identified nonthermal critical fluence $F_{c,a} = 0.59$ mJ/cm$^2$ gives the critical condition over which a nonzero static $|\eta_a|$ will be created according to Eqn. (31). From $|\eta_{c,th}| \approx 0.87$ at $F_{c,a}$, one gets $\tilde{A} = 4.10$ eV·nm$^{-3}$, $a' = 1.29 \times 10^{-2}$ eV·nm$^{-3}$·K$^{-1}$, and $A'_4 = 22.0$ eV·nm$^{-3}$ by fitting the results with Eqn. (31).

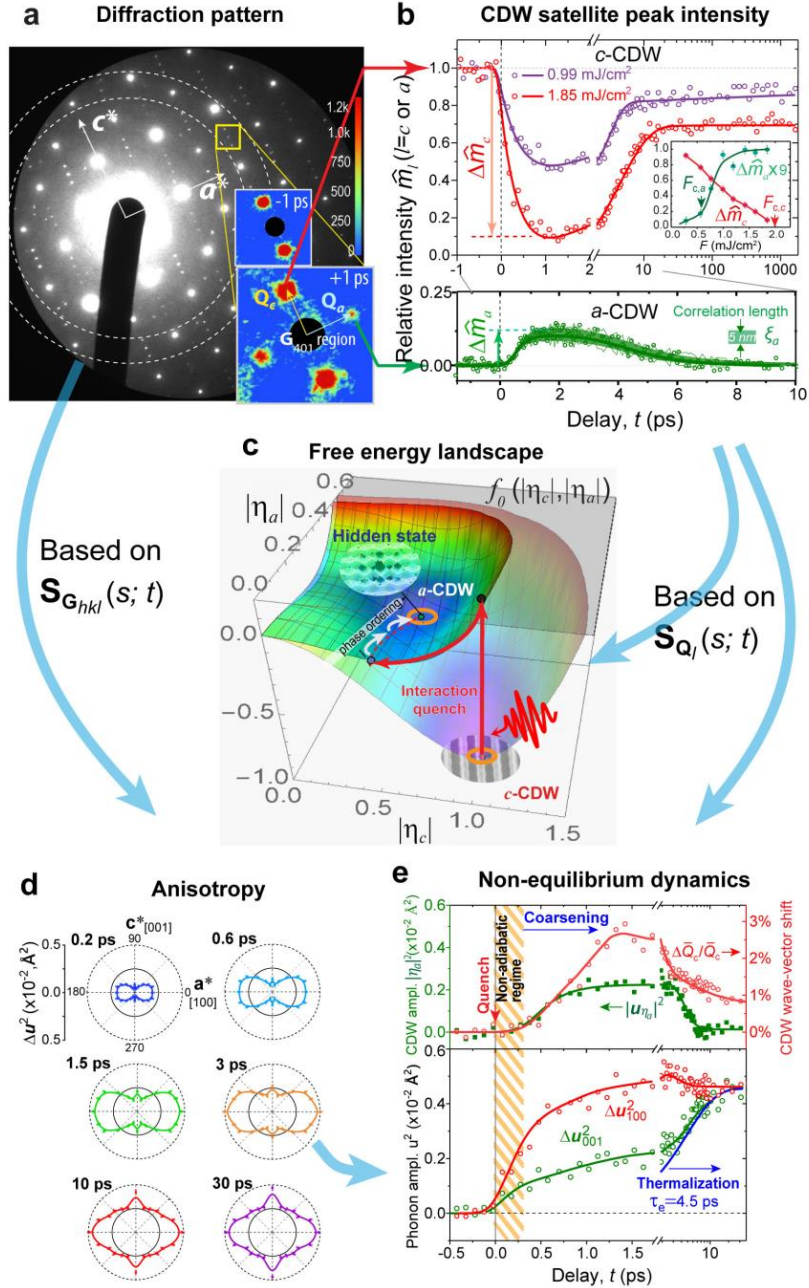

**Figure 8. Nonequilibrium dynamics into a hidden checkerboard order in CeTe$_3$. a**. Diffraction pattern of CeTe$_3$ thin film. The inset shows the scale-up view of the patterns near $G_{401}$ before (-1 ps) and after (+1 ps) applying pump pulses. **b**. The evolution presented in terms of the diffraction integrated intensity for the order parameters. The inset shows the respective changes obtained at the metastable period (~ 1.5 ps) as a function of fluence. **c**. The Landau-Ginzburg free-energy surface obtained for the $F=1.85$ mJ/cm$^2$ case. **d**. Lattice phonon responses deduced by the momentum-dependent Debye-Waller analyses. **e**. (Top) The order parameter field evolution examined via $|u_{0,\eta_a}|^2$ and $\Delta \bar{Q}_c / \bar{Q}_c$ respectively for $a$- and $c$-CDWs. Here, $\bar{Q}_c$ is the mean wavevector including $c$- and $c^+$ contributions.

(Bottom) The vibrational *ms* phonon amplitude changes projected along [001] and [100] were obtained from the Debye-Waller analysis. Panels a, b, d, e are adapted with permission from ref. [23].

*Impulse-adiabatic phenomenology.* We point to the fact that with the above-the-gap excitation, the pump does not couple to the order parameter directly; rather it heats up the carriers first and that suppresses the CDW spectroscopic gap on a shorter timescale than the long-wave response associated with the lattice order parameter. This is consistent with the observation of a carrier spectral weight transfer and the adjustment of the FS topology within 100 fs by photoemission[161, 162]; whereas the overdamped suppression of the associated order parameter amplitude appears on a slower (~ 300 fs) timescale, observed by the scattering techniques (Fig. 8 b)[22, 23, 163]. These hierarchic temporal responses reflect the basic impulse-adiabatic phenomenology for the description of the light-induced phase transition on the free-energy landscape.

The new bidimensional free-energy landscape is now plotted using the refined Landau parameters; see Fig. 8c. We then seek to understand the dynamics of the subsequent order parameter amplitude evolution following the quench presented in Fig. 8b as a potential-driven process on the free energy. Based on the local curvature of the potential surface, the initial dynamics shall appear on a downward trajectory (see arrowed line in red in Fig. 8c) along which the existing order-parameter amplitude is significantly suppressed, manifesting in a nonthermal melting as witnessed in the initial *c*-CDW state evolution. We note during this period there is no detectable change in **Q** (top panel, Fig. 8e). This means the order parameter is temporarily trapped at the saddle point following the local steepest decent. Next, for the nonequilibrium system to establish the broken-symmetry phase with bi-directional components at the new global minimum, the order-parameter dynamics must switch direction. This occurs in the next $\approx$ 1 ps based on the **Q**-shift and the rise of *a*-CDW intensity ($\widehat{m}_a$). However, in the evolution toward the new global minimum the order-parameter field is incoherent driven by fluctuation waves. Since the amplitude mode is gapped, the rate-limiting step is the alignment of the local phase and hence the phase rigidity controls the buildup time. Experimentally, this manifests in the expansion of the static correlation length along with the order parameter amplitude (Fig. 8b). The long-wave modes order the field on increasingly larger scales at the bottom of the free energy surface – a scenario described by coarsening as discussed in Sec. 2 and Fig. 1c. Hence, the simple phenomenological free-energy model explains the stepwise manner in which the hidden state is introduced into the system as observed by the UED experiment.

*Connections between field instabilities and soft modes.* Now we utilize further UED results to look into the microscopic details of the nonequilibrium processes. Of particular interest here is to understand how the rise of the microscopic soft modes is connected to the order-parameter field fluctuations in a nonequilibrium SSB driven by the quench. Here, the nonadiabicity plays a role because of separation of scales. Along the $|\eta_a|$ direction, the order parameter field is initially disordered. Mathematically, a long-wave AM is formed by coherently joining two counter-propagating soft modes. However, initially the order-parameter field is unstable because of the upturned parabolic free energy at initial times[61, 164]. Only the longest fluctuation wave ($|\mathbf{k}| \rightarrow 0$, $\omega \rightarrow 0$) will be relevant to the formation of a single-wavevector incommensurate CDW. The phase transition cannot be said to have happened if the ensemble averaging is taken over a fluctuating order parameter field, where $\langle |\eta_a|(\mathbf{r}, \mathrm{t}) \rangle$ amounts to zero. The rectification to form a static order from the dynamical modes here requires a parametric growth of phase-coherent AM modes.

To understand the initial nonadiabicity pertaining to SSB driven by a quench, we track the scattering weight transfer between $S_\mathbf{G}$ and $S_\mathbf{Q}$. For evaluating **q**-dependent phonon dynamics, one obtains the differential-mean-square lattice fluctuations $\Delta u_{hkl}^2$ by taking the logarithm of the normalized $\widehat{m}_\mathbf{G}$ (after excluding the contributions from the quasi-static CDW-related Bessel function terms; see Eqns. (27—30). This results in a differential DWF, i.e., $l_{hkl}(t) = \frac{e^{-2M_{hkl}(t)}}{e^{-2M_{hkl}(t<0)}} = e^{-2\Delta M_{hkl}(t)}$, which can be taken at different $\mathbf{G}_{hkl}$ to determine the differential fluctuations $\Delta u_{hkl}^2 = 2\Delta M_{hkl}$ projected in the direction of $\mathbf{G}_{hkl}$; see Eqn. (20). The results of this analysis taken from an array of $\mathbf{G}_{hkl}$ along different directions are shown in Fig. 8d. The strong anisotropy in $\Delta u^2$ reflects the underpinning energy landscape, or **q**-dependent electron-phonon coupling that shapes the phonon dispersion curves of the lattice modes upon phase transitions. The large amplitudes of motion reflected in the $\Delta u_{hkl}^2$ along the [100] and [001] directions indeed reveal the predominance of the soft modes directly driven by the landscape changes. Their excitation, within the first 500 fs (see Fig. 7e), is clearly much faster than the generic laser-induced heating (over a few ps).

Now, we turn our attention to the correspondingly determined long-wave parameters, specifically the order parameter amplitude $|u_{0,\eta_a}|^2$ and the momentum wavevector shift (see Fig. 8e). It is quite evident that the build-up of the long-wave modes appears mainly after the soft mode amplitudes have peaked. The results obtained here highlight the essential different dynamics due to separation of scales, and the close relationship between the fluctuation waves (both in the symmetry-breaking and recovery processes) and the respective order-parameter evolution. The empirical results also show nonergodicity between soft modes pertaining to the two CDW systems. A period, where the soft mode amplitudes characterized by the $\Delta u_{100}^2(t)$ and $\Delta u_{001}^2(t)$ along the two perpendicular CDW fields diverge, is witnessed (Fig. 8e, bottom panel). One can easily correlate this period with the timescale where the metastable *a*-CDW phase is created and then destroyed. This reaffirms the out-of-equilibrium phonon dynamics have a key role in sustaining the hidden state. Accordingly, the system thermalization timescale $t_{th}$ is determined to be ~4.5 ps.

*Nonthermal critical point.* Finally, we address how conceptually one can unite the interaction quench with the temperature quench to understand nonthermal SSB as a condensation process on a new free-energy landscape. As is the case under the equilibrium condition, the local temperature of the CDW state is a co-control parameter for the transient free-energy landscape when it is influenced to undergo SSB by the laser interaction quench. Accordingly, we can introduce the concept of the nonthermal critical point $T_c^*$, which is defined by rearranging the Landau-Ginzburg equation:

$$T_{c,a}^* = T_{c,1} - \frac{\tilde{A}}{a'}|\eta_c|^2. \qquad (32)$$

First, one can see how this concept applies to suppress the formation of *a*-CDW under a competitive SSB at equilibrium. Given the degenerate critical point $T_{c,1}$= 540 K[155, 160], with the *c*-CDW chosen as the equilibrium ground state ($|\eta_c|$=1 at $T_b$= 300 K), the experimentally identified $\tilde{A}$ and $a'$ shift $T_{c,a}^*$ to a lower temperature (222 K) than $T_b$. Furthermore, following the BCS behavior, the $|\eta_c|$ is expected to rise as $T_b$ is lowered [155]. This means further cooling the system in the broken-symmetry phase will continue to push the $T_{c,a}^*$ even lower. This explains why under the equilibrium condition the *a*-CDW is suppressed once *c*-CDW becomes the dominant ground state. However, the scenario will change entirely under a nonequilibrium condition where $|\eta_c|$ is to an amplitude below the critical threshold [$|\eta_{c,th}| \approx 0.87$ from Eqn. (12)] driven by an interaction quench. In the case described in Fig. 8b, the $|\eta_c|$ is reduced down to 0.32 at the laser fluence of 1.85 mJ/cm², where the corresponding $T_{c,a}^*$ (508 K) is now well above $T_b$. Therefore, from the perspective of *a*-CDW, the system is now under a deep temperature quench scenario and will undergo condensation to form a new broken-symmetry order in much the same way.

## 6.2 Metamorphosis in vestigial density-wave system: tantalum disulfide.

Here we will approach a different type of light-induced hidden phases involving competitions but not in the same vein as the RTe$_3$ system. The system is 1T-TaS$_2$, which belongs to a broad class of layered transition metal dichalcogenide (TMDC) compounds[165]. Similar to RTe$_3$, these TMDC compounds have isolated 2D metallic layers; see Fig. 7. However, the in-layer atomic structure is triangular rather than square lattices within which the triply degenerate CDWs emerge[165]. Of particular interest here is 1T-TaS$_2$, which hosts nontrivial charge-density wave orders driven by several factors: the instabilities at the Fermi surface, the large lattice distortion possible, and the localized orbital that leads to a Mott-Hubbard gap at the low temperature[152, 165-167]. The hidden CDW state formation in 1T-TaS$_2$ [18] was among the first that was demonstrated in TMDC and since cross-examined by optical[168], resistivity[169], scan tunneling microscopy (STM)[170-172], X-ray[173] and electron scattering[19, 174] techniques. Besides 1T-TaS$_2$[18, 19, 169, 170, 173], light-induced hidden states have also been found in 1T-TaSe$_2$[175], 3R-Ta$_{1+x}$Se$_2$[176, 177], and 1T-TaS$_{1-x}$Se$_x$[174].

Within the Ta layer, the low-temperature ground state forms 13-site clusters in which 12 out of 13 Ta$^{4+}$ ions are distorted towards the central Ta atom to a David-Star shape; see Fig. 6d. It is understood that the unpaired electron in a 13-site cluster with a large spin-orbital coupling gives enough correlation to form a Mott insulator. The $\sqrt{13} \times \sqrt{13}$ superlattice clustering forces the density-wave state to lock-in to the atomic lattice, forming the long-range commensurate charge density order (C-CDW) under ambient pressure. Upon warming, the increasing electronic instabilities deviate the CDW from its perfect commensuration with the lattice, driving a series of discommensurate (DC) state. Upon warming from C-CDW, the ordering first changes to a triclinic state (T-CDW) state at T$_{C-CDW}$ ~ 220 K, then to a hexagonal nearly commensurate density wave (NC-CDW) state at T$_{H-NC}$ ~ 280 K, and finally to an incommensurate CDW (IC-CDW) state at T$_{IC}$ ~ 350 K. The material eventually loses the density-wave order and becomes metallic at T$_M$ ~ 540 K[147-149, 178].

The delicate density-wave ordering is also subject to tuning by chemical doping or applying pressure and can generally lock into a certain type of nearly commensurate ordering before the system entirely loses the David-Star clustering feature. Eventually, melting of the David-Star clusters allows the system to adopt a purely sinusoidal form of single-wave vector IC-CDW[179]. The multi-**Q** effect (Fig. 6b) is pronounced in all forms of NC-CDW[145], in which the CDW deforms from the sinusoidal shape by developing periodic phase slip at distance $\xi_D$[180], forming the domain walls or discommensurate (DC) region. Within the domain, the CDW maintains the commensurate ordering, whereas on average the NC-CDW state has an incommensurate wavevector to avoid the penalty of raising the elastic energy as a whole[152, 179, 181]. This new self-organizing scheme as pointed out by McMillan[179] creates long-range periodic DC textures. In principle, this discreteness effect from the longer-range lock-in scenario will lead to stabilization at every possible topologically compatible DC network, so-called Devil's staircase[182]. In practice, only a few DC structures will develop out of balance between the commensuration energy and instabilities driven by thermal excitations. The triclinic[183] and hexagonal NC[53, 178] states identified in the equilibrium phase transitions belong to these DC states[146, 179].

In these intermediate phases, the C-CDW characteristics are only partially lost by the increase in itinerant instabilities. Therefore, they may be considered as a type of vestigial order[68, 69, 184]. Here, unlike the case for RTe$_3$, the CDW transformation may be modelled on a single order parameter $\eta = |\eta|e^{i\phi}$, where the deviation from $\phi(r) = \mathbf{Q}_c \cdot \mathbf{r}$ locally produces additional free energy gain[179-181]. With McMillan's DC network additional terms in the Landau's free-energy equation are introduced where the CDW phase plays an explicit role in the lock-in energy. The essence of McMillan's model has been shown to describe CDW systems with different types of commensurate orderings; see e.g. Refs. [146, 181].

Taking the essence of McMillan theory and the chief experimental observables from the UED experiments, one can give a similar phenomenological model[114, 185] to capture the essential features in TaS$_2$ CDW phase transition in which the free-energy density is given by

$$f = f_0 + C|\eta|^2(\nabla\phi - \mathbf{Q}_{IC})^2 - D|\eta|^2 cos(\phi - \mathbf{Q}_C \cdot \mathbf{r}) + H.O.. \quad (33)$$

The first term $f_0 = A|\eta|^2 + B|\eta|^4$, where $A = \frac{\alpha}{2}(T - T_M)$ has the expression like Eqn. (1) and is the leading order in the Landau-Ginzburg equation that defines a continuous phase transition. The second term comes from the energy cost of distorting the CDW structure from its ideal IC state set by the susceptibility. The third term reflects the lock-in effect, which favors the commensurate order. The last term contains the additional higher-order CDW interactions and the multi-$\mathbf{Q}$ effects[180]. It is easy to see from Eqn. (33) that the most important factor setting the energy scale is the CDW amplitude $|\eta|$, whereas the lock-in condition set by topology will determine the stable $\mathbf{Q}$ that the vestigial order could settle (discreteness conditions). Here, the coefficients A, B, C, and D are all positive, which allows Eqn. (33) to capture the generic trend of IC-to-C transition[149, 181]. First, let us assume the system is homogeneous. Driven by the lock-in energy gain by the amplitude increases upon lowering the temperature from $T_M$, the $\mathbf{Q}$ evolves continuously and creates a jump in both $\mathbf{Q}$ at T$_{C-CDW}$; see solid line Fig. 9a that is the case for the 2H-TaSe$_2$ phase transition[47]. Now considering the inhomogeneity created by the DC network, relevant to the 1T-polymorph, the CDW system gains additional stabilities by deformation under the discreteness conditions. Here, deviating from the purely sinusoidal form by developing $\Delta\phi(\mathbf{r})$ in the associated DC network, additional jumps by the new vestigial orders are created; see dash line in Fig. 9a. Given that the magnitude of the $\mathbf{Q}$ does not change significantly, the free-energy landscape can be cast in two effective parameters[19]: the amplitude ($|\eta|$) and the orientation angle of $\mathbf{Q}$ ($\varphi$) with respect to $\mathbf{Q}_c$[149] measured in UED. Such a 2D free-energy landscape is schematically depicted in Fig. 9b.

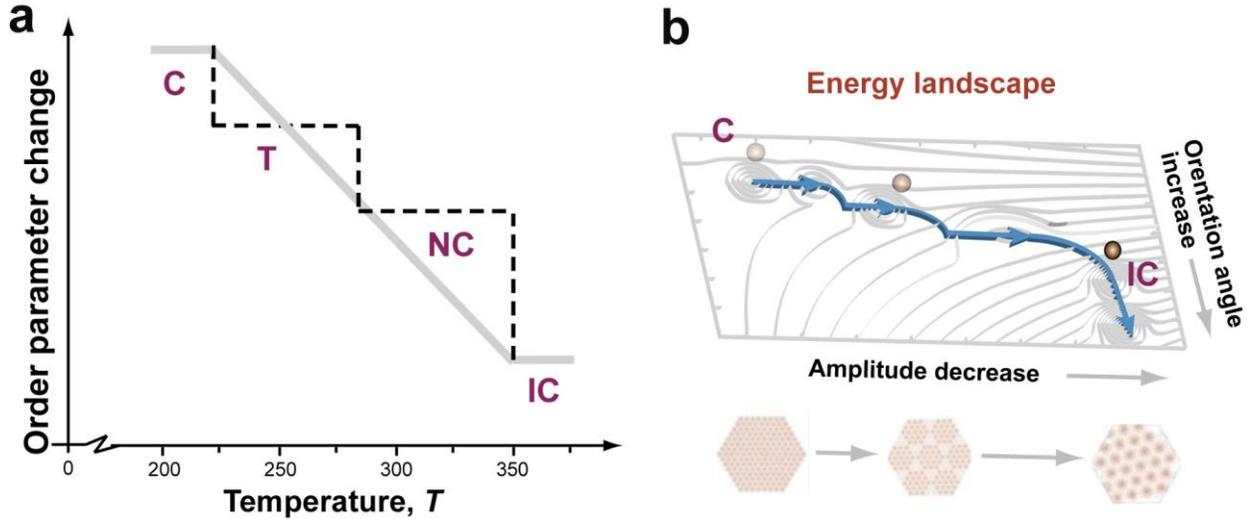

**Figure. 9. Theoretical depiction of the CDW order parameter evolution in TaS$_2$. a**. The temperature dependence of various order parameters. **b**. The two-dimensional landscape hosting the various CDW orders. Panel b is adapted with permission from ref. [19].

More exact models considering the multi-**Q** and high-harmonic effects will be required to precisely determine the experimentally observed vestigial triclinic and hexagonal NC-CDW states[146, 180]. We note, the delicate discommensuration model extended by Shiba and coworkers to include all essential high-harmonics and multi-**Q** interactions based on diffraction [145, 146, 180] can accurately predict the structure of the triclinic and hexagonal NC-CDW states, which have been confirmed by STM experiment[186]. Nonetheless, the phenomenological model given here, while ignoring the multi-**Q** and high-harmonic effects, is sufficient to capture the essential physics of DC phase transitions.

With the phenomenological equation, we are now in a position to understand the basic physics of photoinduced CDW transformation. While not a direct competitive scenario[23], driving down the $|\eta|$ via a laser quench can tip the balance between the commensuration energy gain and the elastic energy cost, triggering the reorganization of the density waves as discussed above. The narrow half-filled band is crucial for the low-temperature Mott phase and the associated cluster-type structure distortions. Therefore, the optical interband excitation from this localized band can lead to the effect of doping poised to shift the energy landscape before carrier thermalization[187]. The experimental phenomenology relevant to the impulse-adiabatic picture of unfolding the order-parameter free-energy landscape[18, 19, 168] is given by recent experiments: trARPES shows the collapse of the Mott-Hubbard gap within 50 fs[187, 188], while the carrier thermalization appears in ~ 200 fs to reduce the Peierls gap[188]; the carrier-lattice equilibrium proceeds on an even longer timescale (ps) which are probed by UED [19, 189-192].

Taking the lead from the interaction quench scenario in CeTe$_3$ problem[23], we can identify the critical thresholds into different DC states as identified by the equilibrium free energy equation. In principle, there is a direct mapping between the equilibrium phase transition and photoinduced phase transition in this manner. By rearranging Eqn. (33), we obtain

$$f = \frac{\alpha}{2}(T - T_{c,l}^*)|\eta|^2 + B|\eta|^4 + H.O.,$$

with a new critical threshold $T_{c,l}^*$ shifted by the interaction quench:

$$T_{c,l}^* = T_{c,l} - \frac{2}{V\alpha}\int d\mathbf{r}[C(\boldsymbol{\nabla}\phi - \mathbf{Q}_{IC})^2 - D\cos(\phi - \mathbf{Q}_C \cdot \mathbf{r})], \qquad (34)$$

where $V$ is the volume of the specimen excited by the laser and $l$ denotes different thermal states. However, the balance between the lock-in energy and the energy cost in deviation from the natural $\mathbf{Q}_{IC}$ is expected to be different in the temperature and interaction-mediated (doping and applying pressure) pathways. Therefore, the intermediate vestigial ordering will likely be different. For example, a stable triclinic phase has not been known under pressure tuning[152, 193] or chemical doping[194, 195].

Nonetheless, this simple picture allows us to better understand the phase diagrams of nonequilibrium phase transition reported by Han et. al. (ref. [19]) and Ravnik et al. (ref. [168]). In these studies, different metastable phases emerge on the ps or longer timescales. The thresholds for creating these states were found to be dependent on the absorbed photon density, rather than the excitation energy. The UED experiments (ref. [19]) gave the results on the CDW amplitude and orientation angle for these states, which can be mapped into the Landau-Ginzburg free-energy landscape assuming they are the stationary conditions. In the spectroscopic investigation (ref. [168]), the time-domain amplitude mode was employed as an indicator for the shifting of the rigidity associated with the free-energy basin supporting the emergence of a new type of CDW orders. These two different views are complementary to each other.

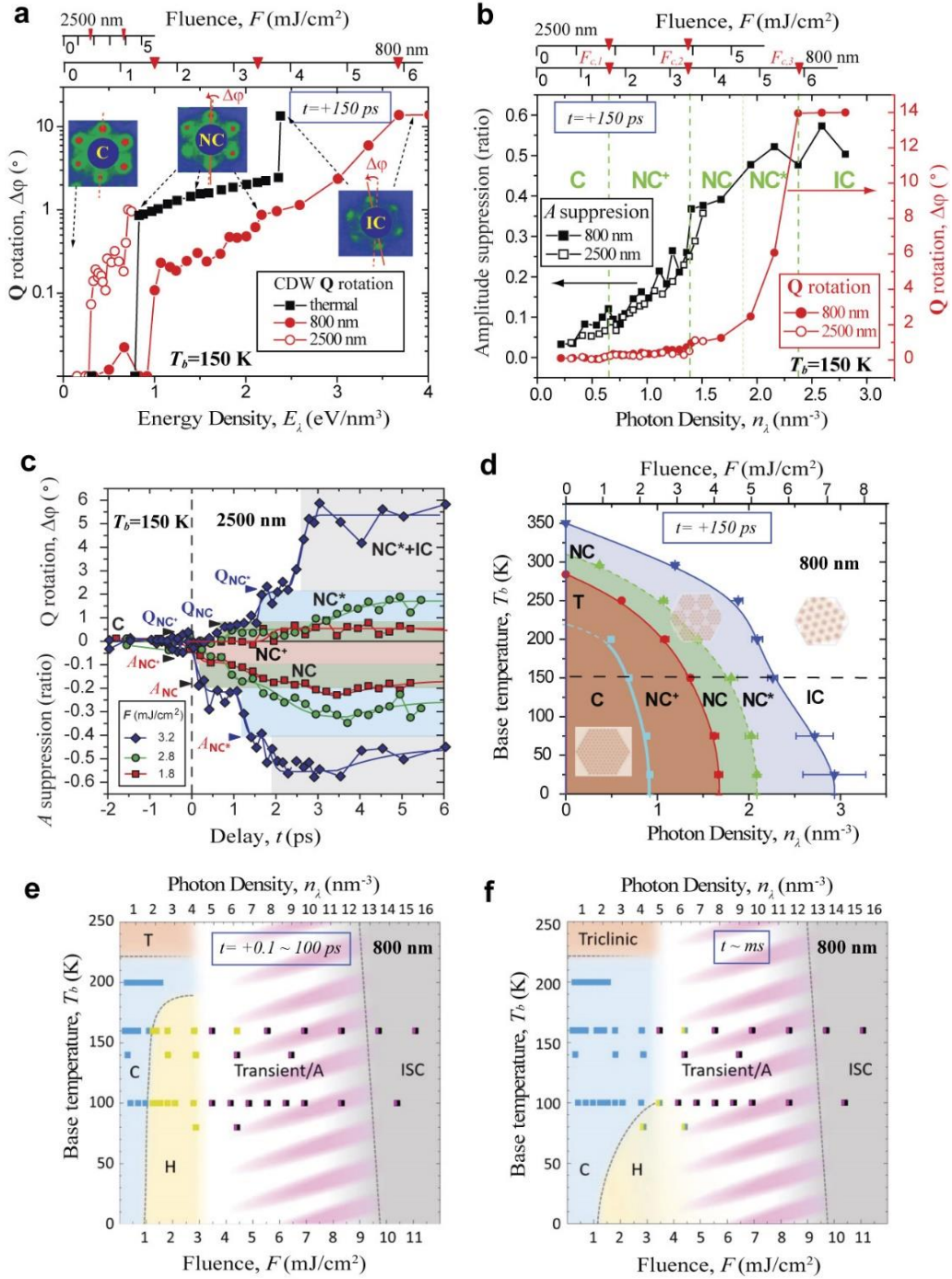

**Figure 10. Photoinduced CDW re-ordering dynamics in 1T-TaS$_2$. a**. Energy-density evolution of the CDW wavevector in thermal and photoinduced phase transitions. **b**. Absorbed photo density evolution of the CDW wavevector and the amplitude pumped by 800 and 2500 nm photons. **c**. The short-time evolution of the CDW order parameters obtained at 2500 nm. **d**. The temperature–photon-density phase diagram obtained at 150 ps pumped by 800 nm photons. e & f. Temperature-fluence phase diagrams obtained for the ultrashort (0.1-100 ps) and ~1 ms timescales pumped by 800 nm photons. Panels a - d are adapted with permission from ref. [19]; panels e-f are adapted with permission from ref. [168].

To understand the role of photo-doping, two different pump laser wavelengths, 800 and 2500 nm, were employed. While in both cases, interband excitation (hence photo-doping) is possible, the excess heat deposited per creation of an electron-hole pair is quite different in the two cases — differs by a factor of 3. In Figs. 10a and b, the respective CDW amplitude and orientation angle changes are recorded at delay $t$=150 ps. The distinct shift in the orientation angle gives a handle to compare phases driven by the two different excitation wavelengths. The relevant changes obtained under the tuning by temperature are included for comparing results based on the applied energy density into the system with the initial state temperature (base temperature, $T_b$) set at 150 K well below the $T_{C-CDW}$ ≈280 K.

We note that from the diffraction results the eventually achieved IC structure (by 800 nm laser) is consistent with that of a thermal state. However, the intermediate phases, identified by the steps in the angular shift, appear to deviate slightly from the known triclinic and near commensurate phases [149, 183]. We also note that in the results obtained from the two different laser excitations the correlation between the amplitude and orientation angle changes is excellent. This allows Han et al.[19] to rescale the applied fluence into a common scale based on the absorbed photon density within the experimental uncertainties, taking into account the difference in the absorption cross-sections from the two laser wavelengths. These results thus are in support of distinct pathways from the non-thermal transitions, irrespective of the applied photon wavelengths. A notable difference is, while the 800 nm laser pulse can bring the system entirely to the eventual IC phase, the 2500 nm excitation fails to reach the same level without causing irreversible change to the CDW system. This observation suggests that thermal energy plays a more important role to fully establish IC-CDW than in the transitions into other vestigial orders.

The phase diagram reported by UED portrays the metastable phases that can persist up to 150 ps (Fig. 10d)[19]. The time-resolved measurements using 2500 nm laser pulses however confirm that the multi-stability nature of the light-induced free-energy landscape already appears on a relatively short timescale[25]. Three different fluences were applied targeting different eventual phases based on the phase diagram established using the data at 150 ps; see Fig. 10c. The phase progression was tracked through $|\eta|$ and $\Delta\varphi$ of the CDW. Without any exception, the phase evolutions progress in a stepwise fashion[19]. First, at a higher photodoping level where one may assume a greater number of free-energy basins will become accessible; see Fig. 9b. However, the steepest descent is only defined by the accessible pathway to the nearest energy basin, and those different basins are separated physically by the length scales set by discreteness conditions (orientation angles) — along which the slope of the free energy is significantly smaller. This explains the stepwise behavior in which the transformation is taken place as the free-energy landscape does not allow a straight path from the initial to the eventual state.

As shown by Fig. 10c the metastable states that could be accessed over a shorter time period are slightly different as the free-energy landscape will continue to evolve over time due to thermalization with the baths. Nonetheless, the key characteristics of these states, as characterized by $|\eta|$ and $\varphi$, do not change significantly. This indicates that the reported vestigial orders are distinct. Furthermore, from the width of the corresponding satellite scattering structure factors, these orders generally possess a long correlation length. As the appearance of the metastable state is relaxational [Eqn (9)], the stronger initial quench due to carrier doping naturally leads to a deeper level of the energy basin to be exposed before the system has time to thermalize. This is an important point to raise because it shows that the impulsive unfolding of the energy landscape is key to set the stage for the order parameter evolution as a relaxational dynamic.

The remarkable ability for the vestigial system to self-organize into different long-range phases in a relatively short timescale shows the evolution is driven largely by the potential effect than the stochastic dynamics. Should the excitation be by a longer pulse, as so demonstrated with a pulse longer than 4 ps in the experiment by Stojchevska et al.[18] and during which the slowly doped system has a sufficient time to interact with the bath, the order parameter evolution would be more thermal-like, namely one would not be able to observe the reported HCDW (ref. [18, 168]) in this case.

There is a growing body of evidences[168, 170, 173] suggesting that the first vestigial order (labelled initially as "T" phase in Ref. [19]) following the laser quench to be one that is identified as HCDW. The key signature to identify a vestigial order is by its deviation from the commensurate ordering $(\mathbf{Q}_c - \boldsymbol{\delta}) \sim \mathbf{Q}_c \Delta\varphi$, thereby defining a characteristic domain size, $\xi_D = 0.69a/\Delta\varphi$ (nm), related by the Fourier analysis[180]. The change in $\varphi$ at the first critical point identified in Figs. 10a&b is ~0.2°, which is much smaller than that of the triclinic phase at equilibrium (~1°). This signifies a much longer-range domain/discomemnsuration structure[180]. Interestingly, this relevant length scale matches the characteristic length scale of the HCDW structure revealed by STM[170], which also suggests the structure as a chiral. A recent X-ray diffraction investigation of HCDW also indicated a $\varphi$-shift and correlated the change with the loss of interlayer dimerization[173] – which the authors attribute as a defining factor introducing the insulating behavior in the C-CDW phase[196], although the relative roles of interlayer stacking and the Mott physics for the insulator-metal transition in $TaS_2$ are still debated[166, 172, 197-200]. The three independent measurements give a similar laser threshold ~ 1 mJ/cm$^2$, for the light-induced phase transition. Therefore, it is very likely the first vestigial order identified here to be the same as HCDW first reported by Stojchevska et al.[18].

All the vestigial states identified by the UED experiments are demonstrated to reverse back to the C-CDW after the pump-probe cycles of 1-10 ms. While the initial discovery of the persistent HCDW is obtained at the lower temperature than studied by UED, with a higher base temperature as investigated by Ravnik et al. the HCDW is shown to be less stable. This may explain why the system is reversible over a ms repetition rate, making it subject to the pump-probe optical measurements. To this end, the time-domain phase diagram established by Ravnik et al. (ref. [168]) also gives a second threshold for transitioning into other types of ordering upon a higher level of excitations — see Fig. 10f. This second fluence threshold is ~ 3 mJ/cm$^2$, which again is consistent with what reported by UED for transition into the second NC vestigial order via 800 nm laser pulses (Fig. 10a)[19]. However, we wish to point out a key difference between the two studies that the irreversibility or disordered states tend to emerge under a higher level of excitation in the STM experiments; whereas in UED studies the system can be driven all the way to the IC state while maintaining the long-range correlation in the photo-doping regime. This notable difference may be explained on the basis of distinction in the sample settings in the two experimental approaches. In the UED investigations, the specimens are typically exfoliated to the thickness less than 50 nm and studied at a higher base temperature. This is in contrast to the STM experiments where the nonequilibrium dynamics are induced at the surface of a bulk sample and typically required a lower temperature to induced a long-live HCDW[168]. Below, we investigate this initial state sensitivity relevant to controlling the evolution of the vestigial orders.

## 6.3 Pump-inhomogeneity-driven dynamics.

Recently from careful analyses of the order parameter responses through time-resolved optical techniques[201] and X-ray diffraction[202], transient stabilization of inverse order parameters[203-206] have been suggested. This possibility is explained on the basis that the broken-symmetry state with local lattice distortion opposite to that of the ground date can be transiently trapped due to the formation of domain walls[201, 207] between the two extrema phases. This scenario was thought to originate from the short penetration depth of the laser pulse relative to the sample thickness. The inhomogeneous excitation could lead to distinctly different phases at different depths within the sample in systems with multiple stable alternative ground states. The externally applied pressure or the strain induced due to sample-substrate mismatch has been used to create a metastable state[169, 208]. It is possible that the light-induced inhomogeneity may be yet another control parameter that plays a role in creating the metastable energy landscape.

We look into this issue by considering near-infrared laser excitations in samples with slab thickness used in the UED experiment (45 nm) and one that is much larger (150 nm) representing the bulk limit. A focus is to consider how different sample thickness settings will alter the excitations and hence influence the interpretation of the results obtained from the different types of pump-probe experiments. In Fig. 11a we show that by simply changing the sample thickness by a factor of ≈3, the excitation of the materials can be altered drastically. The differences are most apparent by comparing the laser absorption intensity profile $I(z)$, calculated for the 1T-TaS$_2$ [209] by solving the Maxwell's equations[210, 211]. In the same sub-surface region (0-45 nm) from the top excited surface ($z = 0$), the 150 nm system shows an exponential-like decay profile with an effective penetration depth of just ≈ 25 nm, which is expected as if the film is a bulk sample. On the other hand, for the 45 nm film, the profile becomes nearly homogeneous. This striking difference reflects the effect of the interferences active in the thinner film where a finite optical penetration leaves a sizeable, reflected component for the waves to interfere internally. This Fabry-Perot effect can sufficiently build up before the ≈50 fs pulse leaves the slab. It can even create an inverted effect as shown in the 45 nm slab where the intensity from the back surface is slightly higher than the one in the middle; see the left panel in Fig. 11b.

To characterize the effect of interferometric modification, we calculate the normalized intensity distribution given in a histogram (see inset of Fig. 11a). The differential RMS value evaluating the dispersion effect is calculated as:

$$\sigma_n = \sqrt{\frac{1}{N}\sum_i \left(\frac{I(z_i)}{\langle I(z)\rangle} - 1\right)^2}.$$

We obtain a $\sigma_n$ of 0.064 for the 45 nm film, whereas for the 150 nm film it is 1.244. It is intriguing to observe how this initial laser intensity profile will drive the diffusion of the carriers and phonons in the systems. The exchange of kinetic energies between the carriers and phonons, which defines respective effective temperatures for electrons, SCP, and WCP subsystems, is described by the three-temperature model (3TM) (Sec. 5 and the Appendix of this paper). We note the typical 3TM[13, 110-113] is conceived mainly for a homogeneous system. To consider diffusions by the hot carriers and phonons excited more significantly at the top surface, we set up the diffusion equations[212] coupling to 3TM and establish the three-temperature diffusion model (3TDM)[213] to allow us capturing the spatial temperature profile evolution. Here, we report on the temperature profiles $T_e(z,t)$ and $T_{lattice}(z,t)$ for the electron and lattice subsystems that will be relevant to addressing the complementary perspectives of the microscopic dynamics probed by the ultrafast scattering, trARPES, THz spectroscopy, and optical reflectivity techniques.

We set up the problem by considering the ultrafast dynamics of electron temperature, $T_e(t)$, made available recently for 1T-TaSe$_2$ by trAPRES experiments[175]. We apply the 3TDM with the 3TM coupling parameters determined based on fitting the experimental $T_e(t)$ data. These parameters are used as the generic coupling constants and applied to the 1T-TaS$_2$ dynamics. We note that from the fitting of the coupling constants between the electron subsystem, SCP and WCP, we derive three characteristic timescales for 'electron-phonon' coupling: 0.5, 1.8 and 40 ps, which lead to a general ps heating dynamics of the lattice consistent with many recent experiments[190-192, 214]. We then report in Figs. 11b&c the transient temperature maps $T_e(z,t)$ and $T_{lattice}(z,t)$ with the mean lattice temperature calculated according to $T_{lattice} = \alpha T_{SCP} + (1-\alpha)T_{WCP}$, where $\alpha$=0.1 as the fraction for SCP is determined also from the fitting. The details of the modeling and the parameters used are listed in the Appendix.

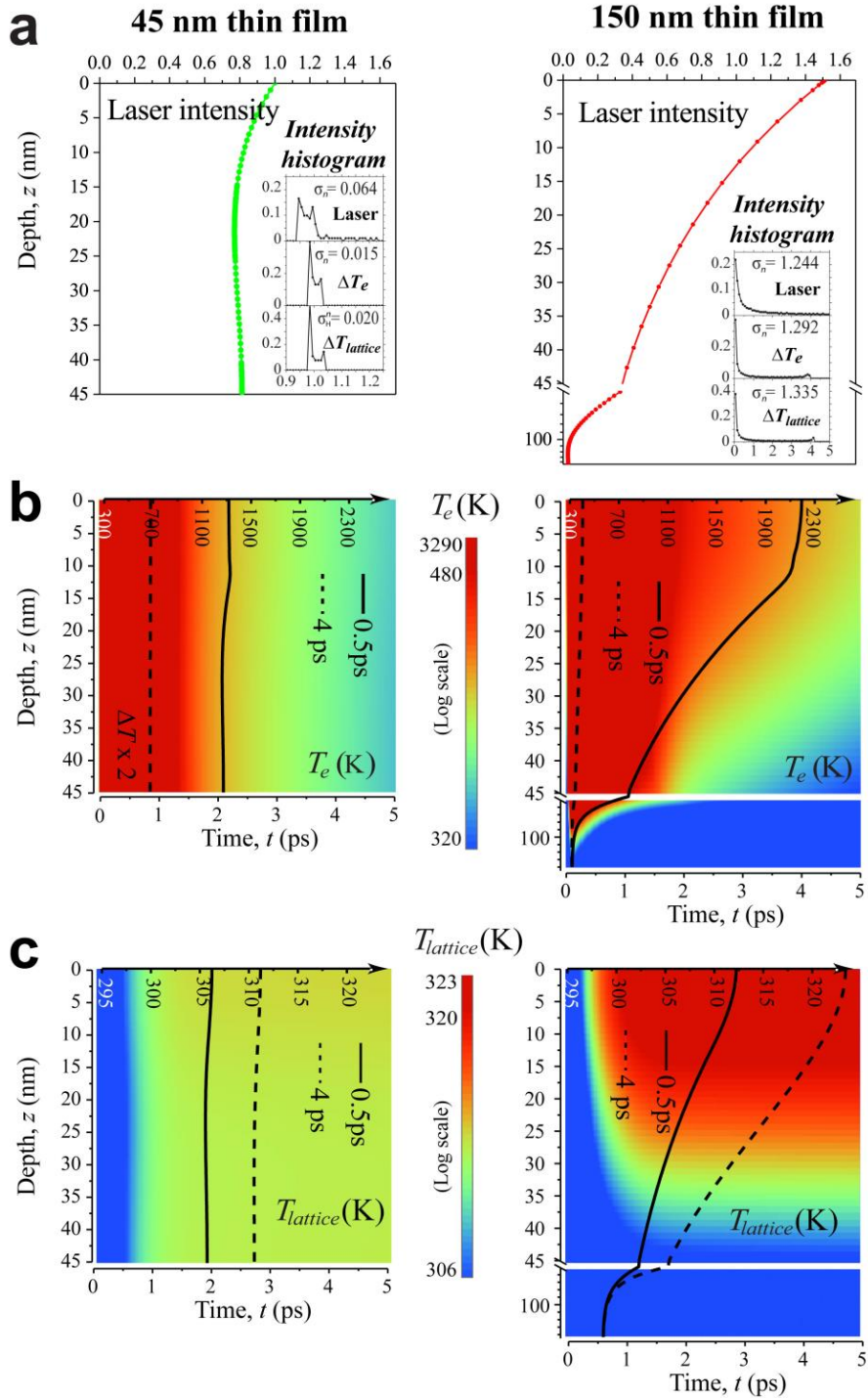

**Figure. 11 Quasi-particle (electrons and phonons) temperature evolutions in photoexcited TaS$_2$ thin films by the three-temperature diffusion model. a**. The laser excitation (800 nm, 50 fs) profile within the 45 and 150 nm thin films. The inset shows the histograms of the intensity and temperature distributions along z taken at 4 ps. **b**. The time

($t$)-depth ($z$) evolution of the electronic temperature, $T_e(z,t)$. **c**. The time ($t$)-depth ($z$) evolution of the lattice temperature, $T_{lattice}(z,t)$.

The full results for the first 5 ps are depicted in Figs. 11 b&c in the dynamical temperature maps of $T_e(t,z)$ and $T_{lattice}(t,z)$. To see the effects of diffusion, one can compare the temperature profile at a given time (the results from 500 fs and 4 ps are highlighted) with the initial $I(z)$. Notably, for the 150 nm film, diffusion strongly modifies the temperature profile already in the sub-ps timescale in the top surface region, which is probed by trARPES and X-ray diffraction. From the results obtained at 500 fs for the 150 nm film, the diffusion effect creates a nearly homogeneously excited top region (≈10 nm) and a sub-surface region where a sharp decay occurs; see the right panel in Fig. 11b. However, over the entire period of observation, the strong inhomogeneous profile over the 150 nm film has not been lifted in the temperature evolutions. Meanwhile, for the 45 nm film the diffusion effects have improved the uniformness of the temperature profile further, as indicated in the $\sigma_n$ calculated for $T_e(t,z)$ and $T_{lattice}(t,z)$ going from 0.064 of $I(z)$ to 0.015 and 0.020 respectively. The high degree of uniformness starting even at the short timescales is expected to be an important aspect in characterizing the critical dynamics associated with the nonequilibrium phase transitions.

We examine how this pump-associated inhomogeneity affects the experimental results. Locally applying the impulse-adiabatic scenario for introducing the free-energy landscape, the strong inhomogeneity scenario is expected to have a direct consequence on the order parameter competitions in the key length scales. This effect has been recently reported for SmTe$_3$ system[202], but the phenomenon has been discussed earlier for TbTe$_3$[201]. The authors posited that the excitation intensity gradient may lead to phase separation along the slab making the domain wall formation between the highly excited region and below. The domain wall formation is said to have helped stabilize the formation of inverse order parameters which were recently hinted also in other TMDC systems[203-206].

We now discuss how the pump-inhomogeneity-driven dynamics affect the dynamical order parameter measurements. It may be now apparent that employing a reasonably thin or small volume specimen will present a better chance to investigate the delicate nonequilibrium critical dynamics. The homogeneous initial condition in the pump profile allows the volume-integrated scattering intensity profile to reveal the true statistics of the nonequilibrium phases in the transient evolution of the state. In contrast, while the surface sensitive probe also samples a relatively homogeneous region, the dispersion effect is less well-defined with the diffusion effects present in nearly all the time. This energy dissipation from the surfaces however might help reduce the bath temperatures in the region that couples to the light-induced state. The study of CeTe$_3$ suggests this may lead to a slower decay of the transiently induced hidden phase[23]. In general, $\sigma_n$ sets a limit on the ability to characterize the width of the phase transition curves and the critical thresholds which are crucial for modeling the nonequilibrium phase transitions as discussed earlier. In addition, the 3TDM investigation here shows that to cross-correlate the results from the different types of measurements, the empirically derived critical thresholds will need to be adjusted to consider the excitation profile differences. Such differences are likely to play a role in properly assigning the optical doping concentration in Figs. 10d-f.

Two different possibilities of additional phase control might result from the pump-induced inhomogeneity. Forming domain walls between competing broken-symmetry phases is one of them. The second control could be the strain effect introduced by the dynamical inhomogeneity, which will be most dominant in a system where different morphologies or lattice structures are created during the phase transition. Such effect could be mediated by the interlayer couplings in layered quantum materials.

## 6.4. UEM experiments

The ultrafast electron microscopy provides a unified setting to probe the order parameter dynamics in both the real space and momentum space. This can potentially address the questions concerning inhomogeneities both in terms of the sample as well as the pump setting as discussed earlier. Central to the multi-messenger approach is to cross-correlate the information obtained about the same illuminated materials in spectral, imaging, and diffraction modalities. Thus far, the discussion of nonequilibrium phase transition presumes a homogeneous dynamical system - one that does not always occur as discussed earlier.

Transmission electron microscope (TEM) provides information of the microscopic features in greater detail than optical approaches. Such information is highly valuable in cross-examining the results provided through spectroscopic or diffraction approaches that typically integrate signals across the region of the entire probe. Unpacking site-specific information requires adjusting only the post-specimen imaging optics of the TEM[131, 215], which, when switched from the diffraction setting to imaging, offers a direct view of the dynamical system in exactly the same pump and probe conditions at the sample plane.

There are multiple ways with which the imaging contrast can be created under a TEM. In Fig. 12a we present the typical exfoliated sample images of $TaS_2$ obtained using the ultrafast electron microscope. The image is obtained in the so-called bright-field mode in which an aperture at the back focal plane of the objective lens (the diffraction plane) passing only the unscattered beam to form the image[216]. In this geometry, the mass-thickness determines the baseline of the intensity level — e.g. see the thinner film at the edge and the small debris left on top of the film are respectively lighter and darker than the general area of the 40 nm film. But even with the same film thickness, the diffraction effect produces fringes, so-called bend contour, from the sample curvature[216].

The diffraction-mediated contrast is sensitive to the inclination of the sample relative to the beam. At the regions of the bright or dark fringes, selected by a tilt angle, the diffraction contrast is enhanced drastically[216]. This is utilized to measure the acoustic waves created as a side product of the laser excitation. The waves traveling back and forth between the front and the back surfaces form a standing wave pattern modulating the sample thickness[217-224]. At the level of exciting density-wave phase transition (~ 1 $mJ/cm^2$ in fluence), the lattice acoustic wave amplitude is expected to be less than 1% of the lattice constant based on heating[147, 225, 226] . However, the intensity modulation from the bend fringes can achieve resolution better than 0.01%, serving as a powerful way to resolve the out-of-plane dynamics.

The dynamical contrast based on the oscillation frequency can offer a measurement of sample thickness to the precision of mono-stacking layer [218, 223]. This is shown in comparing the results from different regions, marked in A, B, C, and D in Fig. 12c; here in the region B the film oscillates in the direction opposite to that of the regions A, C, D. Taking the oscillation out to 400 ps, the oscillation from region D is slightly out of sync with the rest. From the phase difference, one can determine the delay of 6 ps being developed and gives a local film thickness of 0.59 nm (one Ta layer; Fig. 7) higher than the other areas.

Of particular interest is to compare the out-of-plane crystal oscillation with the photoinduced CDW order parameter dynamics, which one obtains by switching to the diffraction mode, as shown in Fig. 12b. Both experiments are conducted under the same sample pump conditions (the ~ 1 $mJ/cm^2$, 50 fs near-infrared pulse illumination in nearly normal incidence at the repetition rate of 1 kHz). It is interesting to see the two dynamics decouple from each other. The CDW order parameter responds to the laser quench within the first 500 fs, and yet it takes more than 5 ps for the acoustic wave amplitude to start building up. The results here show there is no impulsively driven strain wave created over the first few ps, which is anticipated due to the fact that the in-plane heating, which is much more effective here, is largely homogeneous across the layer as predicted for the 45 nm film (Fig. 11). Furthermore, the slow onset and the persistent oscillation of the out-of-plane mode and the absence of such signatures from the CDW order parameter dynamics indicates that the light-induced phase transition behavior is largely a 2D phenomenon — although one fully expects the 3D structural relaxation [226, 227] will take place when the new 2D structure motif is established and the system thermalizes eventually.

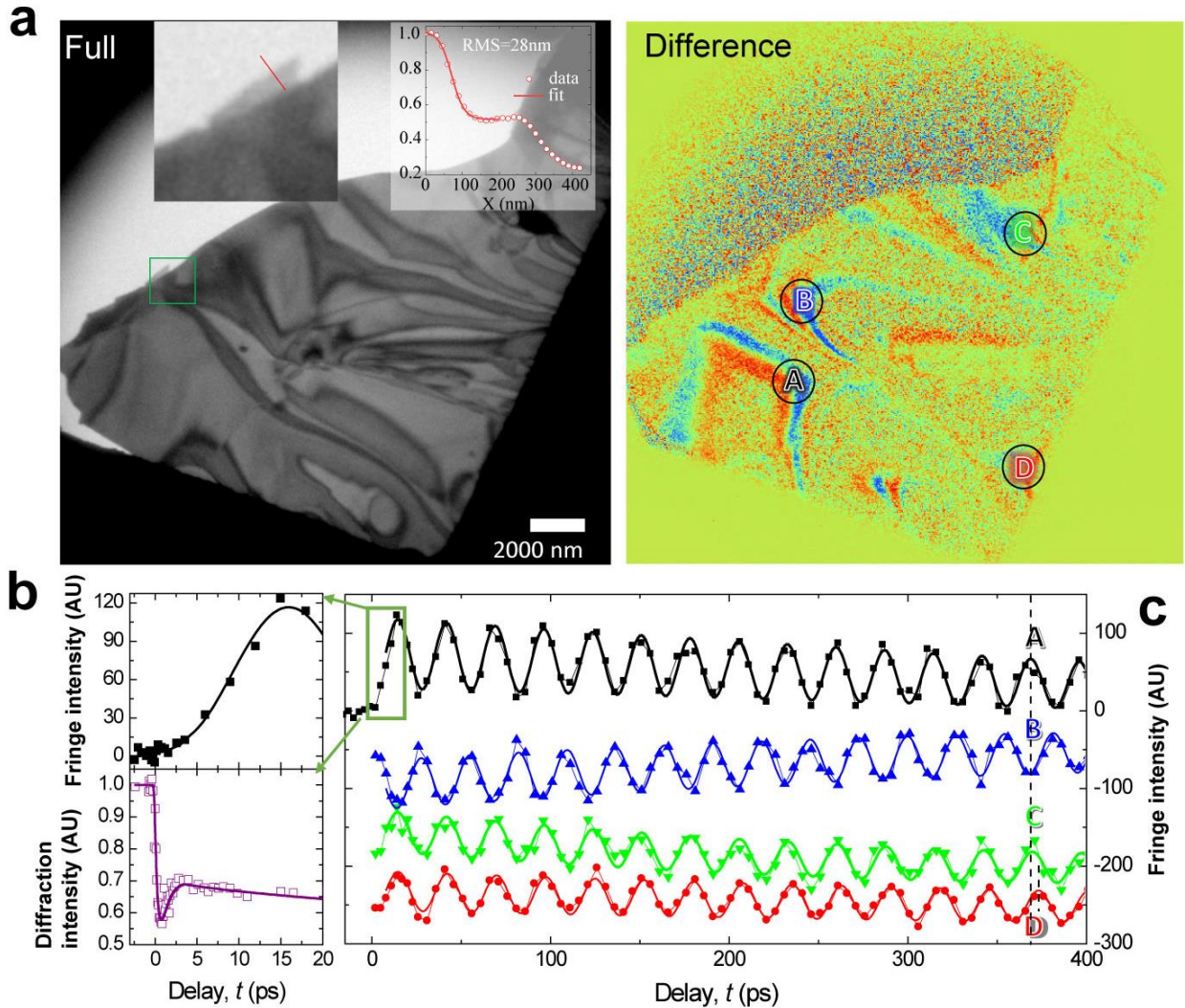

**Figure 12. Ultrafast electron microscopy investigation of 1T-TaS$_2$ thin film. a.** The bright-field image (BFI) of the TaS$_2$ thin film taken under the UEM in full and difference images. The full image is taken at $t= -20$ ps; while the difference image is taken by subtracting the negative-time full image with one obtained at $t=15.5$ ps. **b.** The BFI intensity oscillation versus the CDW satellite intensity evolution in the first 20 ps upon applying 1.5 mJ/cm$^2$ near-infrared laser pulse. **c**. BFI intensity oscillations at selected regions of the image (see **a**). Region A, B & C have the same time period (*T*) of 27.20±0.05ps. A and B are out of phase. Area D has a slightly different time period 27.60±0.08 ps, which is 1.5% higher than the other areas. After 14 cycles of oscillation, we can see a clear offset from area D. The longitudinal speed of sound along the **c**-axis is $v \approx 3$nm/ps [54, 228]. The sample thickness can be estimated by $d = vT/2 \approx 40$ nm. The 1.5% difference in the time periods gives ~0.59 nm difference in thickness, which is almost exactly 1 van der Walls layer of TaS$_2$.

## 6.5 Light-induced states in strained vanadium dioxide nanocrystals.

Vanadium dioxide is a prototypical phase change material in which a strong metal-to-insulator transition (MIT) occurs near room temperature ($T_M$=68 °C)[229, 230], making it a subject of strong interests in applied fields[231, 232]. In MIT, distinct changes in the lattice symmetry from rutile to monoclinic are involved. The equilibrium temperature-stress / temperature-doping phase diagram involving a triple critical point[233-235] between the rutile (R) and the two monoclinic phases ($M_1$ and $M_2$) is depicted in Fig. 13a. However, decoupling phenomena between MIT and structure phase transition (SPT) have also been reported in scenarios with additional stimulation by light[236, 237], or applications of external field or current[238], as well as under strain[239, 240] or driven by the interfacial carrier doping[241, 242]. These unexpected complex phase change behaviors casted in doubt the conventional wisdom of MIT based on a simple extension of the Peierls[243] or Mott[244] physics, but rather a picture where both Peierls and Mott physics are involved, referred to as Peierls-Mott or Mott-Peierls mechanism[245-249].

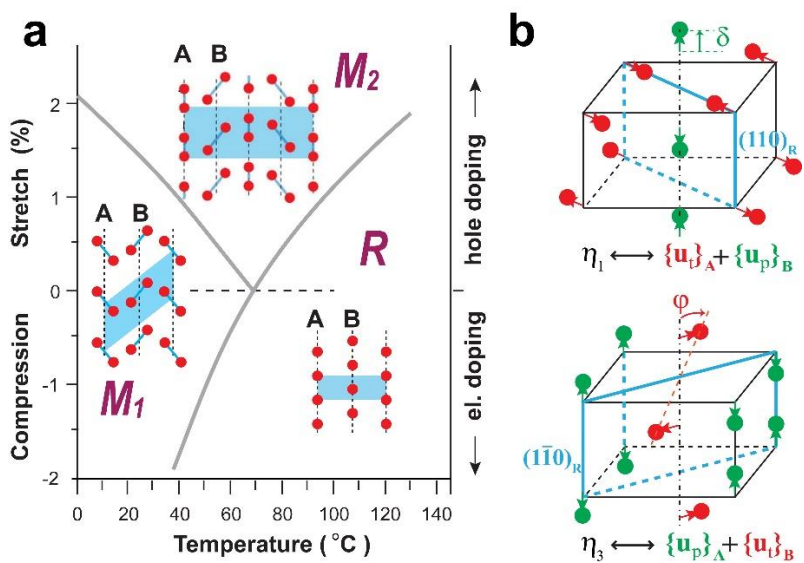

**Figure 13. VO$_2$ structure phase diagram. a.** The schematic phase diagram of VO$_2$. The layout of the structures highlights the vanadium atom distortions in the A and B sublattice chains (oxygen atoms are omitted). The shaded areas represent the unit cell. **b.** The two types of structural distortions $\mathbf{u}_p$ and $\mathbf{u}_t$ along sublattice chains couple to form the structural order parameters $\eta_1$ and $\eta_3$ for describing phase transitions between R to different broken-symmetry phases[230, 250].

The VO$_2$ SPT, which intimately couples to different types of MIT, can be described over the two types of antiferroic structural distortion along $\mathbf{c}_R$[251]: $\mathbf{u}_p$, longitudinal V shift (~ 0.14 Å) forming V-V pairs and $\mathbf{u}_t$, transverse V shift (~ 0.18 Å) twisting the V-V chain away from $\mathbf{c}_R$[230]; see Fig. 13b separately in red and green dots representing the movements. Empirical data have suggested that these two types of distortions are not independent. Two degenerate representations involving the coupled $\mathbf{u}_p$ and $\mathbf{u}_t$ distortion occurring separately in two sub-lattice chains (A and B) are outlined in Fig. 14b as $\eta_1$ and $\eta_3$ (following the convention used in ref. [250]) where $\mathbf{u}_p$ and $\mathbf{u}_t$ are located in the orthogonal $(110)_R$ and $(1\bar{1}0)_R$ planes of the rutile structure. Alternatively, one may also consider $\eta_2$ and $\eta_4$ where the sequence of $\mathbf{u}_p$ and $\mathbf{u}_t$ in the orthogonal $(110)_R$ and $(1\bar{1}0)_R$ planes are switched. However, only one such a pair is required to construct the Landau-Ginzburg free-energy density functional for VO$_2$ written as [250]:

$$f = \frac{a}{2}(T - T_M) \sum_i \eta_i^2 + \frac{1}{4}\sum_{i,j} b_{ij}\eta_i^2\eta_j^2 + \frac{1}{6}\sum_{i,j} d_{ij}\eta_i^2\eta_j^4 \qquad (35)$$

where the 4 known stable phases can be identified as the local minimum states with the coordinates of the free-energy basins at (0,0) for R, $(\eta, \eta)$ for $M_1$, $(\eta, 0)$ for $M_2$, and $(\eta, \eta')$ for T (triclinic phase) [250]. In this representation, $M_2$ state, considered as an intermediate between R and $M_1$[235, 252] under hole doping or tensile stress (Fig. 14a) can transform continuously to $M_1$ or R – by simply a progressive $\eta_3$ or $\eta_1$ distortion, respectively[230, 250].

These structural distortions may be linked to the electronic changes at the MIT. The $\mathbf{u}_t$ shift destabilizes the $\pi^*$ state and provokes an interband charge transfer from $\pi^*$ to the quasi-1D $d_{//}$ band, which becomes half-filled and can then trigger the Peierls instability to open up an insulating gap through the V-V pairing ($\mathbf{u}_p$) in the $M_1$ state[229]. However, electron-electron correlations may not be ignored in this scenario[244, 252]. The NMR studies have found that the electron gas in the metallic rutile state is weakly correlated, while the electrons are localized in the dimerized V-V chains in $M_1$[235].

Given that $VO_2$ also exhibits key characteristics of electron-lattice-coupled competitions found in $RTe_3$ and $TaS_2$, it is intriguing to ask if there will be a light-induced metastable state whose properties are unlike any known thermodynamic phase of $VO_2$? Before addressing this, we first note that recent advances from ultrafast measurements have provided rich literature to cast light on the microscopic mechanism of phase transition induced by light excitations. The ultrafast transient electron dynamics investigated by optical[253-255], THz[256-258] and, trARPES[259, 260] generally suggested a more rapid collapse of the band gap when compared to the structure transformations, which were investigated by the diffraction techniques[236, 261-264]. In addition, the transition thresholds as identified by various spectroscopy techniques were found to be smaller than those of the diffraction approaches[262]. These results were taken as indicators that IMT and SPT are decoupled in light-induced phase transitions. However, a recent re-examination of relevant ultrafast spectroscopic measurements has called into question some of the early claims, citing the differences in the pump-probe repetition rates and sample settings as the causes for discrepancies[121]. These issues remain unsettled and remind us of the challenging topics related to the cross-examinations. At the heart of the debates are not simply the nonequilibrium physics from the Peierls distortions versus the electron correlations[229, 230] but also the issues pertaining to the sample conditions[232, 265], such as the strain, disorder, and interfaces that all play a role in the photoinduced phase transitions.

Here, we give a simple phenomenological nonequilibrium model that might unify the understanding of different recent results from UED, trARPES, and optical measurements. It is based on the impulse-adiabatic free-energy evolution picture as one gives to understand the phase transition of the density wave systems. Upon ultrafast laser excitation, a transient free-energy surface is created in which the $M_1$ structure is no longer the lowest energy basin[250]. Here, the monoclinic phase of $VO_2$ is characterized by two effective distortive parameters: the twisting angle φ and pairing displacement δ (Fig. 13b), which can be directly deduced by analyzing the powder diffraction of $VO_2$ in the UED experiments. The trajectory of the two distortive parameters which are linked to $\eta_1$ and $\eta_3$ (Fig. 13b) can map the phase transition over the new landscape. Because the eventual R phase has an energy basin where both $\eta_1$ and $\eta_3$ are zero[235, 250], the relative dynamics of the transient state can be described on the new free-energy surface as symmetry-recovery by melting the two order parameters – simultaneously or sequentially.

The impulse-driven new energy landscape mediating the phase evolution is justified from the initial rapid suppression of the distortive symmetry-breaking parameters. The events have been recorded by the UED experiments[236, 261, 262], which show the static amplitude of the pairing disappears within the first 300 fs of applying laser pulses beyond a threshold $E_{th}$. Meanwhile, trARPES has reported the Mott-Hubbard gap collapses on an even shorter timescale (< 100 fs)[255, 259, 266], which is understood as driven by the photocarrier doping[255, 266] to both the localized bonding and nonbonding orbitals[266]. Below the threshold dose, the laser excitation typically leads to coherent gap dynamics at 6 THz, observed by the transient THz conductivity[256-258] and optical reflectivity measurements[254, 255]. This coherent phonon signature is effectively the amplitude mode and has been successfully employed as a marker for monitoring the presence of the $M_1$ order[254, 255]. The density functional theory (DFT) calculation[266] puts an optical doping density at $10^{21}$ e-h pairs/cm$^{-3}$ that is needed to entirely soften the landscape supporting the experimental findings.

The DFT found that the characteristic pretransition 6 THz coherent phonon generation upon optical excitation is tied to the perturbative change of the free energy of the monoclinic phase, induced by hole carrier doping of the Mott phase. The V-shift associated with this displacive mode is in the monoclinic $\mathbf{b}_M$-$\mathbf{c}_M$ plane, which is the $(0\bar{1}1)_R$ plane of the rutile structure[266]. Therefore, there is no obvious route to continuously change the $M_1$ phase to the R phase through rectifying the coherent distortive mode excited by hole-doping. Additional (transverse) modes have to be involved as well due to the nonzero contributions from the distortive components, i.e., $\mathbf{u}_p$ and $\mathbf{u}_t$, of the order parameters out of the $(0\bar{1}1)_R$ plane.

We now consider the role of sample settings. The hidden 1D characteristics of $VO_2$ phases give a very prominent lattice constant changes along the crystalline $\mathbf{c}_R$ direction during the symmetry-breaking phase transition. End-clamping the $VO_2$ nanobeams, grown along $\mathbf{c}_R$, upon cooling from the R phase is equivalent to applying tensile stress, which is known to induce the intermediate $M_2$ phase [235, 250, 252]. This occurs because applying tensile stress or hole doping tips the balance of the cooperative symmetry-breaking between $\eta_1$ and $\eta_3$, which is maintained during the thermal R-$M_1$ transition under ambient pressure. On the other hand, in studying ultrafast $VO_2$ phase transition the stress is transiently induced by the rapid heating across $T_M$ where the lattice expands disruptively. Stress relief is a key step involved in the transition between $M_1$ and R phases[264]. This calls into consideration of the feedback effect from the internal pressure created during phase evolution on the external control parameter. Crystal cracking upon rapid heating has also been a well-known problem[267] in studying bulk or epitaxially grown materials in the pump-robe studies[25]. To remediate these effects, employing free-standing nanobeams or microbeams[268, 269] or small-volume[270] $VO_2$ crystals can not only preserve the intrinsic first-order transition characteristics but also maintain the integrity over repeated experimental cycles without causing the sample to crack.

Taking the transient stress into account, Tao et al. conducted the UED experiments using 31 nm $VO_2$ crystalline grains deposited on the non-epitaxial Si membrane (9 nm)[262]. The sample excitation involves both 800 nm and 2000 nm laser pulses — a setting similar to the study of the $TaS_2$ system to verify the photodoping effect driving the nonequilibrium phase transitions. The powder diffraction patterns of $VO_2$ phase transitions are presented in Fig. 14a. The inhomogeneous crystalline grain size distribution leads to a dispersion effect manifested in the line-broadening in the diffraction, which can be attributed to the surface-strain size effect. Given the high sensitivity of $VO_2$ phase transition to the strain, the size dispersion also leads to a broadening of $T_M$ in the transition curves. The broadening effect is naturally extended to the pump-induced state under both pump wavelengths; see Fig. 14b for the transition curves taken based on the diminishments of the monoclinic reflections, such as $(30\bar{2})_M$.

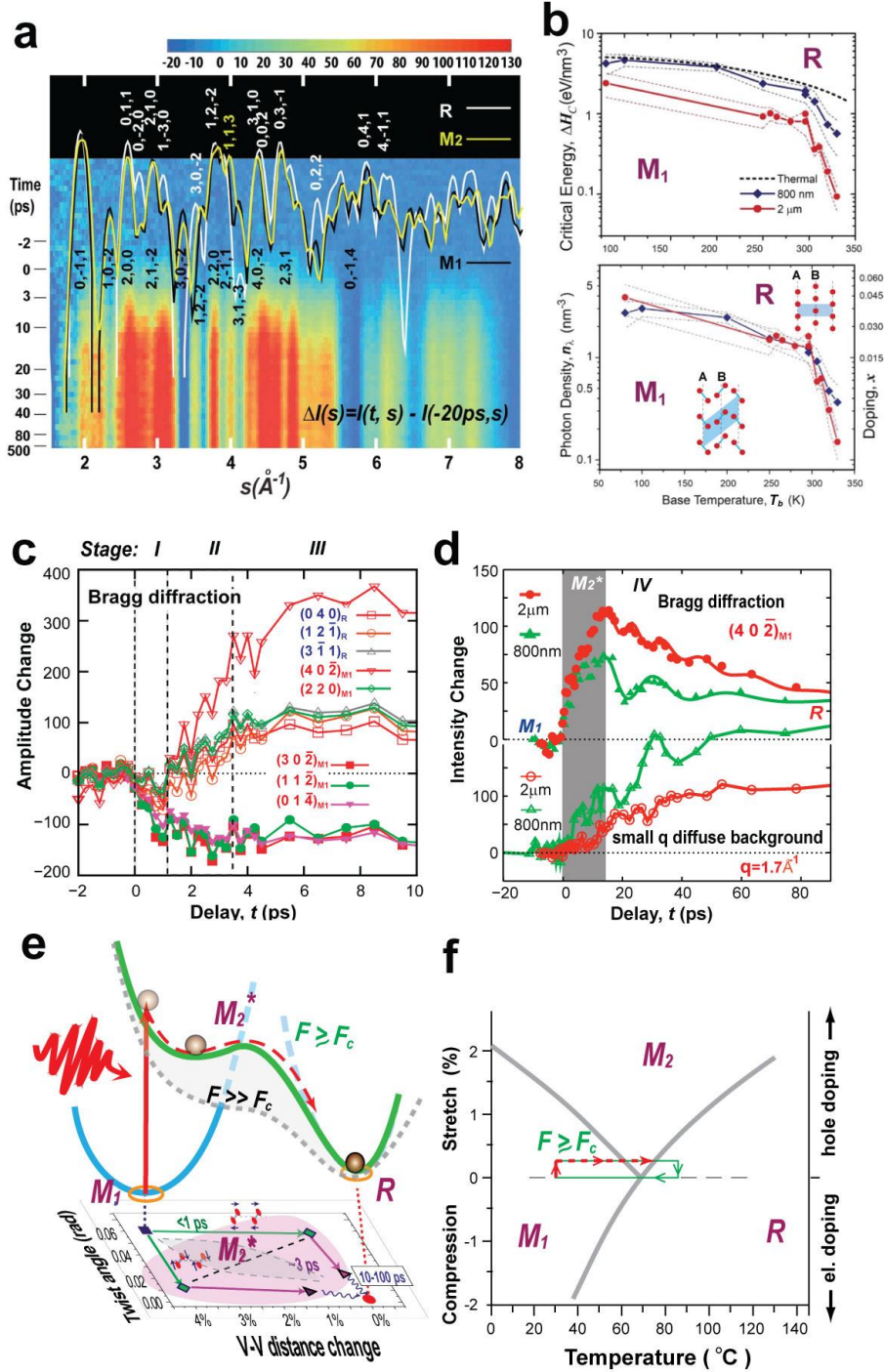

**Figure 14. VO$_2$ structure phase diagram. a.** The diffraction curves modelled after different VO$_2$ phases and the corresponding diffraction difference map highlighting the change in intensities relative to the unpumped $M_1$ phase (t<0). **b.** The empirical energy density ($\Delta H$) – temperature ($T_b$) and photon density ($n_\lambda$) – temperature ($T_b$) phase diagrams for VO$_2$ phase transitions obtained under two different pumping wavelengths. **c.** Selected diffraction evolution at the short time. **d.** The $(40\bar{2})_M$ diffraction evolution at a longer time showing a cross-over behavior. **e.** Schematic free-energy changes that drive a non-straightforward path from $M_1$ to R state. The intermediate structure domain is that of the $M_2$, but not exact; hence noted as $M_2^*$. **f.** The structure pathway under the hole doping, involving $M_2$. Panels a-e are adapted with permission from ref. [262].

It was shown that a stable R phase can be introduced at the +150 ps time period by ultrafast pulses of 800 nm and 2000 nm wavelengths, respectively. The transient phase transitions occur despite that the calculated absorbed energy density is much lower than what is typically required for thermodynamically heating up the crystal lattices from the base temperature ($T_b$) to $T_M$; see Fig. 14b top panel. Similar to the case of TaS$_2$ this nonthermal scenario is reconciled by joining the scales for phase transition using the absorbed phonon density $n_\lambda$ rather than the deposited energy $\Delta H$ for comparison; see Fig. 14b bottom panel. Incidentally, the threshold behavior set by the disappearance of the monoclinic features is at the level of $4\times10^{21}$ e-h pairs/cm$^{-3}$ at low temperatures. This is in the range of what was presented for the carrier-doping-induced collapse of the Mott-Hubbard gap — a proof that the applied photon dose would be sufficient to transform the monoclinic free-energy surface despite of being sub-thermal, paving the way for photoinduced phase transitions.

However, from the free-energy perspective the heating effect must also be considered, especially in this cooperative system where a strong shift in $T_M$ is known from applying pressure or doping (Fig. 13a). Hence, similar to the case of TaS$_2$, the effective critical temperature $T_M^*$ for the phase transition (Eqn. (34)) could be downshifted by the photo-doping effect. This explains why a lower laser dose is needed for the transition into R; whereas by the same reasoning, increasing $T_b$ shall also decrease the required laser dose — an effect observed in Fig. 14b closer to $T_M$.

For a sufficiently small crystalline specimen, one may argue that the system may self-manage the elastic stress and favors a homogenous state to reduce the free-energy cost as the microbeam experiments have repeatedly confirmed[232]. This makes the experiments by Tao et al. [262] interesting with regard to the dynamics of stress relief[264] and its implication for the photoinduced phase transition. There is a better chance of seeing a more stable free-energy basin or saddle point from the order parameter dynamics if the pump fluence is placed just above the critical threshold. This allows the relaxation dynamics to map the intermediate landscape with less heating-related blurring induced by coupling to the stochastic bath set up by laser pumping. The chosen fluences in the comparative experiments are set at 20% above the mean threshold, corresponding to the two sigmas of the transition window (Figs. 14a, c &d) for two wavelengths.

The line-scan data of the raw diffraction images with the pre-pumped state pattern subtracted (diffraction difference image) to show the pump-induced state evolution are depicted in Fig. 14a as the color map. The general trend of ultrafast evolution of the multi-step changes in the representative group of symmetry-breaking and recovery Bragg peaks are shown (see the downward moving and up-warding moving raw peak intensities respectively in Fig. 14c where an initial sub-ps decrease of diffuse background offsets the scale), which, in a broad stroke, are quite similar to the other UED investigations of systems with larger gain sizes[237] or single crystals[261]. Here, the universal collapse in intensities of pairing-related peaks and a slower increase in intensities of the higher symmetry peaks of the rutile state are quite pronounced. We take these signals as an evidence showing the route from the initial state of M$_1$ to the eventual state of R is not direct. This reminds us of the same phenomenology in the previous two systems where we also see the interaction quench leads to a different order-parameter basin or saddle point before turning toward that of the eventual state.

The gentler quench (here for 800 nm it is at ~ 8.5 mJ/cm$^2$, which is 2-5 times smaller than the deep quench investigations[237, 259, 271]) does give more details of the trajectory reflecting order parameter evolution immediately following the intensity drop of the low-symmetry peaks; here a slower second decay in intensities appears on a similar time scale to the rise of the high-symmetry Bragg peaks after their delayed onset. However, the high-symmetry groups will continue to evolve. On this longer evolution (Fig. 14d), the behaviors from the two different excitations will diverge after 10 ps where the higher photon energy pulse leads to a quicker turning over in the $(40\bar{2})_M$ (a high-low symmetry mixed peak; see Fig. 14d) dynamics to settle on a more even level characteristic of the R phase than the trajectory taken by the 2000 nm case.

An acoustic modulation of the peak intensity characteristic of the breathing mode of the nanoparticles was also identified driven by impulsive heating on the ps timescale(Fig. 15d). The frequency changes evidence a shift in the elastic speed of sound from one characteristic of the monoclinic phase to the slower one of the rutile phase, occurring interestingly on roughly the same timescale as the $(40\bar{2})_M$ intensity changes. These results seem to indicate that the acoustic potential of the broken-symmetry state only changes into that of the melted phase after the transient stress is released, as unveiled by acoustic modulations. The less distinct turning over behavior under the 800 nm excitation leads one to believe the transient state characteristics would be even less distinct if a higher excitation fluence were applied. An increase in the diffuse background observed here is consistent with more heat being generated by the 800 nm pulses. The transient heating, while largely impacting the system after the impulse period, will modify the adiabatic free-energy surface with an increased base temperature along the path that makes the trajectory more like the thermodynamic one with a larger stochastic blurring effect; see the deep quench scenario ($F \gg F_c$) in Fig. 14e. The recent ultrafast thermal diffuse scattering studies of $VO_2$ phase transition under a deep quench have shown the transition pathway to be largely stochastically driven[271].

As the photo-doping effect is involved, it is intriguing to consider the role of the $M_2$ known to be introduced under a moderate level of hole doping. However, more excessive hole doping will drive the system eventually to the R phase where the $M_2$ state may pose as a saddle point. A simple structural model based on the homogeneous system evolution suggests this is a more likely scenario. The structural model applies a continuous change in (φ, δ) to fit the experimental results with the coordinates of the $M_1$ phase as the starting point, as presented in Fig. 14e. Unsurprisingly, the data cannot be reconciled with a direct path from $M_1$ to R[262]. The best fit to the results requires the two sub-lattice to act differently as is the case of the $M_1$-$M_2$ transition where only one order parameter ($\eta_3$) is to be melted. The short period of metastability and limited gains in strength signifies that the lattice instabilities normally associated with the $M_2$ symmetry are gaining, but it is unlikely to be consolidated into a long-range order. The digression into the $M_2$ structural domain is simply because, when induced, the $M_2$ basin/saddle point is closer to the initial ground state than the R basin on the free-energy surface. To further transition into R, the additional order parameter ($\eta_1$) needs to be melted. This situation occurs, as indicated by the findings, as an energetically more favorable path after the system has entered the domain of $M_2$ but not before.

One cannot rule out yet other vestigial orders, as predicted by the free-energy equation with all four independent order parameters considered[250] may be involved. However, the gateway picture proposed here based on a non-straightforward order parameter trajectory is different from an isostructural solution identified as a monoclinic metallic phase[237, 263]. Some modification of the structure distortion prototypical to $M_1$ might still contribute to the difference map (Fig. 15a) as the low-frequency background. Future more precise experiments allowing a better momentum resolution to distill out different symmetry-breaking contributions shall help address this topic.

## 7. Summary and future perspectives

In this topical review, we discussed a set of results that provide important insights into ultrafast nonthermal control of quantum materials, in particular, the photoinduced phase transitions to thermodynamically inaccessible states by ultrashort optical excitation in $RTe_3$, $TaS_2$ and $VO_2$. Here, one utilizes light to shift the electronic interactions, break or restore the crystal symmetries to change the balance between competing phases to stabilize novel quantum phases out of equilibrium. Their nonequilibrium behaviors as probed by ultrafast electron scattering have revealed the interplays between the co-control parameters introduced by the pump responsible for the free energy: the local effective temperature and the interaction quench. Useful concepts in conjunction with the Landau-Ginzburg paradigm of description, such as the nonthermal critical point and the local effective temperatures, are introduced to facilitate the understanding of both the heating and interaction effects necessarily to account for the observed evolution of the order-parameter dynamics. In order to realize robust ultrafast control of nonthermal states, the studies here show it is desirable to minimize energy absorption while promoting a controlled modification of the free-energy landscape through different frequency optical quench. The studies also unveil the need for understanding the role of dissipation and properly considering pump inhomogeneity to reach a cross-platform understanding of ultrafast nonthermal controls reported using different pump-probe experimental settings.

As the present paper focuses on the quench dynamics and the relevant nonequilibrium landscape for their description, we intentionally leave out key fields of photoinduced phase transitions mediated by different pump schemes; other nonequilibrium routes to stabilize novel phases of matter include nonlinear phononics and Floquet control under periodic driving; e.g., see recent reviews [272, 273]. In these studies, instead of driving the system through above-the-gap carrier excitations by applying brief high-frequency photons, lower-energy or off-resonance laser fields are preferred to avoid excessive absorption. Furthermore, exploring nonequilibrium universities under a soft quench to the proximities of the nonthermal critical point shall reveal the salient features of the transient energy landscape inaccessible at equilibrium. These topics are ripe for future investigations.

The experiments discussed above demonstrate how the advance of ultrashort electron pulses into brighter and more coherent territories will have a far-reaching impact on nonequilibrium quantum material sciences. While the FEL-based X-ray sources, including the latest development of seeded FEL, will remain the ultimate powerhouse for achieving high resolutions, the ultrafast high-brightness electron sources will revolutionize the research on mesoscopic sciences with integrated multi-messenger approaches. Given the high-brightness electron source and FEL ultrafast instrumentation share many core technologies, such as the precise synchronization between the source and the pump optical fields, the RF accelerator and optical controls, and the detector technologies, the two fields will likely advance together. Indeed, the transformation from integrating the RF accelerator technologies into the UED and the next-generation UEM instrumentation has been tremendous. From the source brightness perspective, the utilities of the electron sources, while still technically challenging to implement to the fullest, remain largely under-explored. This is evident in comparing the performances of UED and UEM plotted in terms of time resolution ($\Delta t_{pp}$) and particle numbers in Fig. 15, summarized from the recent literatures. Here the comparisons are hard to be precise given the broad varieties of system configurations. Nonetheless, for a given type of instrument the trends that $\Delta t_{pp}$ largely follows $N_{e,0}^{1/2}$ as discussed in ML-FMM simulations are quite clear. This gives some confidence in giving projections on the future faces of the technologies with the knowledge from the ML-FMM.

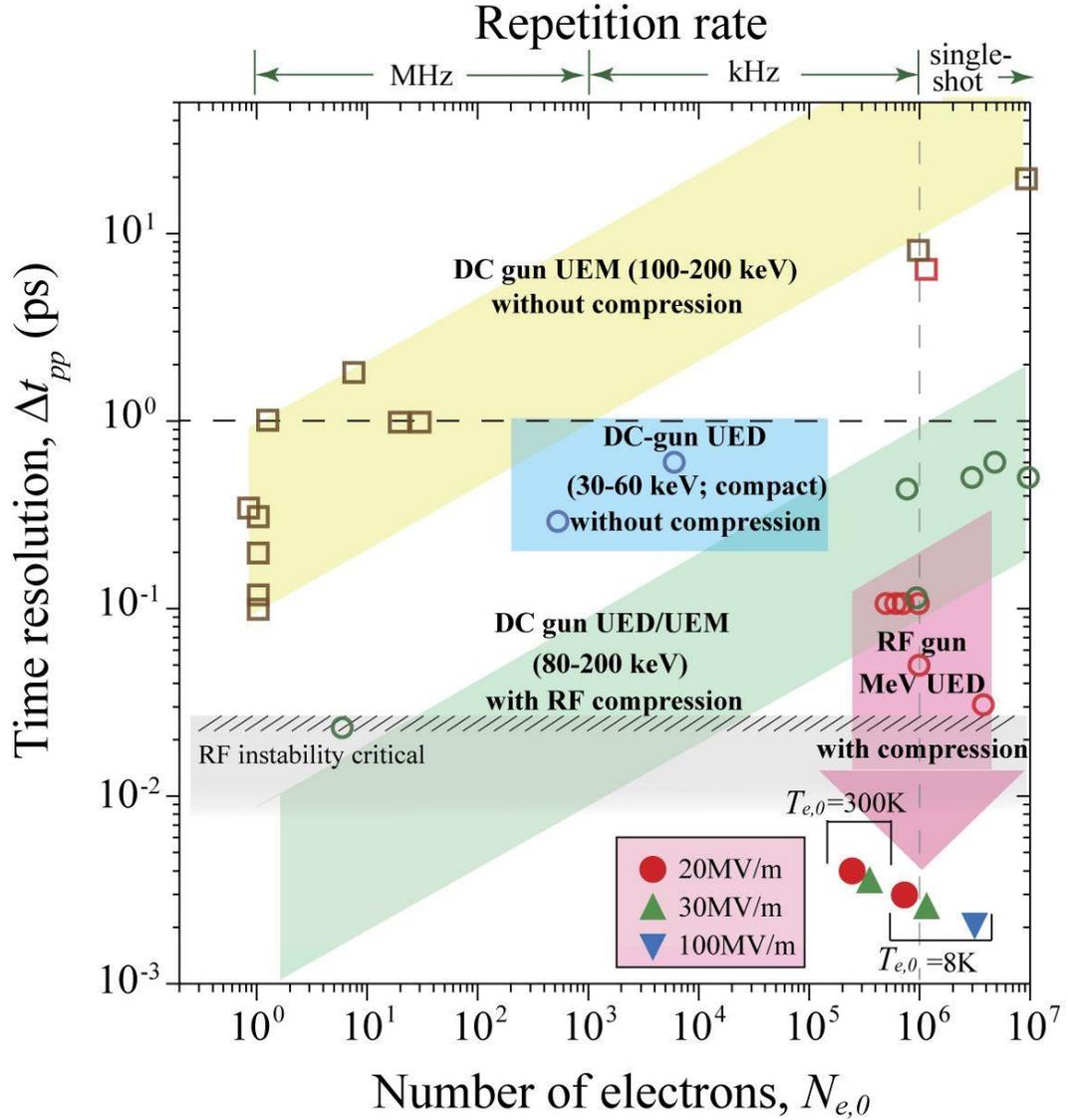

**Figure 15. Performance of UED/UEM instrumentation**. The data are extracted from the recent literatures of the photoemission keV DC gun UED and UEM with ([88, 89, 92, 93]) and without ([84, 93, 96-98, 100, 101, 103-105, 108]) pulse compression and the MeV RF gun UED systems ([82, 83, 86, 87, 91, 99, 206]). The $T_{e,0}$ represents the source effective temperature, a property of the energy spread in the cathode region.

    With the mesoscopic material applications in mind, the future of ultrafast electron sciences shall lie in the high-flux regime with potentially the ability to conduct single-shot experiments. This is because the intrinsic high electron scattering cross-section sets a relatively low bar of the required $N_{e,0}$ of $10^6$-$10^8$ based on applications), which are accessible by both the photoemission DC and RF guns at their respective virtual cathode limit of emission. The implementation of pulse compression schemes easily boosted the temporal resolution into sub-ps regimes at such high-flux limits; see Fig. 15. However, we want to emphasize that in both types of instrumentation, the current levels of performance are strongly influenced by the RF system and the beamline (both optical and electron) stabilities that affect the arrival time and the precision of the time and energy focusing at the samples. The performance can be further improved by a

factor of 10 before reaching the physical limits set by the beam brightness. With a complete control of the electron pulse parameters under advanced compression and streaking schemes, it is possible to obtain pulses with few femtosecond durations with MeV RF UED system and few tens of fs for the keV DC gun systems, operating near the virtual cathode limit. Clearly, the facility-based MeV UED systems will ultimately be the prime options for most dose-hungry and high-temporal-resolution-prone experiments, and likely rival many FEL-type experiments in these directions in the years to come. Putting the ultimate temporal resolution and dose aside, the keV DC gun system, however, will excel in the momentum and energy resolutions. Armed with the matured technologies developed for the current electron microscopes at the familiar beam energy scale, we envision in the decade ahead, a major push will be in obtaining robust resolutions for a true multi-messenger approach that combines diffraction, imaging, and spectroscopy under a single platform. To reach the resolution limits set by the source brightness of the DC-gun designs ($\lesssim 50$ fs, $\lesssim 1$ nm, and $\lesssim 100$ meV levels respectively close to the virtual cathode limit), the development of ingenious schemes to integrate beam stabilization, pump-probe synchronization and implementing suitable electron optics/detector technologies will be the main focus.

# Appendix

**Three-temperature diffusion model (3TDM) calculation.** The 3TDM considers the microscopic energy relaxations following the electronic laser quench in the quantum materials. This model extends on the conventional three-temperature model (3TM)[13, 110-113], where one typically assumes the absorbed photon energy is deposited into the electronic sub-system and internally thermalizes instantaneously. The subsequent relaxation involves energy transfer from the electrons to a subset of phonons strongly coupled to the initial electronic excitation, referred to as strongly coupled phonons (SCP), and also to the rest of the lattice modes that are less efficiently coupled, referred to as weakly coupled phonons (WCP). The dynamics of energy relaxation are described by a set of coupled differential equations with the local temperature $T_i$ and specific heat $C_i$ prescribed to each subsystem. In the simplest form of 3TM without considering the diffusion effects, one may write the coupled rate equation $C_i \partial_t T_i = -G_{i-j}(T_i - T_j) - G_{i-k}(T_i - T_k)$, where $G_{i-j}$ is the coupling constant between the subsystems $i$ and $j$. In this picture, the characteristic timescale for energy transfer $(i \rightarrow j)$ is given by: $\tau_{ij}^{-1} = G_{i-j}/C_i$. Hence, a larger specific heat typically slows down the dynamics of energy transfer from the subsystem. This schematic three-temperature model is given in Fig. A1 where the extent of partition of specific heat between the SCP and WCP is given by the fraction α.

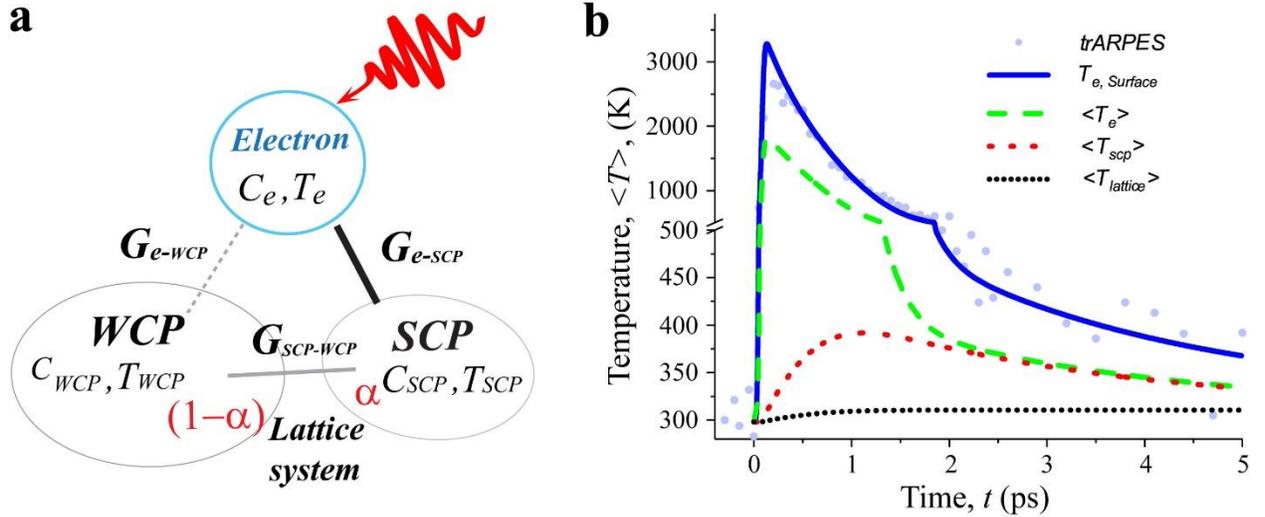

**Figure A1. The effective temperature calculations based on the three-temperature diffusion model. a.** The schematic picture of the three-temperature model with the initial excitation energy from laser first stored in the electronic subsystem, then transferred to the lattice subsystem which is sub-divided into two types of phonon manifolds (SCP and WCP) due to the coupling hierarchy. **b.** The temperatures obtained for different subsystems using the three-temperature diffusion model (3TDM) in thin film geometry[213]. The 3TDM calculates the surface electronic temperature evolution, $T_e(t)$ in solid blue line, for 1T-TaSe$_2$ at 0.69 mJ/cm$^2$ and compared with the data (light blue dots) taken from the trARPES (see Supplementary Materials from ref. [175]). The parameters used in the calculations are listed in Table A1. Alternatively, under the same pump condition, $\langle T_i(t) \rangle$ are obtained for a 45 nm 1T-TaS$_2$ thin film with the bracket denoting averaging, relevant to the study by the ultrafast scattering techniques; see text for discussion.

However, such a model is not complete for an inhomogeneous system driven also by the diffusion effects. This is exemplified in the trARPES and the grazing incidence X-ray diffraction experiments conducted predominantly over the surface regions where the energy dissipation into the bulk interior is important for considering the nonequilibrium phase transition. To this end, the 3TDM considers the non-Fourier thermal diffusion of carriers and phonons[212]. The full description of the 3TDM is given in ref. [213]. Here, we apply the 3DTM to give key predictions about the dynamical behaviors involving different sample thickness settings typical for the UED and trARPES experiments. The calculation is based on a near-infrared (800 nm) pump pulse with a duration of 50 fs at 45° incidence and S-polarization.

The goal here is to illustrate how varying the excitations in the materials (based on solving the Maxwell equation; see Sec. 6.3) may lead to drastically different nonequilibrium temperature relaxation profile in two different thin films (45 and 150 nm). The results will impact the evolution of the developed nonequilibrium phases therein, which are probed by UED and trARPES. For establishing a common baseline, the detailed $T_e(t)$ dataset made available recently by trARPES on 1T-TaSe$_2$ is employed as the target for refining the 3TDM model. The key parameters that reproduce the dynamics are given in Table A.1. Here, the model considers 150 nm and the $T_e(t; z = 0)$ is calculated to reproduce the data; see Fig. A1. Given the three-way dynamical couplings, it is generally difficult to directly isolate the relaxation times from the data, but based on the refined 3TDM parameters and $\tau_{ij}^{-1} = G_{i-j}/C_i$, we may give nominal values of $\tau_{el-SCP}$=0.5 ps, $\tau_{el-WCP}$=1.8 ps, and $\tau_{SCP-WCP}$=40 ps, which are in general agreements with the reports of electron-phonon coupling times of 0.5-4 ps for these systems[18, 19, 188-192]. The same calculation is then conducted for the 45 nm film, where as discussed in Sec. 6.3, and a significantly more homogenous profile can be obtained due to the interference effect. With the UED experiments in mind, here we give the temperatures (noted in bracket) averaged over the entire slab. The lattice temperature is calculated as $T_{lattice} = \alpha T_{SCP} + (1 - \alpha) T_{WCP}$, where $\alpha = 0.1$ is determined from fitting. The time and depth-dependent temperature profiles for the two different films are given in Figs. 11b&c.

To consider the temperature profile away from surface regions, the out-of-plane electron thermal conductivity $K_e$ plays an important role. The nominal value used in the simulation is derived from the out-of-plane resistivity $\rho_\perp$ =1.0×10$^{-5}$ Ωm but the reported value for $\rho_\perp$ differed by an order of magnitude in different measurement geometries; see refs. [274] and [275]. The impact of different $\rho_\perp$ on the excited material temperature profile is evaluated for the 150 nm film. Increasing $K_e$ by 10 ($\rho_\perp$ =1.0×10$^{-6}$ Ωm), the $\sigma_n$ calculated for $T_e$ decreases by 40% whereas decreasing $K_e$ by 10 ($\rho_\perp$ =1.0×10$^{-4}$ Ωm), σ for $T_{lattice}$ increases by 20%. The impacts on the lattice temperature track generally with $T_e$ changes. These calculations are taken at 4 ps when $T_e$ and $T_{lattice}$ reach a fair level of thermalization. The calculation here shows that the different $\rho_\perp$ values reported in the literature have some effects on the pump inhomogeneity profile. In general, reducing $\rho_\perp$ will lead to a higher homogeneity in the temperature profiles. This is relevant as one generally expects that $\rho_\perp$ will be smaller in the pumped state – a phenomenon that should be resolvable by studying long-time relaxation dynamics with the help of 3TDM modeling.

**Table A1. Parameters for three-temperature diffusion model**

| Name | Meaning | Parameters used for 1T-TaS$_2$ |
|---|---|---|
| $C_e = \gamma T_e$ | Electron heat capacity | $\gamma$ =12.96 Jm$^{-3}$ K$^{-2}$ (ref.[276]) |
| $\tau_e$ | Electron relaxation time | 0.2 ps (ref. [214]) |
| $K_e = \frac{LrtzT_e}{\rho_\perp}$ ¶ | Out-of-plane electron thermal conductivity | $\rho_\perp$ =1.0×10$^{-5}$ Ωm (refs. [274] and [275]) |
| $K_l$ # | Out-of-plane lattice thermal conductivity | 0.5 (W m$^{-1}$ K$^{-1}$) (ref. [274] and [277]) |
| $C_{tot}$ | Total nuclear heat capacity | 1.85×10$^6$ (J m$^{-3}$ K$^{-1}$) (ref. [190]) |
| $\alpha$ | Strongly coupled phonon fraction | 0.1 (Ref. [213]) |
| $G_{el-scp.}$ | Coupling constant between electron and strongly coupled phonons | 2.5×10$^{16}$ Wm$^{-3}$ K$^{-1}$ |
| $G_{el-wcp.}$ | Coupling constant between electron and weakly coupled phonons | 1.0×10$^{14}$ Wm$^{-3}$ K$^{-1}$ |
| $G_{scp.-wcp.}$ | Coupling constant between strongly coupled phonons and weakly coupled phonons | 6.0×10$^{16}$ Wm$^{-3}$ K$^{-1}$ |

¶ The calculation of electronic thermal conductivity follows the Wiedemann-Franz law. Lrtz is the Lorentz number, and Lrtz=2.44×10$^{-8}$ WK$^{-2}$

# The out-of-plane lattice thermal conductivity is taken to be 1/10 of the in-plane value reported in Ref. [277], based on the understanding that the two typically differs by an order of magnitude[274].

## Acknowledgements

The authors acknowledge the many helpful discussions with M. Berz, K. Chang, P.M. Duxbury, M. Eckstein, I.R. Fisher, J. Freericks, Y.A. Gerasimenko, R.H. Haglund, T.R.T. Han, M.G. Kanatzidis, R. Kandle, A.F. Kemper, S. Lund, M. Maghrebi, S.D. Mahanti, K. Makino, C.D. Malliakas, D. Mihailovic, R. Murdick, K. Nasu, J. Portman, N. Sepulveda, T. Sun, Z. Tao, S. Wall, J. Williams, B. Zerbe, H. Zhang, M. Zhang, F. Zhou. The work was funded by the U.S. Department of Energy, Grant DE-FG0206ER46309. The experimental facility was supported by U.S. National Science Foundation, Grant DMR 1625181.